\documentclass[letter, 12pt]{article}
\usepackage{a4wide, amsmath, amsfonts, amssymb, geometry, shuffle, mathtools}

\usepackage{bm}
\usepackage{fancyhdr, dsfont}
\usepackage[latin1]{inputenc}
\usepackage[pdftex]{graphicx} 
\usepackage{graphicx}
\usepackage{float}
\usepackage{array}
\usepackage[sc,small]{caption}
\usepackage{nccmath}
\usepackage{enumitem}
\usepackage[usenames,dvipsnames]{xcolor}

\usepackage{jheppub}

\usepackage{color}
\definecolor{dgreen}{rgb}{0,0.70,0.30}
\definecolor{gold}{rgb}{0.85,.66,0}
\definecolor{purple}{rgb}{1.0,0.3,0.6}
\definecolor{mathgreen}{rgb}{0,.5,0}

\usepackage{tikz}
\usetikzlibrary{calc} \usetikzlibrary{patterns} \usetikzlibrary{decorations.pathreplacing} \usetikzlibrary{decorations.markings} \usetikzlibrary{decorations.pathmorphing} \usetikzlibrary{positioning}

\allowdisplaybreaks

\restylefloat{figure}
\usepackage{epstopdf}

\def\beq{\begin{equation}}
\def\eeq{\end{equation}}
\let\Re\relax
\let\Im\relax
\DeclareMathOperator{\Re}{Re}
\DeclareMathOperator{\Im}{Im}

\newcommand{\dd}{\mathrm{d}}
\newcommand{\te}{\textrm}
\newcommand{\ap}{\alpha'}
\newcommand{\vph}{\varphi}
\newcommand{\pp}{\mathfrak{p}}

\newcommand{\nn}{\nonumber}
\newcommand{\tildeF}{J}

\newcommand{\vecb}{\left(\begin{array}{c}}
\newcommand{\vece}{\end{array}\right)}
\newcommand{\ccb}{\left(\begin{array}{cc}}
\newcommand{\cce}{\end{array}\right)}
\newcommand{\cccb}{\left(\begin{array}{ccc}}
\newcommand{\ccce}{\end{array}\right)}
\newcommand{\ccccb}{\left(\begin{array}{cccc}}
\newcommand{\cccce}{\end{array}\right)}
\newcommand{\cccccb}{\left(\begin{array}{ccccc}}
\newcommand{\ccccce}{\end{array}\right)}
\newcommand{\ccccccb}{\left(\begin{array}{cccccc}}
\newcommand{\cccccce}{\end{array}\right)}

\newcommand{\CP}{\mathbb{CP}}

\newcommand{\QQ}{\mathbb Q}

\newcommand{\be}{\begin{equation}}
\newcommand{\ee}{\end{equation}}
\newcommand{\bea}{\begin{eqnarray}}
\newcommand{\eea}{\end{eqnarray}}

\renewcommand{\varphi}{\hat{\omega}}

\title{Coaction and double-copy properties of configuration-space integrals at genus zero}
\author[a,b]{Ruth Britto,}\emailAdd{britto@maths.tcd.ie}
\author[c]{Sebastian Mizera,}\emailAdd{smizera@ias.edu}
\author[d]{Carlos Rodriguez,}\emailAdd{carlos.rodriguez@physics.uu.se}
\author[d]{Oliver Schlotterer}\emailAdd{oliver.schlotterer@physics.uu.se}
\affiliation[a]{School of Mathematics and Hamilton Mathematics Institute, Trinity College, \\ Dublin 2, Ireland}
\affiliation[b]{Institut de Physique Th{\'e}orique, Universit\'e Paris Saclay, 
CEA, CNRS, \\ F-91191 Gif-sur-Yvette cedex, France}
\affiliation[c]{Institute for Advanced Study, Einstein Drive, Princeton, NJ 08540, USA}
\affiliation[d]{Department of Physics and Astronomy, Uppsala University, 75108 Uppsala, Sweden}

\abstract{
We investigate configuration-space integrals over punctured Riemann spheres from the viewpoint of the motivic Galois coaction and double-copy structures generalizing the Kawai--Lewellen--Tye (KLT) relations in string theory.
For this purpose, explicit bases of twisted cycles and cocycles are worked out whose orthonormality simplifies the coaction.
We present
methods to efficiently perform and organize the expansions of configuration-space integrals in the inverse string tension $\alpha'$ or the dimensional-regularization parameter $\epsilon$ of Feynman integrals. Generating-function techniques open up a new perspective on the coaction of multiple polylogarithms in any number of variables and analytic continuations in the unintegrated punctures. We present a compact recursion for a generalized KLT kernel and discuss its origin from intersection numbers of Stasheff polytopes and its implications for correlation functions of two-dimensional conformal field theories. We find a non-trivial example of correlation functions in $(\mathfrak{p},2)$ minimal models, which can be normalized to become uniformly transcendental in the $\mathfrak{p} \to \infty$ limit.
}

\preprint{TCDMATH 21-06\\ \null\hfill IPhT-t21/030\\ \null\hfill UUITP-08/21}
\arxivnumber{2102.06206}

\begin{document}

\setcounter{tocdepth}{2}
\maketitle
\setcounter{page}{3}


\numberwithin{equation}{section}




\section{Introduction}

Recent studies of scattering amplitudes revealed a wealth of mathematical structures that initiated
a fruitful crosstalk between particle phenomenology, string theory, algebraic geometry and number theory.
Iterated integrals such as multiple polylogarithms and multiple zeta values (MZVs) became a common theme
of Feynman integrals and low-energy expansions of string amplitudes. In a broad spectrum of physical settings, 
dramatic simplifications and striking connections between seemingly unrelated theories have been found on the basis of the Hopf-algebra structures of polylogarithms and MZVs.

Most prominently, amplitudes in a variety of theories were observed to exhibit universal stability properties
under the motivic Galois coaction of polylogarithms \cite{Goncharov:2001iea, Goncharov:2005sla}. These observations
support the coaction conjecture or {\it coaction principle} \cite{Schnetz:2013hqa, Brown:2015fyf, Panzer:2016snt, Schnetz:2017bko}  which states that certain
classes of amplitude building blocks close under
the motivic Galois coaction. So far, the coaction principle was found to apply to disk integrals in open-string tree-level amplitudes \cite{Schlotterer:2012ny}, periods in $\phi^4$ theory \cite{Panzer:2016snt}, the anomalous magnetic moment of the electron \cite{Schnetz:2017bko}, six-point amplitudes in ${\cal N}=4$ super Yang--Mills theory \cite{Caron-Huot:2019bsq}, various families of Feynman integrals \cite{Abreu:2014cla,Abreu:2015zaa,Abreu:2017enx,Abreu:2017mtm,Abreu:2019eyg,Tapuskovic:2019cpr} and related  hypergeometric functions \cite{Brown:2019jng, Abreu:2019xep}. 

The primary goal of this work is to extend the coaction principle in string tree-level amplitudes to more general configuration-space integrals at genus zero where not all of the punctures on the Riemann sphere are integrated over. This relates to the incarnation of the coaction principle in generalized hypergeometric functions through the similarity of their representations as Euler-type integrals amenable to the formalism of \cite{AomotoKita}. 
In the context of both string scattering \cite{Mizera:2017cqs, Mizera:2019gea} and hypergeometric integrals (see for instance \cite{Oprisa:2005wu, Puhlfuerst:2015gta} for earlier work on their connections), the underlying generalized disk
integrals are dual pairings of twisted homologies and cohomologies. 
For a given homology representative $\gamma$  and cohomology representative  $\omega$ in these spaces, 
the coaction of the dual pairing given by the integral $\int_\gamma \omega$ is conjectured to take the form \cite{Abreu:2017enx,Abreu:2017mtm}
\begin{equation}
\Delta \int_\gamma \omega = \sum_{a,b=1}^{d} c_{ab} \int_\gamma \omega_a \otimes \int_{\gamma_b} \omega \,,
\label{intr.1}
\end{equation}
where the $\{ \omega_a \}$ and $\{\gamma_b\}$ respectively 
generate the twisted (co-)homology group of dimension $d$.
The coefficients $c_{ab}$ are rational functions fixed by the choice of bases. 
In this paper, we will present a natural construction of such bases in the case of the generalized disk integrals associated to tree-level string scattering, with the nice property that the coefficients $c_{ab}$ form the identity matrix.

The master formula (\ref{intr.1}) 
can be viewed as a generating function of coaction identities for polylogarithms and MZVs. In the string-theory incarnation of these integrals, the coaction acts order by order in the expansion
with respect to the inverse string tension $\alpha'$, or more precisely with respect to the dimensionless quantities $2\alpha' k_i\cdot k_j$
with lightlike momenta $k_i$.
For  hypergeometric functions associated to dimensionally-regularized Feynman integrals, however, the analogous expansion is with respect to the dimensional-regularization parameter $\epsilon$.
The formal analogy between $\alpha'$ and $\epsilon$ has already been noticed by comparing differential equations of Feynman integrals and configuration-space integrals of string amplitudes at genus zero \cite{Henn:2013pwa, Broedel:2013aza} and at genus one \cite{Adams:2018yfj, Mafra:2019ddf, Mafra:2019xms}, as well as in the context of twisted cohomology \cite{Mastrolia:2018uzb,Frellesvig:2019kgj,Frellesvig:2019uqt,Abreu:2019wzk,Mizera:2019vvs,Mizera:2020wdt,Frellesvig:2020qot}. The discussion of this work only applies to the genus-zero case while leaving important extensions to non-polylogarithmic integrals to the future.

The main results in this work are:
\begin{itemize}[leftmargin=*]
\item To give explicit pairs of orthonormal bases $\{ \gamma_a \}$ and $\{\omega_b\}$ in (\ref{intr.1}) for generalized disk integrals over
any number of punctures, while leaving an arbitrary number of additional punctures unintegrated.
\item To describe systematic methods of generating the uniformly transcendental $\alpha'$- or $\epsilon$-expansions of the basis integrals
$\int_{\gamma_a} \omega_b$ in terms of multiple polylogarithms and MZVs.
\item To organize the multiple polylogarithms and MZVs contributing to the $d  \times d $ matrix $\int_{\gamma_a} \omega_b$ into matrix products
\begin{equation}
\int_{\gamma_a} \omega_b(z_1,z_2,\ldots,z_\ell) = \!  \! \! \! \! \sum_{c_1,c_2,\ldots,c_\ell=1}^{d}  \! \! \!  \!  \!
 \mathbb G(1)_{   ac_1 } \mathbb G(z_\ell)_{ c_1c_2} \mathbb G(z_{\ell-1})_{c_2c_3}\ldots
\mathbb G(z_2)_{c_{\ell-1}c_\ell}  \mathbb G(z_1)_{   c_\ell b}
\label{intr.2}
\end{equation}
Each factor of $\mathbb G(z_j)$ is by itself a matrix-valued series in $\alpha'$ or $\epsilon$, with
polylogarithms at the same argument $z_j$ in its coefficients (such that $\mathbb G(1)$
is a series of MZVs similar to those in open-string tree amplitudes \cite{Schlotterer:2012ny})
and letters to be spelt out below.
\item To refine the coaction formula (\ref{intr.1}) to the individual factors in (\ref{intr.2}), 
\begin{equation}
\Delta \mathbb G(z_j) =  \mathbb G(z_j) \times {\rm ad}_{\rm L}\big(
\mathbb G(1)  \mathbb G(z_\ell)  \mathbb G(z_{\ell-1}) \ldots
\mathbb G(z_{j+1})
\big) \mathbb G(z_j)
\label{intr.3}
\end{equation}
where the operation ${\rm ad}_{\rm L}$ will be defined below and
the contributions from MZVs obey the particularly simple special case
$\Delta \mathbb G(1) = \mathbb G(1) \otimes \mathbb G(1)$.
\item To explore the analytic continuation between configurations changing the order of unintegrated punctures on the real axis. Such deformations can be compactly described by braid matrices acting on a vector of disk integrals and are relevant to the study of monodromies and discontinuities of polylogarithmic Feynman integrals \cite{AIHPA_1967__6_2_89_0,Abreu:2014cla,Abreu:2017ptx,Bourjaily:2020wvq,Corcoran:2020epz}.
\end{itemize}

Another place in physics where identical integrals appear is in the context of conformal field theories in the Coulomb gas formalism \cite{Dotsenko:1984ad,Dotsenko:1984nm}. On the one hand, their conformal blocks are integrals of the type $\int_{\gamma_a} \omega_b$, where a subset of punctures is fixed while the remaining ones are integrated. 
On the other hand, the full correlation functions are given by sphere integrals, schematically $\int_{{\cal C}^{(n,p)} } \bar \omega_a \omega_b$.
The integration domain ${\cal C}^{(n,p)} $ is the
configuration space of $p$ punctures on a sphere with $n{-}p$ points removed. 

We point out an interesting phenomenon in which correlation functions of $(\pp, \pp')$ minimal models in the $\pp \to \infty$ limit (with $\pp'$ fixed and finite) behave as either the $\alpha' \to 0$ or $\alpha' \to \infty$ limit of string amplitudes, depending on whether charges of conformal primary operators decay or grow in this limit. For $(\pp,2)$ models specifically, we find examples of correlation functions exhibiting the uniform-transcendentality principle in the large-$\pp$ expansion, familiar from the 
$\alpha'$-expansion of superstring amplitudes and 
$\epsilon$-expansion of Feynman integrals.

The punctured sphere also naturally appears in the context of gauge-theory scattering. In particular, in the multi-Regge limit of planar ${\cal N}=4$ super Yang--Mills theory, it arises as a kinematic configuration space where the punctures are associated to the momenta of external scattering states. Motivated by this observation, amplitudes for arbitrary number of loops and legs are given in terms of single-valued multiple polylogarithms~\cite{Dixon:2012yy,DelDuca:2016lad,DelDuca:2019tur}. Similar functional dependence can be seen in the high-energy limit of dijet scattering for generic gauge theories~\cite{DelDuca:2013lma,DelDuca:2017peo}.

At this stage one may take inspiration from string theory, where the case of sphere integrals with three unintegrated punctures form the backbone of closed-string tree-level amplitudes. These sphere integrals are related to the disk integrals of open strings 
in two complementary ways: 
\begin{itemize}[leftmargin=*]
\item By the Kawai--Lewellen--Tye (KLT) relations \cite{Kawai:1985xq}, the sphere integrals
$\int_{{\cal C}^{(n,n-3)} } \bar \omega_a \omega_b$ boil down to bilinears in disk integrals
$\int_{\gamma_c }  \omega_a   \int_{\gamma_d} \omega_b$ weighted by trigonometric
functions of $\alpha'$ built from inverse intersection numbers \cite{Mizera:2017cqs}.
\item At the level of the MZVs in their $\alpha'$-expansion, closed-string integrals
$\int_{{\cal C}^{(n,n-3)} } \bar \omega_a \omega_b$
are single-valued images \cite{Schnetz:2013hqa, Brown:2013gia} of disk integrals
\cite{Schlotterer:2012ny,Stieberger:2013wea,Stieberger:2014hba, Schlotterer:2018zce, Vanhove:2018elu,Brown:2019wna} $\int_{\gamma_a} \omega_b$ of open strings with suitably chosen integration contours $\gamma_a$.
\end{itemize}
Another key achievement of this work is to generalize both the KLT relations and
the single-valued map between disk and sphere integrals to ${\cal C}^{(n,p)} $
with $p <n{-}3$, i.e.\ more than three unintegrated punctures. In these cases, the coefficients in the $\alpha'$-expansions augment
single-valued MZVs by single-valued polylogarithms in one variable \cite{svpolylog} ($p=n{-}4$) 
or multiple variables \cite{Broedel:2016kls, DelDuca:2016lad} ($p\leq n{-}5$). An independent approach to the generalized KLT kernel at $p=n{-}4$
relating the momentum-kernel formalism \cite{momentumKernel} to the single-valued map 
can be found in \cite{Vanhove:2018elu}. 

For any number of integrated punctures $p$ and unintegrated
ones $n{-}p$, we will spell out the explicit form of the KLT-relations
between ${\cal C}^{(n,p)} $-integrals and products of generalized
disk integrals and their complex conjugates. For a convenient choice
of bases for the twisted integration cycles of the disk integrals, we
present an efficient recursion for the generalized ``KLT kernel'' that 
determines the coefficients in their bilinears. The generalized KLT kernel
is again the inverse of an intersection matrix with trigonometric functions
in its entries which we derive from adjacency properties of Stasheff polytopes \cite{10.2307/1993608}.
Our results furnish an explicit realization of several of the general mathematical concepts relating double copy, single-valued integration and string amplitudes \cite{Brown:2018omk, Brown:2019wna}. Many all-multiplicity statements in this work are left as conjectures, and we hope that the ideas of the references set the stage to find rigorous proofs.

This work is organized as follows: The basic definitions of the configuration-space integrals under investigation and the explicit form of their orthonormal bases of cycles $\{ \gamma_a \}$ and forms $\{\omega_b\}$ are given in section \ref{sec:2}. We then discuss the structure of and practical tools for the $\alpha'$-expansions of $\int_{\gamma_a} \omega_b$ in section \ref{sec:3} and introduce their polylogarithmic building blocks $\mathbb G(z_j)$ in (\ref{intr.2}).
In section \ref{sec:4}, the coaction (\ref{intr.1}) of the integrals is translated into that of the generating series $\mathbb G(z_j)$ of polylogarithms, and we 
derive the operation ${\rm ad}_{\rm L}$ in (\ref{intr.3}) in detail. Section \ref{sec:5} is dedicated to the analytic continuation of $\int_{\gamma_a} \omega_b$ in the unintegrated punctures.

In section \ref{sec:6}, complex integrals $\int_{ {\cal C}^{(n,p)}}\overline\omega_a \omega_b$ are discussed from the perspectives of the single-valued map, intersection numbers and compact recursions for a KLT kernel. Finally, the implications for correlation functions of minimal models in the Coulomb-gas formalism can be found in section \ref{sec:7}.
Further details and examples of $\alpha'$-expansions and analytic continuations are relegated to two appendices.
 
\section{Orthonormal bases of forms and cycles}
\label{sec:2}

In this section we introduce orthonormal bases of differential forms and integration cycles. In order to do so, we start with reviewing the relevant notation and explaining why such bases are needed in the first place.
We discuss the well-established case of
a single integration variable to set
the stage for our general formula
and verify orthonormality using intersection theory.

Let us consider a genus-zero Riemann surface, $\CP^1 = \mathbb{C} \cup \{\infty\}$. The arena in which the integrals of our interest are defined is the configuration space of $p$ points on a sphere with $n{-}p$ punctures:
\be
\mathcal{C}^{(n,p)} = \mathrm{Conf}_{p}(\CP^1 - \{ n{-}p \;\mathrm{ points} \}) \ .
\ee
In other words, out of the total $n$ punctures, $p$ are dynamical and are allowed to be moved/integrated, while $n{-}p$ are frozen in their positions. This space has $p$ complex dimensions. We assume $1 \leq p \leq n{-}3$ and denote the inhomogeneous coordinates of each puncture by $z_i$ for $i=1,2,\ldots,n$. As the integrals of our interest are conformally invariant, we will work in the $\mathrm{SL}(2,\mathbb{C})$-frame with
\be
(z_1,z_{n-1},z_n) = (0,1,\infty) \ .
\label{conv1}
\ee
We will use the convention in which $z_2,z_3,\ldots,z_{p+1}$ are the integrated punctures. In these coordinates we can write explicitly
\begin{align}
{\cal C}^{(n,p)} = \{ (z_2,z_3,\ldots, z_{p+1}) \in \mathbb{C}^{p} \,|\, z_i \neq z_1, z_{i+1}, z_{i+2}, \ldots, z_{n-1} \;\mathrm{for\; all}\; i=2,3,\ldots,p{+}1\},\label{C-np}
\end{align}
since we fixed one puncture to infinity.
We next introduce the generalized Koba--Nielsen factor
\begin{align}
{\rm KN}^{(n,p)} 
&= \prod_{2 \leq i \leq p+1} 
\left(|z_{1i}|^{s_{1i}}
\prod_{i < j \leq n-1} |z_{ij}|^{s_{ij}} 
\right)
\notag \\
&= \Big( \prod_{2\leq i < j }^{p+1} |z_{ij}|^{s_{ij}} \Big) 
\Big(\prod_{\ell=2}^{p+1} |z_\ell|^{s_{1\ell}} |1{-}z_\ell|^{s_{\ell,n-1}} \Big)
\Big( \prod_{k=2}^{p+1}\prod_{m=p+2}^{n-2} |z_{km}|^{s_{km}} \Big) \ ,
\label{conv2}
\end{align}
where differences between positions of punctures are denoted by 
\beq
z_{ij} = z_i{-}z_j
\eeq
and $s_{ij}$ are real variables that might take different meanings depending on the physical application. In the context of string perturbation theory at genus zero, for instance, we can take them to be the dimensionless Mandelstam invariants
\beq
s_{ij} = 2\ap k_i \cdot k_j 
\label{conv2a}
\eeq
for light-like momenta $k_i$ and inverse string tension $\ap$. The naming comes from the fact that in the case $p=n{-}3$, where all but three punctures are integrated, \eqref{conv2} reduces to the Koba--Nielsen factor in the integrand of string tree-level amplitudes.
Note that our definition (\ref{conv2}) omits the $z_{ij}$ for pairs of unintegrated punctures, $i,j=1,p{+}2,p{+}3,\ldots,n$, since they could be universally pulled out of all the integrals at fixed $n,p$. We also assume that $s_{ij}$ are generic real numbers or formal variables.

\subsection{Main ingredient: Disk integrals} \label{sec:2.0}

We are interested in the  matrices of contour integrals
$F^{(n,p)}_{ab}$, defined by
\beq
F^{(n,p)}_{ab}   =  \langle \gamma_a^{(n,p)} | \omega_b^{(n,p)} \rangle =
\int_{\gamma_a^{(n,p)}}
{\rm KN}^{(n,p)}\, \omega_b^{(n,p)} \ ,
\label{conv5}
\eeq
where $\gamma^{(n,p)}_a$ and $\omega^{(n,p)}_b$ denote integration cycles and holomorphic $p$-forms corresponding to bases of twisted homology and cohomology groups, respectively, for the twist 1-form given by $\dd \log {\rm KN}^{(n,p)}$. Through  $\gamma^{(n,p)}_a$ and $\omega^{(n,p)}_b$, the integrals $F^{(n,p)}_{ab}$
depend on punctures or cross-ratios $z_{p+2},\ldots,z_{n-2}$ and the 
Mandelstam invariants (\ref{conv2a}).
The integrals in (\ref{conv5}) are of the form exhibited in the coaction formula (\ref{intr.1}), where in the integrand we have now  explicitly separated the twist factor ${\rm KN}^{(n,p)}$, and the remaining single-valued form is now denoted by $\omega_b^{(n,p)}$.

The indices $a,b$ in (\ref{conv5}) run from $1$ to the dimensions $d^{(n,p)}$ of the associated twisted (co-)homologies \cite{aomoto1987gauss,Mizera:2019gea}\footnote{More generally, the Poincar{\'e} polynomial of ${\cal C}^{(n,p)}$ is given by ${\mathtt P}^{(n,p)}(t) = \prod_{k=n-p-1}^{n-2} (1 + kt)$, which follows from a simple extension of the arguments given in \cite{arnold1969cohomology}. The dimension of the only non-trivial $p$-th twisted cohomology is equal to $(-1)^p{\mathtt P}^{(n,p)}(-1) = \tfrac{(n-3)!}{(n-3-p)!}$, which is smaller than that of the ordinary (untwisted) $p$-th cohomology, $\tfrac{1}{p!}\partial_t^{p} {\mathtt P}^{(n,p)}(0) = \tfrac{(n-2)!}{(n-2-p)!}$, which in turn is even smaller than the total number of possible real cycles (chambers in the real slice of ${\cal C}^{(n,p)}$) \cite{zaslavsky1975facing} given by ${\mathtt P}^{(n,p)}(1) = \tfrac{(n-1)!}{(n-1-p)!}$.}
\beq
d^{(n,p)} = \frac{ (n{-}3)! }{(n{-}3{-}p)!} \, ,
\label{conv3}
\eeq
which, up to a sign, are the Euler characteristics of the configuration spaces ${\cal C}^{(n,p)}$.

The twisted cycles $\gamma^{(n,p)}_a$ can be taken to be regions of the real section of $\mathcal{C}^{(n,p)}$, whose boundaries are 
  contained in the union of hyperplanes $\left\{z_{ij}=0\right\}$ appearing in the Koba--Nielsen factor ${\rm KN}^{(n,p)}$.  The unintegrated punctures $z_1,z_{p+2},z_{p+3},\ldots,z_{n-1}$ can be assigned a fixed order on the real axis. We will always take 
  \beq
  0=z_1<z_{p+2}<z_{p+3}<\cdots<z_{n-2}<z_{n-1}=1\,,
  \label{order-unintegrated}
\eeq
except for the discussions of analytic continuations in section \ref{sec:5}.

Twisted cohomologies give a geometric description of the equivalence classes of integrands $\omega_b^{(n,p)}$, up to total derivative terms:
\be\label{equiv}
\omega_b^{(n,p)} \;\cong\; \omega_b^{(n,p)} + (\dd+\dd\log\mathrm{KN}^{(n,p)}\wedge)\xi
\ee
for any $(p{-}1)$-form $\xi$. Both sides of (\ref{equiv}) integrate to the same result, since boundary terms as $z_i \rightarrow z_j$ are suppressed by the Koba--Nielsen factor, and can hence be treated as being equivalent. The representatives of the twisted cohomology classes are holomorphic $p$-forms with poles only at $z_{i}=z_{j}$.
We will often strip the overall differential, so that the differential forms in (\ref{conv5}) are written as 
\beq
\omega^{(n,p)}_b= \vph^{(n,p)}_b \prod_{k=2}^{p+1} \dd z_k \ ,
\eeq
where the functions $\vph^{(n,p)}_b$ are Laurent polynomials in the variables $z_{ij}$. Let us see how the equivalence relations \eqref{equiv} translate to these functions. The simplest case would be to consider any closed form $\xi$ ($\dd\xi = 0$), which can be written generally as
\be
\xi = \sum_{i=2}^{p+1} \hat{\xi}_{i} \prod_{\substack{k=2\\ k\neq i}}^{p+1} \dd z_k \qquad\text{with}\qquad \partial_{i} \hat{\xi}_i = 0 \quad\forall \ i=2,3,\ldots,p{+}1 \ .
\ee
Here we introduced the short-hand notation $\partial_i = \partial/\partial z_i$.
Together with \eqref{equiv}, it implies that any $\vph^{(n,p)}_b$ can be shifted by terms of the form
\beq
(\partial_i \log \mathrm{KN}^{(n,p)} )\, \hat{\xi}_i \;=\; \Big( \sum_{\substack{j=1\\  j \neq i}}^{n-1} \frac{s_{ij}}{z_{ij}} \Big)\, \hat{\xi}_i
\label{kndrv}
\eeq
for any $i$. Throughout this work the symbol $\cong$ will denote equality up to such equivalence relations (relations with $\dd\xi \neq 0$ will not be needed in our applications).
 
We would like to choose bases of cycles $\gamma^{(n,p)}_a$ and cocycles $\omega^{(n,p)}_b$, for $1 \leq a,b \leq d^{(n,p)}$,
to yield orthonormal field-theory limits 
\beq
\lim_{\alpha' \rightarrow 0} F^{(n,p)}_{ab}  = \delta_{ab} \, .
\label{conv6}
\eeq
If the condition (\ref{conv6}) is satisfied, a coaction formula of the following form is claimed \cite{Abreu:2018nzy, Abreu:2019xep}:
\beq
\Delta F^{(n,p)}_{ab} = \sum_{c=1}^{d^{(n,p)}} F^{(n,p)}_{ac} \otimes F^{(n,p)}_{cb}  \, ,
\label{conv7}
\eeq
consistent with the coaction of terms in the $\ap$-expansion. 
At $p=n{-}3$, this specializes to the results of \cite{Schlotterer:2012ny, Drummond:2013vz} on the $\ap$-expansion of open-string tree-level amplitudes. As a practical advantage of orthonormal field-theory limits (\ref{conv6}), they minimize the number of
terms in the coaction: One can identify (\ref{conv7}) as a special case of the master formula (\ref{intr.1}) with $c_{ab} = \delta_{ab}$ and therefore $d^{(n,p)}$ in place of the $(d^{(n,p)})^2$ summands that would arise for generic bases of $\gamma^{(n,p)}_a$ and $\omega^{(n,p)}_b$.
Moreover, the (factorially growing) numbers of terms in the expressions below for $\omega^{(n,p)}_b$ are tailored to
remove kinematic poles from the entire $\ap$-expansion of $F^{(n,p)}_{ab}$ and to simplify the expressions at each order.
With this motivation in mind, we now propose a pair of bases at general $n$ and $p$ satisfying the condition (\ref{conv6}).

\subsection{One integrated puncture} 
\label{sec:2.1}

As a warm-up, consider first the case of $p=1$ with a single integration variable, $z_2$, and we have $d^{(n,1)}=n{-}3$. The integrals $F^{(n,1)}_{ab}$ are then closely related\footnote{The difference is the absence of gamma-function prefactors in this work. The coaction for gamma functions can easily be incorporated as desired according to the treatment in \cite{Abreu:2019xep}.} to Lauricella functions $F_D^{n-4}$, for which a coaction was given in \cite{Brown:2019jng, Abreu:2019xep}.
By the ordering (\ref{order-unintegrated}) of the unintegrated punctures on the real line, it is thus natural to choose the following basis of integration contours for $z_2$, which are simply the intervals bounded by consecutive finite punctures,
\begin{align}\label{p1contours}
\gamma^{(n,1)}_{1} &= \{ 0<z_2<z_3\} \, ,
\qquad
\gamma^{(n,1)}_{n-3}=\{ z_{n-2}<z_2<1 \}  \, ,
\\
\gamma^{(n,1)}_{a} &= \{ z_{a+1}<z_2<z_{a+2}\} \quad\textrm{for}\quad 2\leq a \leq n{-}4\,.
\notag
\end{align}
Now we would like to identify a set of forms $\omega^{(n,1)}_b= \dd z_2 \vph^{(n,1)}_b$ that are Laurent polynomials in the variables $z_{2i}$ and satisfy the duality condition (\ref{conv6}) with this set of contours. The functions $\vph^{(n,1)}_b$ can be chosen to have only simple poles, as follows.
\begin{align}
\vph^{(n,1)}_1 &= \frac{s_{21}}{z_{21}} \, , \qquad
\vph^{(n,1)}_{n-3} = \frac{s_{21}}{z_{21}}
+ \sum_{j=3}^{n-2} \frac{s_{2j}}{z_{2j}}\, , \label{p1example}
\\
\vph^{(n,1)}_b&= \frac{s_{21}}{z_{21}}+\sum_{j=3}^{b+1} \frac{s_{2j}}{z_{2j}} \quad\textrm{for}\quad 2\leq b \leq n{-}4\, .
\notag
\end{align}
From the pole structure of these $\omega^{(n,1)}_b$, it is now easy to see that they are dual to the set of contours in (\ref{p1contours}). Contributions to the $\alpha' \rightarrow 0$ limit of the integral $F^{(n,1)}_{ab}$ arise only when the poles coincide with the endpoints of integration. The logarithmic divergence at such an endpoint, say $z_i$, is regulated by the Koba--Nielsen factor, resulting in a contribution of $s_{2i}^{-1}$, cancelling the numerators in the differential forms.
Thus the contributions from the poles are either absent or cancel pairwise except when $a=b$. 
Adding a Koba--Nielsen derivative to \eqref{p1example} yields an alternative set of cohomology representatives,
\begin{align}
\vph^{(n,1)}_1 &\cong \sum_{j=3}^{n-1} \frac{s_{j2}}{z_{j2}} \,
, \qquad
\vph^{(n,1)}_{n-3} \cong 
\frac{s_{n-1,2}}{z_{n-1,2}} \, , \label{ibpequiv}
\\
\vph^{(n,1)}_b &\cong \sum_{j=b+2}^{n-1} \frac{s_{j2}}{z_{j2}}
\quad\textrm{for}\quad 2\leq b \leq n{-}4\ ,
\notag
\end{align}
which we will sometimes find more convenient in specific calculations below.

\subsection{The general case} 
\label{sec:2.2}

For the general case $(n,p)$ of (\ref{conv5}), we select the basis of twisted cycles to correspond to regions labeled by distinct real orderings of the $p$ integrated variables $z_{i_1},z_{i_2},\ldots,z_{i_p}$ among the $(n{-}p)$ unintegrated variables in their fixed order (\ref{order-unintegrated}). We write 
\beq
\gamma^{(n,p)}_{\vec{A},\vec{i}}=(1,A_1,i_1,A_2,i_2,A_3,\ldots,A_p,i_p,A_{p+1},n{-}1,n)
\label{gen.1} \, ,
\eeq
where $\vec{A}=(A_1,A_2,\ldots,A_{p+1})$ represents a partition of the ordered list of unintegrated variables $z_{p+2},\ldots,z_{n-2}$ into possibly empty parts $A_j$.
Each sequence $\ldots,A_k,i_k,A_{k+1},\ldots$ in \eqref{gen.1}
with $A_k=(a_{k1},a_{k2},\ldots,a_{k \ell_k})$ translates into the range $z_{a_{k  \ell_k}}<z_{i_k}<z_{a_{k+1,1}}$ for the associated integration variable $z_{i_k}$ (with $z_{i_{k-1}}<z_{i_k}$ and $z_{i_k}<z_{i_{k+1}}$ in case of $A_{k}= \emptyset$ and $A_{k+1}= \emptyset$, respectively).
Thus there are $\binom{n-3}{p}$ values of $\vec{A}$ and $p!$ values of $\vec{i}=(i_1,i_2,\ldots,i_p)$ corresponding to permutations of $(2,3,\ldots,p{+}1)$.
These cycles correspond to the bounded chambers of the hyperplane arrangement defined by $\{ z_{ij}=0\}$.

The dual cocycle satisfying the condition of orthonormal field-theory limits (\ref{conv6}), which can be understood as a recursive application of the case with $p=1$ to successive integration variables,
reads
\begin{align}
\vph^{(n,p)}_{\vec{A},\vec{i}}= \sum_{j_1\in\{1,A_1\}} \frac{ s_{i_1,j_1}}{z_{i_1,j_1}}
 \sum_{j_2\in\{1,A_1,i_1,A_2\}} \frac{ s_{i_2,j_2}}{z_{i_2,j_2}} \ldots
  \sum_{j_p\in\{1,A_1,i_1,A_2,\ldots \atop{\ldots,A_{p-1},i_{p-1},A_p \} }}  \frac{ s_{i_p,j_p}}{z_{i_p,j_p}}\, .
\label{gen.2}
\end{align}
As in the $p=1$ case, it is clear that the divergences contributed from endpoint singularities of the integral result in the orthonormality required for the condition \eqref{conv6}.
Similar to \eqref{ibpequiv}, one can attain alternative cohomology representatives of \eqref{gen.2} by adding Koba--Nielsen derivatives. The following $p{+}1$ choices without double poles follow from adding derivatives in $z_{i_{k+1}},\ldots,z_{i_p}$ with $k=0,1,\ldots,p$:
\begin{align}
\vph^{(n,p)}_{\vec{A},\vec{i}}&\cong  \sum_{j_1\in\{1,A_1\}} \frac{ s_{i_1,j_1}}{z_{i_1,j_1}}
 \sum_{j_2\in\{1,A_1,i_1,A_2\}} \frac{ s_{i_2,j_2}}{z_{i_2,j_2}} \ldots
  \sum_{j_k\in\{1,A_1,i_1,A_2,\ldots \atop{\ldots,A_{k-1},i_{k-1},A_k \} }}  \frac{ s_{i_k,j_k}}{z_{i_k,j_k}}
    \notag \\
 &\ \ \ \ \times  
 \sum_{j_{k+1}\in\{A_{k+2},i_{k+2},A_{k+3},\ldots \atop{\ldots,A_{p},i_{p},A_{p+1},n-1 \} }}  \frac{ s_{j_{k+1},i_{k+1}}}{z_{j_{k+1},i_{k+1}}}
 \ldots
 \sum_{j_p\in\{ A_{p+1},n-1 \} }  \frac{ s_{j_p,i_p}}{z_{j_p,i_p}}
  \, .
\label{gen.2alt}
\end{align}
In case of double-integrals $p=2$, the twisted cycles \eqref{gen.1} and the dual functions \eqref{gen.2} become
\begin{align}
\gamma^{(n,2)}_{(A_1,A_2,A_3),(i_1,i_2)}&=(1,A_1,i_1,A_2,i_2,A_3,n{-}1,n)
\notag \\
\vph^{(n,2)}_{(A_1,A_2,A_3),(i_1,i_2)}&= \sum_{j_1\in\{1,A_1\}} \frac{ s_{i_1,j_1}}{z_{i_1,j_1}}
 \sum_{j_2\in\{1,A_1,i_1,A_2\}} \frac{ s_{i_2,j_2}}{z_{i_2,j_2}} \label{p2expl}  \\
 &\cong\sum_{j_1\in\{1,A_1\}} \frac{ s_{i_1,j_1}}{z_{i_1,j_1}}
 \sum_{j_2\in\{A_3,n-1\}} \frac{ s_{j_2,i_2}}{z_{j_2,i_2}}
 \notag \\
 &\cong 
 \sum_{j_1\in\{A_2,i_2,A_3,n-1\}} \frac{ s_{j_1,i_1}}{z_{j_1,i_1}}
 \sum_{j_2\in\{A_3,n-1\}} \frac{ s_{j_2,i_2}}{z_{j_2,i_2}}
 \, , \notag
\end{align}
where the last two lines contain the alternative representatives
\eqref{gen.2alt} with $k=0,1$.

\subsection{Verification via intersection numbers}
\label{sec:2.3}
 
More systematically, we can verify orthonormality \eqref{conv6} with the above cocycles using intersection numbers.  The $\alpha' \to 0$ limit of $F_{ab}^{(n,p)}$ is computed by intersection numbers of twisted cocycles,
\beq
\lim_{\alpha' \rightarrow 0} F^{(n,p)}_{\vec{A},\vec{i};\vec{B},\vec{j}}  \;=\; \lim_{\alpha' \rightarrow 0} \int_{\gamma_{\vec{A},\vec{i}}^{(n,p)}}
{\rm KN}^{(n,p)}\, \omega_{\vec{B},\vec{j}}^{(n,p)} \;=\;  \langle \nu_{\vec{A},\vec{i}}^{(n,p)} | \omega_{\vec{B},\vec{j}}^{(n,p)} \rangle \, ,
\label{conv6b}
\eeq
since the forms constructed from the $\vph_{\vec{B},\vec{j}}^{(n,p)}$ in \eqref{gen.2} are logarithmic. Here the  $\nu_{\vec{A},\vec{i}}^{(n,p)}$ form a basis of dual cocycles that correspond to $\gamma_{\vec{A},\vec{i}}^{(n,p)}$ from \eqref{gen.1}, in the sense that each $\nu_{\vec{A},\vec{i}}^{(n,p)}$ has logarithmic singularities with unit residues along the boundaries of $\gamma_{\vec{A},\vec{i}}^{(n,p)}$. In the terminology of \cite{Arkani-Hamed:2017tmz},
the $\nu_{\vec{A},\vec{i}}^{(n,p)}$ are the canonical forms associated to the positive geometries described by $\gamma_{\vec{A},\vec{i}}^{(n,p)}$, and indeed any region bounded by hyperplanes is a positive geometry for which a canonical form exists. 
We can write out the latter as
\begin{align}
\gamma_{\vec{A},\vec{i}}^{(n,p)} 
&= \{ z_{b_{i_1}} < z_{i_1} < z_{c_{i_1}} \} \times \{ z_{b_{i_2}} < z_{i_2} < z_{c_{i_2}} \} \times \cdots \times \{ z_{b_{i_p}} < z_{i_p} < z_{c_{i_p}} \} \, ,
\end{align}
such that
\begin{align}
\int_{\gamma_{\vec{A},\vec{i}}^{(n,p)} }\Big( \prod_{k=2}^{p+1} \dd z_k \Big) = \int^{z_{c_{i_1}}}_{z_{b_{i_1}}}\dd z_{i_1}
 \int^{z_{c_{i_2}}}_{z_{b_{i_2}}}\dd z_{i_2}
 \ldots
  \int^{z_{c_{i_p}}}_{z_{b_{i_p}}}\dd z_{i_p} \, ,
\end{align}
i.e.\ for each integrated puncture $z_{i_k}$, the indices $b_{i_k}$ and $c_{i_k}$ label the variables adjacent to it in the ordering \eqref{gen.1}.\footnote{Note that in case of adjacent integration variables $z_2,z_3$ bounded by $z_b<z_{2}<z_3 < z_c$, only one of $z_2,z_3$ appears
among the integration limits $z_{b_i},z_{c_i}$, i.e.
\[
\int_{z_b<z_2<z_3<z_c} \dd z_2 \, \dd z_3 = \int^{z_c}_{z_b} \dd z_3 \int^{z_{3}}_{z_b} \dd z_2 = \int^{z_c}_{z_b} \dd z_2 \int^{z_{c}}_{z_2} \dd z_3 \,.
\]
Hence, the choice of $z_{b_i},z_{c_i}$ is in general not unique, but each parametrization of simplices such 
as $z_b<z_2<z_3<z_c$ lead to the same expression for
the forms $\nu_{\vec{A},\vec{i}}^{(n,p)}$ in (\ref{nu-forms}) related by partial fraction.}
 This gives a natural cocycle counterpart:
\begin{gather}
\nu_{\vec{A},\vec{i}}^{(n,p)} =  \hat \nu_{\vec{A},\vec{i}}^{(n,p)}  \prod_{k=2}^{p+1} \dd z_k \label{nu-forms} \\
\hat \nu_{\vec{A},\vec{i}}^{(n,p)} = \left( \frac{1}{z_{i_1,b_{i_1}}} - \frac{1}{z_{i_1,c_{i_1}}} \right)\left( \frac{1}{z_{i_2,b_{i_2}}} - \frac{1}{z_{i_2,c_{i_2}}} \right)\cdots \left( \frac{1}{z_{i_p,b_{i_p}}} - \frac{1}{z_{i_p,c_{i_p}}} \right)\, . \notag
\end{gather}
Since both bases $\nu_{\vec{A},\vec{i}}^{(n,p)}$ and $\omega_{\vec{A},\vec{i}}^{(n,p)}$ are logarithmic, the evaluation of intersection numbers can be carried out on the support of critical points of ${\rm KN}^{(n,p)}$ \cite{Mizera:2017rqa} given by solutions of the equations:
\be\label{crit-pts}
\partial_{k} \log {\rm KN}^{(n,p)} = \sum_{\substack{j=1\\ j\neq k}}^{n-1} \frac{s_{kj}}{z_{kj}} = 0 \, , \qquad \mathrm{for~}k = 2,3,\ldots, p{+}1 \, .
\ee
For generic values of the kinematic variables, the equations \eqref{crit-pts} have exactly $d^{(n,p)}$ solutions \cite{aomoto1987gauss,Mizera:2019gea}. Let us denote the $a$-th solution by $(z_2^{(a)}, z_3^{(a)}, \ldots, z_{p+1}^{(a)})$ with $a=1,2,\ldots,d^{(n,p)}$.
The right-hand side of \eqref{conv6b} can then be computed as
\be\label{inter-delta-omega}
\langle \nu_{\vec{A},\vec{i}}^{(n,p)} | \omega_{\vec{B},\vec{j}}^{(n,p)} \rangle \;=\; (-1)^p \sum_{a=1}^{d^{(n,p)}} \frac{\hat{\nu}_{\vec{A},\vec{i}}^{(n,p)} \hat{\omega}_{\vec{B},\vec{j}}^{(n,p)}}{\det J^{(n,p)}} \bigg|_{z_k = z_k^{(a)}} \, ,
\ee
where $J^{(n,p)}_{kl}$ is a Hessian matrix with entries
\be
J^{(n,p)}_{kl} = \partial_k \partial_l \log {\rm KN}^{(n,p)} = \begin{dcases}
\qquad\frac{s_{kl}}{z_{kl}^2} &\qquad \mathrm{for}\quad k\neq l \, ,\\
-\sum_{\substack{j=1\\ j\neq k}}^{n-1} \frac{s_{kj}}{z_{kj}^2} &\qquad \mathrm{for}\quad k = l \, ,
\end{dcases}
\ee
for $k,l=2,3,\ldots,p{+}1$. We stress that this formula can be only used for logarithmic forms, as otherwise it is valid only asymptotically in the $\alpha' \to \infty$ limit \cite{Mizera:2017rqa,Mizera:2019vvs}. We checked numerically for all values of $(n,p)$ up to and including $(10,7)$ that this formula gives rise to the identity matrix, i.e.,
\be\label{orthonormality}
\langle \nu_{\vec{A},\vec{i}}^{(n,p)} | \omega_{\vec{B},\vec{j}}^{(n,p)} \rangle = \delta_{(\vec{A},\vec{i}),(\vec{B},\vec{j})}\, ,
\ee
which confirms \eqref{conv6}. The largest checks required summing over $d^{(10,7)}=5040$ critical points for each entry of the $5040 \times 5040$ matrix $\langle \nu_{a}^{(10,7)} | \omega_{b}^{(10,7)} \rangle $. This high-multiplicity computation was made possible by following \cite{Sturmfels:2020mpv} to interpret $\log \mathrm{KN}^{(n,p)}$ as a log-likelihood function in algebraic statistics and extremizing it according to \eqref{crit-pts} using the \texttt{Julia} package \texttt{HomotopyContinuation.jl} \cite{10.1007/978-3-319-96418-8_54}.

\subsection{String amplitudes from many integrated punctures}
\label{sec:2.4}

For the maximum number $p=n{-}3$
of integrations, the integrals in (\ref{conv5}) agree with the basis of disk integrals in open-superstring amplitudes
obtained in \cite{Mafra:2011nv} (with permutations $\rho_a, \rho_b$ acting on $2,3,\ldots,n{-}2$, i.e.\ $a,b=1,2,\ldots,(n{-}3)!$), 
\begin{align}
F^{(n,n-3)}_{ab}   &=
\int_{\gamma_a^{(n,n-3)}} \Big( \prod_{j=2}^{n-2} \dd z_j \Big) \, \prod_{1\leq i<j}^{n-1} |z_{ij}|^{s_{ij}}\, \vph_b^{(n,n-3)}  
 \notag\\ 
\gamma^{(n,n-3)}_{a}   &=\{
0{<}z_{\rho_a(2)} {<}z_{\rho_a(3)} {<} \ldots {<}z_{\rho_a(n-2)}{<}1\} \, , \ \ \ \ \rho_a \in S_{n-3}  
\label{conv7a} \\ 
\vph^{(n,n-3)}_{b}   &= \frac{ s_{1\rho_b(2)} }{ z_{\rho_b(2),1} }
\Big(  \frac{ s_{1\rho_b(3)} }{ z_{\rho_b(3),1} } {+}  \frac{ s_{\rho_b(2),\rho_b(3)} }{ z_{\rho_b(3),\rho_b(2)} } \Big)
\cdots  \notag \\ 
& \ \ \ \ \  \cdots \times 
\Big(  \frac{ s_{1\rho_b(n-2)} }{ z_{\rho_b(n-2),1} } {+} 
\ldots{+}  \frac{ s_{\rho_b(n-3)\rho_b(n-2)} }{ z_{\rho_b(n-2),\rho_b(n-3)} } \Big)
  \, , \ \ \ \ \rho_b \in S_{n-3} \, .
\notag
\end{align}
As pointed out in \cite{Gomez:2013wza}, this representation of the 
integrand for open superstrings can be readily exported
to ambitwistor string theories, and the equations
\eqref{crit-pts} are known in this case as the scattering equations \cite{Cachazo:2013gna}.
The conjectural patterns among the MZVs in the $\alpha'$-expansion
\cite{Schlotterer:2012ny} to be reviewed below imply the coaction formula (\ref{conv7}) \cite{Drummond:2013vz}.

In the case of $p=n{-}4$ integrations, the integrals (\ref{conv5}) are relabellings of the auxiliary functions
$\hat F^\sigma_\nu$ studied in \cite{Broedel:2013aza} to extract open-string $\ap$-expansions from the
Drinfeld associator (also see \cite{Terasoma, Drummond:2013vz, AKtbp}) and in \cite{Vanhove:2018elu} to identify closed-string integrals as single-valued correlation functions.

\section{Structure of the $\alpha'$-expansion}
\label{sec:3}

This section is dedicated to the $\alpha'$-expansion of the integrals $F^{(n,p)}_{ab}$ in (\ref{conv5})
which is used to test the coaction property (\ref{conv7}) order by order in $\alpha'$. 
We will focus on the situation where the 
unintegrated punctures are ordered on the real axis according to 
\beq
0=z_{1}<z_{p+2}<z_{p+3}<\ldots < z_{n-2}<z_{n-1}=1
\label{expa.1}
\eeq
and discuss the analytic continuation to different regions in section \ref{sec:5}. As will be detailed below, the coefficients in the Taylor expansion of $F^{(n,p)}_{ab}$ with respect to the $s_{ij}$ are $\mathbb Q$-linear
combinations of MZVs and multiple polylogarithms in $z_{p+2},\ldots,z_{n-2}$, defined respectively by
\begin{align}
\zeta_{n_1,n_2,\ldots,n_r} &= \sum_{0<k_1<k_2<\ldots <k_r}^{\infty} k_1^{-n_1}k_2^{-n_2} \ldots k_r^{-n_r} \, , 
\label{expa.2} \\
G(a_1,a_2,\ldots,a_w;z) &= \int^z_0 \frac{ \dd t }{t-a_1} G(a_2,\ldots,a_w;t) 
 \, ,
\label{expa.3}
\end{align}
where $n_j \in \mathbb N ,\ n_r\geq 2$ and $a_j,z \in \mathbb C$,
and the recursive definition of polylogarithms starts with $G(\emptyset;z) =1$.
MZVs and polylogarithms are assigned (transcendental) weight 
$n_1+n_2+\ldots+n_r$ and $w$, respectively, and $r$ in (\ref{expa.2})
is referred to as the depth of an MZV. The endpoint divergences of 
$G(\ldots,0;z) $ are shuffle-regularized with the assignment
\beq
G(\underbrace{0,0,\ldots,0}_{n};z) = \frac{1}{n!} (\log z)^n\, .
\label{shufreg}
\eeq
For instance, shuffle regularization can be used to reduce depth-one polylogarithms $G(0,\ldots,0,1,0,\ldots,0;z)$
to linear combinations of
\beq
G(1;z) = \log(1{-}z) \, , \ \ \ \
G(\underbrace{0,0,\ldots,0}_{p-1},1;z) = -  {\rm Li}_p(z)  \, , 
\ \ \ \ p \geq 2
\eeq
multiplying powers of $\log z$.
The appearance of MZVs in the $\ap$-expansion of $F^{(n,p)}_{ab}$ will be traced
back to the case $p=n{-}3$ relevant to string amplitudes: The polynomial
structure of $F^{(n,n-3)}_{ab}$ in the $s_{ij}$ at any multiplicity $n$ can be generated from the Drinfeld
associator \cite{Broedel:2013aza, AKtbp} or Berends--Giele recursions \cite{Mafra:2016mcc} (also see \cite{Oprisa:2005wu, Stieberger:2007jv, Boels:2013jua, Puhlfuerst:2015gta} for relations to hypergeometric functions at $n \leq 7$ points). The polylogarithms
in turn are determined by the KZ equations of the $F^{(n,p)}_{ab}$ which take
the schematic form \cite{aomoto1987gauss,Terasoma,Mizera:2019gea}
\beq
\partial_j F^{(n,p)}_{ab} = \sum_{c=1}^{d^{(n,p)}}  \bigg\{ \frac{ (e^{(n,p)}_{j1})_{bc} }{z_{j1}} 
+  \frac{ (e^{(n,p)}_{j,n-1})_{bc} }{z_{j,n-1}} + \sum_{m=p+2 \atop{m \neq j}}^{n-2} \frac{ (e^{(n,p)}_{jm})_{bc} }{z_{jm}} \bigg\}  
F^{(n,p)}_{ac} \, , 
\label{conv9}
\eeq
where $j=p{+}2,p{+}3,\ldots,n{-}2$ and $\partial_j = \frac{ \partial }{\partial z_j}$. The
entries of the $d^{(n,p)}\times d^{(n,p)}$ braid matrices $e^{(n,p)}_{jm}$ are linear in 
$s_{ij}$ which will allow us to solve (\ref{conv9}) perturbatively in $\alpha'$. The linear
appearance of $\alpha'$ on the right-hand side of (\ref{conv9}) is analogous to the
$\epsilon$-form of the differential equation for dimensionally regulated Feynman integrals
\cite{Henn:2013pwa, Adams:2018yfj}.

Given the ordering (\ref{expa.1}) of the unintegrated punctures,
it will be convenient to solve (\ref{conv9}) with the following choice
of fibration bases for the polylogarithms in the $\alpha'$-expansion: 
The labels in a factor of $G(a_1,a_2,\ldots,a_w;z_j)$ with $p{+}2\leq j \leq n{-}2$
are taken from $a_k \in \{0,1,z_{j+1},\ldots,z_{n-2}\}$. For example, in the case of $(n,p)=(6,1)$, the integral   $F^{(6,1)}_{ab}$
will feature products of MZVs, $G(a_k {\in} \{0,1\};z_4)$ and $G(a_k {\in} \{0,1,z_4\};z_3)$. 
As previewed in (\ref{intr.2}), these polylogarithms turn out to enter the $\ap$-expansions through certain 
matrix-valued generating series that will be specified below, denoted by 
$\mathbb G^{(6,1)}_{\{0,1\}}(z_4)$, $\mathbb G^{(6,1)}_{\{0,1,z_4\}}(z_3)$,
and more generally $ \mathbb G^{(n,p)}_{\{0,1,z_{j+1},z_{j+2}, \ldots ,z_{n-2}\}}(z_{j}) $. 
The main result of this section is a factorized
form of the $\alpha'$-expansion,
\begin{align}
F^{(n,p)}(z_{p+2},z_{p+3},\ldots,z_{n-2}) &= \mathbb P^{(n,p)}\mathbb M^{(n,p)}
\mathbb G^{(n,p)}_{\{0,1\}}(z_{n-2}) \mathbb G^{(n,p)}_{\{0,1,z_{n-2}\}}(z_{n-3}) \ldots  \label{apexp.35} \\
&\ \ \ \ \times
\mathbb G^{(n,p)}_{\{0,1,z_{p+4},\ldots,z_{n-2}\}}(z_{p+3})
\mathbb G^{(n,p)}_{\{0,1,z_{p+3},z_{p+4},\ldots,z_{n-2}\}}(z_{p+2}) \, ,
\notag
\end{align}
where $\mathbb P^{(n,p)}$, $\mathbb M^{(n,p)}$ are constant series involving MZVs. 
We suppress the indices $a,b$ of $F^{(n,p)}_{ab}$ and the $d^{(n,p)} \times d^{(n,p)}$ matrices on the right-hand
side, with matrix-multiplication between neighboring factors $\mathbb P^{(n,p)},\mathbb M^{(n,p)}$ and $\mathbb G^{(n,p)}$.

\subsection{MZVs in string amplitudes and general genus-zero integrals}
\label{sec:3.1}

The $n$-point integrals (\ref{conv7a}) seen in string amplitudes with $p=n{-}3$ integrated
punctures solely involve MZVs in their $\ap$-expansion \cite{Terasoma, Brown:2009qja} 
without any polylogarithms
at argument $z \neq 1$. The factorized form (\ref{apexp.35}) of the $\ap$-expansion then reduces to
\cite{Schlotterer:2012ny}
\beq
F^{(n,n-3)}= \mathbb P^{(n)}\mathbb M^{(n)} \, , \ \ \ \ \ \ 
 \mathbb P^{(n)} =  \mathbb P^{(n,n-3)} \, , \ \ \ \ \ \ 
 \mathbb M^{(n)} = \mathbb M^{(n,n-3)}\, ,
\label{expa.5}
\eeq
where $\mathbb P^{(n)}$ and $\mathbb M^{(n)}$ comprise different types of MZVs and
decompose as follows \cite{Schlotterer:2012ny}, 
\begin{align}
\mathbb P^{(n)}  &= \mathds{1} + \zeta_2 P^{(n)}_2 + \zeta^2_2 P^{(n)}_4+ \zeta_2^3 P^{(n)}_6 + \zeta_2^4 P^{(n)}_8+
{\cal O}(s_{ij}^{10})\,,
\label{expa.6} \\
\mathbb M^{(n)}  &= \mathds{1} + \zeta_3 M^{(n)}_3 + \zeta_5 M^{(n)}_5+\frac{1}{2} \zeta_3^2 M^{(n)}_3 M^{(n)}_3
+ \zeta_7 M^{(n)}_7 \notag \\
&\ \ \ \ + \zeta_3\zeta_5 M^{(n)}_5 M^{(n)}_3
+ \frac{1}{5} \zeta_{3,5}[ M^{(n)}_5, M^{(n)}_3]
+{\cal O}(s_{ij}^{9})\, .
\label{expa.7}
\end{align}
The entries of the $(n{-}3)! \times (n{-}3)!$ matrices $P^{(n)}_w =  P^{(n,n-3)}_w$ and  $M^{(n)}_w = M^{(n,n-3)}_w$
are degree-$w$ polynomials in the $s_{ij}$ with rational coefficients, and the leading term $\mathds{1}$ stands  for the $(n{-}3)! \times (n{-}3)!$ unit matrix, reflecting the orthonormal field-theory limits of (\ref{conv7a}). The decomposition (\ref{expa.5})--(\ref{expa.7}) determines the coefficients of arbitrary MZVs in terms of matrix products of those of the primitives, i.e.\
$\zeta_{2k+1}M^{(n)}_{2k+1}$ and $\zeta_{2}^kP^{(n)}_{2k}$.
For example, we find
\beq
F^{(n,n-3)} \, \Big|_{\zeta_2 \zeta_3} = P^{(n)}_2 M^{(n)}_3 \, , \ \ \ \ \ \ 
F^{(n,n-3)} \, \Big|_{\zeta_{3,5}} = \frac{1}{5} [ M^{(n)}_5, M^{(n)}_3] \, .
\label{expa.8}
\eeq
We are employing the conjectural $\mathbb Q$-bases of \cite{Blumlein:2009cf} for MZVs, see e.g.\ \cite{GilFresan, Jianqiang} for a general
account of the relations and various other aspects of MZVs. The non-intuitive prefactor $\frac{1}{5}$ in the coefficient of $\zeta_{3,5}$ can be understood by passing to the $f$-alphabet description of MZVs \cite{Brown:2011ik} (or strictly speaking, of motivic MZVs \cite{Goncharov:2005sla, BrownTate}):
Based on a non-canonical isomorphism $\phi$, (motivic) MZVs can be mapped to a comodule with
commuting generator $f_2$ and non-commuting generators $f_3,f_5,f_7,\ldots$ such that\footnote{We will informally omit the superscript of motivic MZVs $\zeta^{\mathfrak m}_{n_1,\ldots,n_r}$ in (\ref{expa.9}) and below. Examples of $\phi( \zeta_{n_1,n_2,\ldots,n_r})$ at higher weight can be found in \cite{Brown:2011ik, Schlotterer:2012ny}, but the conventions in the references differ from ours by a swap $A \otimes B \to B \otimes A$ and therefore by a reversal $f_{2k_1+1}f_{2k_2+1} \ldots f_{2k_r+1}  \mapsto f_{2k_r+1} \ldots f_{2k_2+1} f_{2k_1+1}$. The conventions for ordering the entries of the coaction in this work follow for instance those of \cite{Goncharov:2005sla, Duhr:2012fh, Abreu:2017enx,Abreu:2017mtm}.}
\beq
\phi(\zeta_2) = f_2 \, , \ \ \ \ \phi(\zeta_{2k+1}) = f_{2k+1} \, , \ \ \ \ \phi(\zeta_{3,5}) = - 5 f_3 f_5 \, , \ \ \ \ {\rm etc.} 
\label{expa.9}
\eeq
The isomorphism $\phi$ is constructed such that the product of MZVs is mapped to a shuffle of the non-commutative $f_{2k+1}$, and the coaction of (motivic) MZVs translates into deconcatenation,
\begin{align}
\phi(\zeta_{A} \zeta_B) &= \phi(\zeta_{A}  ) \shuffle \phi( \zeta_B) 
\label{expa.10} \\
\Delta f_2^N f_{2k_1+1}f_{2k_2+1} \ldots f_{2k_r+1} &=
\sum_{j=0}^r  f_2^N f_{2k_1+1}f_{2k_2+1} \ldots f_{2k_j+1}\otimes
f_{2k_{j+1}+1} \ldots f_{2k_r+1}\, .
\label{expa.11}
\end{align}
In this setup, the all-order structure of the matrices in (\ref{expa.5}) was proposed to be \cite{Schlotterer:2012ny}
\begin{align}
\mathbb P^{(n)}  &= \mathds{1}+ \phi^{-1} \sum_{k=1}^{\infty} f_2^k P_{2k}^{(n)}
\label{expa.12} \\
\mathbb M^{(n)}  &= \phi^{-1}\sum_{r=0}^{\infty} \sum_{k_1,k_2,\ldots, k_r =1}^{\infty} f_{2k_1+1} f_{2k_2+1} \ldots f_{2k_r+1} M^{(n)}_{2k_1+1} M^{(n)}_{2k_2+1}\ldots M^{(n)}_{2k_r+1} 
\label{expa.13}
\end{align}
which by (\ref{expa.11}) implies the coaction formula (\ref{conv7}) at $p=n{-}3$ \cite{Drummond:2013vz}.

As a necessary condition for (\ref{conv7}) to carry over to general $p \leq n{-}3$, the same statements 
are claimed to carry over to the MZV-dependent parts $\mathbb P^{(n,p)}$ and $\mathbb M^{(n,p)}$
of (\ref{apexp.35}). We propose
that
\begin{align}
\mathbb P^{(n,p)}  &= \mathds{1}+ \phi^{-1} \sum_{k=1}^{\infty} f_2^k P_{2k}^{(n,p)}  \,,
\label{expa.14} \\
\mathbb M^{(n,p)}  &= \phi^{-1}\sum_{r=0}^{\infty} \sum_{k_1,k_2,\ldots, k_r =1}^{\infty} f_{2k_1+1} f_{2k_2+1} \ldots f_{2k_r+1} M^{(n,p)}_{2k_1+1} M^{(n,p)}_{2k_2+1}\ldots M^{(n,p)}_{2k_r+1}    \, ,
\label{expa.15}
\end{align}
where the entries of the $d^{(n,p)} \times d^{(n,p)}$ matrices $P^{(n,p)}_w $ 
and  $M^{(n,p)}_w$ are again degree-$w$ polynomials in the $s_{ij}$ with rational coefficients.
Note that (\ref{expa.14})--(\ref{expa.15}) is equivalent to
\beq
\Delta \mathbb P^{(n,p)}   = \mathbb P^{(n,p)}   \otimes \mathds{1}  \, , \ \ \ \ \ \ 
\Delta \mathbb M^{(n,p)}   = \mathbb M^{(n,p)}   \otimes \mathbb M^{(n,p)}  \, .
\label{expa.16}
\eeq
In the following, we will spell out examples of the $P^{(n,p)}_w $, $M^{(n,p)}_w$ at $p\neq n{-}3$
and describe methods to compute them in general cases. Explicit results for the $P^{(n)}_w $, 
$M^{(n)}_w$ at $n\leq7$ are available for download on the website \cite{wwwap}, and code for generating
all-multiplicity results can be obtained from \cite{wwwbgrec}. 

Note that the image of MZVs of depth $r \geq 2$ under the $\phi$-map in (\ref{expa.15}) depends on a choice of reference basis. We follow the conventions of \cite{Brown:2011ik, Schlotterer:2012ny} to assign vanishing coefficients of $f_w$ to the $\phi$-image of those higher-depth MZVs at weight $w$ in the (conjectural) $\mathbb Q$-bases of \cite{Blumlein:2009cf} (say $\zeta_{3,5},\zeta_{3,7},\zeta_{3,3,5},\ldots$). Still, the form of (\ref{expa.12}) to (\ref{expa.15}) does not depend on these choices, only the $s_{ij}$-dependence in the entries of $P^{(n,p)}_{w}$ and $M^{(n,p)}_{w}$ depends on the reference bases for MZVs at weight $w$.

\subsection{Warm-up example $(n,p)=(5,1)$}
\label{sec:3.2}

In order to illustrate the origin of (\ref{apexp.35}) and exemplify the explicit form of
the series $\mathbb G^{(n,p)}_{\{0,1,\ldots\}}(z_j)$, we shall now give a detailed derivation
of the $\alpha'$-expansion of $F^{(5,1)}_{ab}$. The two-dimensional bases of cocycles (\ref{p1example})
and cycles (\ref{gen.1}) are
\beq
\gamma_1^{(5,1)} = \{0<z_2<z_3\} \, , \ \ \ \ \ \ \gamma_2^{(5,1)} = \{z_3<z_2<1\} \,,
\label{51.2}
\eeq
as well as
\beq
\vph_1^{(5,1)} = \frac{ s_{21} }{z_{21}} \cong  \frac{ s_{32} }{z_{32}} +  \frac{ s_{42} }{z_{42}}  \, , \ \ \ \ \ \ 
\vph_2^{(5,1)} = \frac{ s_{21} }{z_{21}}+ \frac{ s_{23} }{z_{23}} \cong    \frac{ s_{42} }{z_{42}} \, .
\label{51.3}
\eeq
We have discarded Koba--Nielsen derivatives $\partial_2 {\rm KN}^{(5,1)} =( \frac{ s_{12} }{z_{21}}
+\frac{ s_{23} }{z_{23}}+\frac{ s_{24} }{z_{24}} ) {\rm KN}^{(5,1)} $ in passing between different
representations of $\vph_b^{(5,1)}$ in the twisted cohomology. The same integration-by-parts identities
allow us to determine the $2\times 2$ braid matrices\footnote{Note that the soft limit $s_{23}\rightarrow 0$ of $e^{(5,1)}_{31}=e^{(5,1)}_{0}$ and $e^{(5,1)}_{34}=e^{(5,1)}_{1}$ followed by relabelling $s_{24}\rightarrow s_{23}$ reproduces the four-point
instances of the arguments of the $2\times2$ Drinfeld associator in \cite{Broedel:2013aza}.
See \cite{AKtbp} for a discussion of this method in the framework of twisted de Rham theory. The $z_3$-derivatives of $F^{(5,1)}_{ab}$ have been simplified using partial fractions and integration by parts in order to attain the form on
the right-hand side of (\ref{51.4}) and to identify 
the expressions (\ref{51.5}) for the braid matrices.}
\beq
e^{(5,1)}_{31} = \ccb s_{12}{+}s_{23} &-s_{12} \\ 0 &0 \cce \, , \ \ \ \ \ \
e^{(5,1)}_{34} = \ccb 0 &0 \\ -s_{24} &s_{24}{+}s_{23}   \cce \,,
\label{51.5}
\eeq
in the KZ equation (\ref{conv9})
\beq
\partial_3 F^{(5,1)}_{ab} = \sum_{c=1}^2 \bigg\{ \frac{ (e^{(5,1)}_{31})_{bc} }{z_{31}} + \frac{ (e^{(5,1)}_{34})_{bc} }{z_{34}} \bigg\} F^{(5,1)}_{ac} \, .
\label{51.4}
\eeq
One can solve (\ref{51.4}) through the generating series of polylogarithms $G(a_k \in \{0,1\};z_3)$
\begin{align}
\mathbb G^{(5,1)}_{\{0,1\}}(z_3) &=  \mathds{1} + \sum_{a_1 \in \{0,1\}} G(a_1;z_3)E^{(5,1)}_{a_1,z_3} 
+ \sum_{a_1,a_2 \in \{0,1\}} G(a_2,a_1;z_3)E^{(5,1)} _{a_1,z_3}E^{(5,1)} _{a_2,z_3} + {\cal O}(s_{ij}^3)
\notag \\
&= \sum_{r=0}^{\infty}  \sum_{a_1,a_2,\ldots \atop{\ldots,a_r \in \{0,1\}} } G(a_r,\ldots,a_2,a_1;z_3)
E^{(5,1)}_{a_1,z_3}E^{(5,1)}_{a_2,z_3} \ldots E^{(5,1)}_{a_r,z_3} \, .
\label{apexp.24}
\end{align}
with the transpose of the braid matrices (\ref{51.5})
\beq
E^{(5,1)} _{0,z_3} = (e^{(5,1)}_{31})^t=
\ccb  s_{12} {+} s_{23} &0 \\ {-}s_{12} &0  \cce
\, , \ \ \ \ \ \ 
E^{(5,1)} _{1,z_3} = (e^{(5,1)}_{34})^t =
\ccb  0& {-}s_{24} \\ 0 &s_{23} {+} s_{24} \cce\,,
\label{apexp.LE}
\eeq
which may multiply arbitrary $z_3$-independent matrices from the right.
In order to tailor these constant matrices to the target integrals $F^{(5,1)}_{ab}$,
we determine their asymptotics\footnote{While (\ref{in51.3}) follows from the rescaling $z_2 = x z_3$ of the integration variable with $x \in (0,1)$, one needs an additional change of
variables $z_2 \rightarrow 1{-}z_2$ in the derivation of (\ref{in51.alt}).} as $z_3\rightarrow 0$ and $z_3\rightarrow 1$,
\begin{align}
F^{(5,1)}_{1b}(z_3\rightarrow 0) &=  \delta_{b,1} |z_3|^{s_{12}+s_{23}}  \frac{ \Gamma(1{+}s_{12}) \Gamma(1{+}s_{23} ) }{\Gamma(1{+}s_{12}{+}s_{23}) }\,, \label{in51.3}\\
F^{(5,1)}_{2b}(z_3\rightarrow 1)&=  \delta_{b,2} |1{-}z_3|^{s_{23}+s_{24}}  \frac{ \Gamma(1{+}s_{23}) \Gamma(1{+}s_{24} ) }{\Gamma(1{+}s_{23}{+}s_{24}) } \, .  \label{in51.alt}
\end{align}
Finally, it remains to expand the $F^{(5,1)}_{2b}$ associated with the integration domain $z_2 {\in} (z_3,1)$ around
$z_3\rightarrow 0$ in order to expand the entire $2\times 2$ matrix of $F^{(5,1)}_{ab}$ in terms of polylogarithms with the same basepoint. In presence of the pole $z_{21}^{-1}$ of $\vph_1$, the $\alpha'$-expansion of 
$F^{(5,1)}_{21}$ does not commute with the limit $z_3 \rightarrow 0$. Hence, as detailed in appendix \ref{app:A.1}, we instead infer the
$\alpha'$-expansion via monodromy relations \cite{BjerrumBohr:2009rd, Stieberger:2009hq} involving cycles where the $\ap$-expansions
commute with the limit $z_3 \rightarrow 0$ and obtain
\beq
\mathbb P^{(5,1)}  \mathbb M^{(5,1)} = \ccb
 \frac{  \Gamma(1{+}s_{12})\Gamma(1{+}s_{23}) }{\Gamma(1{+}s_{12}{+}s_{23})} &0 \\
 \frac{ s_{12}}{s_{12}{+}s_{23} } \Big\{ \frac{ \Gamma(1{+}s_{24}) \Gamma(1{+}s_{12}{+}s_{23} ) }{\Gamma(1{+}s_{12}{+}s_{23}{+}s_{24}) }
- \frac{ \Gamma(1{+}s_{23}) \Gamma(1{-}s_{12}{-}s_{23} ) }{\Gamma(1{-}s_{12}) } \Big\} 
& \frac{  \Gamma(1{+}s_{12}{+}s_{23})\Gamma(1{+}s_{24}) }{\Gamma(1{+}s_{12}{+}s_{23}{+}s_{24})}
\cce \, .
\label{expa.21}
\eeq
Note that the factor of $(s_{12}{+}s_{23})^{-1}$ in the $(2,1)$-entry is cancelled by the difference of Euler beta functions, 
and we obtain a regular Taylor expansion around $\ap = 0$,
\begin{align}
&\frac{ s_{12}}{s_{12}{+}s_{23} } \bigg\{ \frac{ \Gamma(1{+}s_{24}) \Gamma(1{+}s_{12}{+}s_{23} ) }{\Gamma(1{+}s_{12}{+}s_{23}{+}s_{24}) }
- \frac{ \Gamma(1{+}s_{23}) \Gamma(1{-}s_{12}{-}s_{23} ) }{\Gamma(1{-}s_{12}) } \bigg\} \label{expa.22} \\
&={-} \zeta_2 s_{12}(s_{23}{+}s_{24}) + \zeta_3 s_{12}( s_{24}^2{+}s_{23}s_{24}{+}s_{12}s_{24}{-}s_{12}s_{23}) + {\cal O}(s_{ij}^4)\, , \notag
\end{align}
which is consistent with the $z_3\rightarrow1$ limit (\ref{in51.alt}). Taking (\ref{expa.21}) as a formal
initial value $z_3\rightarrow 0$, the $\ap$-expansion of $F^{(5,1)}_{ab}$ at generic $z_3 \in (0,1)$ is obtained
by right-multiplication with the series (\ref{apexp.24}) in polylogarithms
\beq
F^{(5,1)}(z_3) = \mathbb P^{(5,1)}  \mathbb M^{(5,1)} \mathbb G^{(5,1)}_{\{0,1\}}(z_3)
\label{expa.23}
\eeq
with matrix multiplication between the three factors. The individual $P^{(5,1)}_{2k},M^{(5,1)}_{2k+1}$ may
be obtained from (\ref{expa.21}) by extracting the coefficients of $\zeta_{2}^k,\zeta_{2k+1}$ in the Taylor expansion of 
\beq
F^{(4,1)}(s_{12},s_{23}) = \frac{ \Gamma(1{+}s_{12}) \Gamma(1{+}s_{23}) }{\Gamma(1{+}s_{12}{+}s_{23})} = \exp\Big( \sum_{k=2}^{\infty} \frac{ \zeta_k }{k} (-1)^k \big[  s_{12}^k {+} s_{23}^k {-} (s_{12}{+}s_{23})^k \big]\Big) \, ,
\label{apexp.36}
\eeq
i.e.\ they are determined by the single four-point integral ($d^{(4,1)} = 1$). In (\ref{expa.21}) and later expressions for initial values of $F^{(n,p)}$, we already incorporate a central conjecture on the structure of the $\ap$-expansion by writing the left-hand side as a matrix product of $\mathbb P^{(5,1)}$ and $\mathbb M^{(5,1)}$. Like this, the appearance of $\zeta_2$ is claimed to follow the expansions in (\ref{expa.14}) and (\ref{expa.15}) which we have verified order by order in $\ap$. It would be interesting to find an all-order argument based on the right-hand side of (\ref{expa.21}).

Given that MZVs are recovered from polylogarithms at unit argument via 
\beq
\zeta_{n_1,n_2,\ldots,n_r} = (-1)^r G( \underbrace{0,0,\ldots,0}_{n_r-1},1,\underbrace{0,\ldots,0}_{n_{r-1}-1},1,\ldots,
\underbrace{0,\ldots,0}_{n_1-1},1;1)  \, ,
\label{expa.25}
\eeq
one can check that (\ref{expa.23}) is consistent with both (\ref{in51.3}) and
(\ref{in51.alt}), validating our procedure to determine the formal initial value of $z_3=0$
from monodromy relations. The coaction properties of (\ref{expa.23}) extending
our conjecture (\ref{expa.16}) for $\Delta \mathbb P^{(n,p)},\Delta \mathbb M^{(n,p)}$ are discussed in
the later section \ref{sec:4}, and the explicit form of the $\alpha'^{\leq 2}$-orders can be
found in appendix \ref{app:A.2}.

\subsection{Warm-up example $(n,p)=(6,1)$}
\label{sec:3.3}

We shall now illustrate the selection of fibration bases for polylogarithms in two
variables by analyzing and solving the differential equations of $F_{ab}^{(6,1)}$.
The bases of master contours
\beq
\gamma_1^{(6,1)} = \{0<z_2<z_3\} \, , \ \ \ \ \gamma_2^{(6,1)} = \{z_3<z_2<z_4\} 
 \, , \ \ \ \ \gamma_3^{(6,1)} = \{z_4<z_2<1\} 
\label{61.2}
\eeq
and dual cocycles (see (\ref{p1example}))
\begin{align}
\vph_1^{(6,1)} &= \frac{ s_{21} }{z_{21}} \cong  \frac{ s_{32} }{z_{32}} +  \frac{ s_{42} }{z_{42}}+  \frac{ s_{52} }{z_{52}}  \notag \\
\vph_2^{(6,1)} &= \frac{ s_{21} }{z_{21}}+ \frac{ s_{23} }{z_{23}} \cong    \frac{ s_{42} }{z_{42}}  +  \frac{ s_{52} }{z_{52}}
\label{61.3} \\
\vph_3^{(6,1)} &= \frac{ s_{21} }{z_{21}}+ \frac{ s_{23} }{z_{23}}  +  \frac{ s_{24} }{z_{24}} \cong    \frac{ s_{52} }{z_{52}} 
 \notag
\end{align}
give rise to the following $3\times 3$ braid matrices
\begin{align}
e^{(6,1)}_{31} &= \cccb s_{12}{+}s_{23} &-s_{12}  &0 \\ 0 &0 &0 \\ 0 &0 &0 \ccce \, , 
&e^{(6,1)}_{41} &= \cccb s_{24} &s_{12} &-s_{12} \\s_{24} &s_{12} &-s_{12}  \\ 0 &0&0 \ccce \notag \\
e^{(6,1)}_{35} &= \cccb 0 &0 &0 \\ -s_{25} &s_{25} &s_{23}  \\
-s_{25} &s_{25} &s_{23}   \ccce \, , 
&e^{(6,1)}_{45} &= \cccb 0&0&0\\0&0&0\\ 0 &-s_{25} &s_{24}{+}s_{25} \ccce \label{61.5}
 \\
e^{(6,1)}_{34} &= \cccb 0 &0 &0 \\ -s_{24} &s_{24}{+}s_{23} &-s_{23} \\ 0 &0 &0 \ccce \notag 
\end{align}
in the KZ equations (\ref{conv9})
\begin{align}
\partial_3 F^{(6,1)}_{ab} &= \sum_{c=1}^3 \bigg\{ \frac{ (e^{(6,1)}_{31})_{bc} }{z_{31}} + \frac{ (e^{(6,1)}_{35})_{bc} }{z_{35}} 
+ \frac{ (e^{(6,1)}_{34})_{bc} }{z_{34}} \bigg\} F^{(6,1)}_{ac} 
\label{61.4} \\
\partial_4 F^{(6,1)}_{ab} &= \sum_{c=1}^3 \bigg\{ \frac{ (e^{(6,1)}_{41})_{bc} }{z_{41}} + \frac{ (e^{(6,1)}_{45})_{bc} }{z_{45}} 
+ \frac{ (e^{(6,1)}_{34})_{bc} }{z_{43}} \bigg\} F^{(6,1)}_{ac}  \, .
\label{61.4two}
\end{align}
A convenient strategy is to focus on the differential equation (\ref{61.4}) in $z_3$ and
to solve it in terms of polylogarithms $G(a_j \in \{0,1,z_4\};z_3)$,
\beq
\mathbb G^{(6,1)}_{\{0,1,z_{4}\}}(z_3) 
= \sum_{r=0}^{\infty}  \sum_{a_1,a_2,\ldots ,a_r\atop{ \in \{0,1,z_{4}\}} } G(a_r,\ldots,a_2,a_1;z_3)
E^{(6,1)}_{a_1,z_3}E^{(6,1)}_{a_2,z_3} 
\ldots E^{(6,1)}_{a_r,z_3} \, .
\label{apexp.00}
\eeq
The formal initial value with respect to $z_3=0$ multiplying $\mathbb G^{(6,1)}_{\{0,1,z_{4}\}}(z_3)$ 
from the left is still a function of $z_4$ which 
obeys the differential equation (\ref{61.4two}). The latter at $z_3=0$ is solved by
\begin{align}
\mathbb G^{(6,1)}_{\{0,1\}}(z_4) &= \sum_{r=0}^{\infty}  \sum_{a_1,a_2,\ldots \atop{\ldots,a_r \in \{0,1\}} } G(a_r,\ldots,a_2,a_1;z_4)
E^{(6,1)}_{a_1,z_4}E^{(6,1)}_{a_2,z_4} \ldots E^{(6,1)}_{a_r,z_4} 
\label{axp.24}
\end{align}
with a left-multiplicative factor that does not depend on $z_3$ or $z_4$. Hence,
the dependence of $F_{ab}^{(6,1)}$ on $z_3,z_4$ stems from $\mathbb G^{(6,1)}_{\{0,1\}}(z_4)\mathbb G^{(6,1)}_{\{0,1,z_{4}\}}(z_3) $
multiplying a formal $z_3,z_4 {\rightarrow} 0$ limit from the right,
and the combinations of braid matrices in (\ref{apexp.00}) and (\ref{axp.24}) are
\begin{align}
 E^{(6,1)}_{0,z_4}&=(e^{(6,1)}_{41}{+}e_{34}^{(6,1)})^t
 = \cccb 
 s_{24}& 0& 0\\s_{12}& s_{12} {+} s_{23} {+} s_{24}& 0\\ {-}s_{12}& {-}s_{12} {-} s_{23}& 0
  \ccce  \label{apexp.28}\\
 E^{(6,1)}_{1,z_4}&=(e_{45}^{(6,1)})^t
  = \cccb
  0& 0& 0\\0& 0& -s_{25}\\0& 0& s_{24} {+} s_{25}
    \ccce  \, , \ \ \ \ \ \ 
E^{(6,1)}_{0,z_3}=(e_{31}^{(6,1)})^t
 = \cccb 
 s_{12} {+} s_{23}& 0& 0\\ {-}s_{12}& 0& 0\\0& 0& 0 \ccce
 \notag \\
E^{(6,1)}_{z_4,z_3}&=(e_{34}^{(6,1)})^t
 = \cccb 
 0& {-}s_{24}& 0\\0& s_{23} {+} s_{24}& 0\\0& {-}s_{23}& 0
  \ccce   \, , \ \ \ \ \ \ 
E^{(6,1)}_{1,z_3}=(e_{35}^{(6,1)})^t
 = \cccb
 0& {-}s_{25}& {-}s_{25}\\0& s_{25}& s_{25}\\0& s_{23}& s_{23}
   \ccce \, .\notag
\end{align}
The initial values are determined by the asymptotics
\begin{align}
F^{(6,1)}_{1b}(z_3\rightarrow 0,z_4) &= \delta_{b,1} |z_4|^{s_{24}} \frac{ \Gamma(1{+}s_{12}) \Gamma(1{+}s_{23} ) }{\Gamma(1{+}s_{12}{+}s_{23}) } 
\notag
\\
F^{(6,1)}_{2b}(z_3,z_4\rightarrow z_3) &= \delta_{b,2} |z_3|^{s_{12}} |1{-}z_3|^{s_{25}} \frac{ \Gamma(1{+}s_{23}) \Gamma(1{+}s_{24} ) }{\Gamma(1{+}s_{23}{+}s_{24}) } 
\label{in61.2}
\\
F^{(6,1)}_{3b}(z_3,z_4 \rightarrow 1 ) &= \delta_{b,3} |1{-}z_3|^{s_{23}} \frac{ \Gamma(1{+}s_{24}) \Gamma(1{+}s_{25} ) }{\Gamma(1{+}s_{24}{+}s_{25}) } \, ,
\notag
\end{align}
and one can again use monodromy relations as explained in appendix \ref{app:A.1} to also infer the $z_3 \rightarrow 0$ asymptotics of
$F^{(6,1)}_{2b}$ and $F^{(6,1)}_{3b}$ (contours different from (\ref{61.2}) are necessary in intermediate steps whose limits $z_3 \rightarrow 0$ commute with their $\alpha'$-expansions).
One arrives at the formal limit
\beq
F^{(6,1)}_{ab}(z_3 \rightarrow 0,z_4)  = \cccb
|z_4|^{s_{24}} \frac{ \Gamma(1{+}s_{12}) \Gamma(1{+}s_{23})  }{ \Gamma(1{+}s_{12}{+}s_{23}) }  &0 &0 \\
\frac{ s_{12} \hat F^{(5,1)}_{11}}{s_{12}{+}s_{23}} - K^{(6,1)} &\hat F^{(5,1)}_{11} &\hat F^{(5,1)}_{12} \\
\frac{ s_{12} \hat F^{(5,1)}_{21}}{s_{12}{+}s_{23}}  &\hat F^{(5,1)}_{21} &\hat F^{(5,1)}_{22}
\ccce_{\!ab} \, ,
 \label{in61.21}
\eeq
where the hat notation on the right-hand side stands for changes of arguments, 
\beq
\hat F^{(5,1)}_{ab}=F^{(5,1)}_{ab}(z_4)\Big| \begin{smallmatrix} 
s_{12} \rightarrow s_{12}{+}s_{23} \\
s_{23} \rightarrow s_{24}\\
s_{24} \rightarrow s_{25}
 \end{smallmatrix} \, ,
\label{charg}
\eeq
and the $(a,b)=(2,1)$ entry of (\ref{in61.21}) involves
\begin{align}
K^{(6,1)} &= \frac{ \sin (\pi s_{12}) }{\sin (\pi(s_{12}{+}s_{23})) }|z_4|^{s_{24}}  \frac{ \Gamma(1{+}s_{12}) \Gamma(1{+}s_{23})  }{ \Gamma(1{+}s_{12}{+}s_{23}) } \notag \\
&=   |z_4|^{s_{24}} \frac{ s_{12} }{s_{12}{+}s_{23}} \frac{ \Gamma(1{+}s_{23}) \Gamma(1{-}s_{12}{-}s_{23}) }{\Gamma(1{-}s_{12}) } \, .
 \label{in61.22}
\end{align}
By importing the formal $z_4\rightarrow 0$ limit of $\hat F^{(5,1)}_{ab}$ from
(\ref{expa.21}) with the above replacement rules for the $s_{ij}$, we arrive at
\begin{align}
\mathbb P^{(6,1)}  \mathbb M^{(6,1)} &= \cccb
 \frac{ \Gamma(1{+}s_{12}) \Gamma(1{+}s_{23})  }{ \Gamma(1{+}s_{12}{+}s_{23}) } &0 &0 \\
 \hat K^{(6,1)}_{21} & \frac{ \Gamma(1{+}s_{12}{+}s_{23}) \Gamma(1{+}s_{24})  }{ \Gamma(1{+}s_{12}{+}s_{23}{+}s_{24}) } &0 \\
 \hat K^{(6,1)}_{31} &\hat K^{(6,1)}_{32} &
 \frac{ \Gamma(1{+}s_{12}{+}s_{23}{+}s_{24}) \Gamma(1{+}s_{25})  }{ \Gamma(1{+}s_{12}{+}s_{23}{+}s_{24}{+}s_{25}) }
\ccce
\label{general.29}
\end{align}
with (cf.\ (\ref{apexp.39b}))
\begin{align}
\hat K^{(6,1)}_{21} &= \frac{s_{12} }{s_{12}{+}s_{23}}
\bigg\{ \frac{ \Gamma(1{+}s_{24}) \Gamma(1{+}s_{12}{+}s_{23} ) }{\Gamma(1{+}s_{12}{+}s_{23}{+}s_{24}) }
- \frac{ \Gamma(1{+}s_{23}) \Gamma(1{-}s_{12}{-}s_{23} ) }{\Gamma(1{-}s_{12}) } \bigg\}\, ,
\label{expa.29} \\
\hat K^{(6,1)}_{31} &= \frac{s_{12}}{s_{12}{+}s_{23}{+}s_{24}}
\bigg\{ \frac{ \Gamma(1{+}s_{25}) \Gamma(1{+}s_{12}{+}s_{23}{+}s_{24} ) }{\Gamma(1{+}s_{12}{+}s_{23}{+}s_{24}{+}s_{25}) }
- \frac{ \Gamma(1{+}s_{24}) \Gamma(1{-}s_{12}{-}s_{23}{-}s_{24} ) }{\Gamma(1{-}s_{12}{-}s_{23}) } \bigg\}\, ,
\notag \\
\hat K^{(6,1)}_{32} &= \frac{s_{12}{+}s_{23} }{s_{12}{+}s_{23}{+}s_{24}}
\bigg\{ \frac{ \Gamma(1{+}s_{25}) \Gamma(1{+}s_{12}{+}s_{23}{+}s_{24} ) }{\Gamma(1{+}s_{12}{+}s_{23}{+}s_{24}{+}s_{25}) }
- \frac{ \Gamma(1{+}s_{24}) \Gamma(1{-}s_{12}{-}s_{23}{-}s_{24} ) }{\Gamma(1{-}s_{12}{-}s_{23}) } \bigg\} \, ,
\notag
\end{align}
i.e.\ the $3\times 3$ matrices $P^{(6,1)}_{2k},M^{(6,1)}_{2k+1}$ are again determined by the 
four-point integral (\ref{apexp.36}). The factor of $|z_4|^{s_{24}}$ in (\ref{in61.21}) has been 
replaced by $1$ in the formal $z_4 \rightarrow 0$ limit
since all the regularized polylogarithms in
\beq
|z_4|^{s_{24}} = 1 + \sum_{w=1}^{\infty} s_{24}^w G(\underbrace{0,0,\ldots,0}_{w};z_4)
\eeq
are later on generated by (\ref{axp.24}).
The denominators $(s_{12}{+}s_{23})^{-1}$
and $(s_{12}{+}s_{23}{+}s_{24})^{-1}$ on the right-hand side of (\ref{expa.29}) are cancelled by the differences of Euler beta functions as in (\ref{expa.22}) such that all entries of the matrices $P^{(6,1)}_{2k},M^{(6,1)}_{2k+1}$ determined from (\ref{general.29}) are indeed polynomials in $s_{ij}$.

By the above arguments, the $\ap$-expansion of $F^{(6,1)}_{ab}$ exhibits a 
matrix multiplicative structure
\beq
F^{(6,1)}(z_3,z_4) = \mathbb P^{(6,1)}\mathbb M^{(6,1)}
\mathbb G^{(6,1)}_{\{0,1\}}(z_4)\mathbb G^{(6,1)}_{\{0,1,z_4\}}(z_3)
\label{apexp.27}
\eeq
similar to (\ref{expa.23}), where the building blocks are given by
(\ref{apexp.00}), (\ref{axp.24}), (\ref{general.29}) and (\ref{expa.29}). This representation
realizes the integration of the KZ form $\Omega^{(6,1)}$ in $\dd F^{(6,1)} = \Omega^{(6,1)} F^{(6,1)}$
along the path $(0,0) \rightarrow (0,z_4) \rightarrow (z_3,z_4)$, and the alternative choice of path 
$(0,0) \rightarrow (z_3,0) \rightarrow (z_3,z_4)$ is discussed in section \ref{sec:5}.

\subsection{General result}
\label{sec:3.4}

The structural results (\ref{expa.23}) and (\ref{apexp.27}) on the $\ap$-expansion of $F^{(5,1)}$ and $F^{(6,1)}$
can be readily generalized to higher multiplicity: The KZ equations (\ref{conv9}) can be solved by
the matrix product (\ref{apexp.35}), where the $z_j$-dependent building blocks
\beq
\mathbb G^{(n,p)}_{\{0,1,z_{j+1},z_{j+2},\ldots,z_{n-2}\}}(z_j) 
= \sum_{r=0}^{\infty}   \! \sum_{a_1,a_2,\ldots ,a_r\atop{ \in \{0,1,z_{j+1},z_{j+2},\ldots,z_{n-2}\}} }  \! \! \! \! G(a_r,\ldots,a_2,a_1;z_j)
E^{(n,p)}_{a_1,z_j}E^{(n,p)}_{a_2,z_j} 
\ldots E^{(n,p)}_{a_r,z_j} 
\label{apexp.26}
\eeq
involve the following combinations of braid matrices
\beq
E^{(n,p)}_{z_k,z_j} = (e_{jk}^{(n,p)})^t \quad \forall \ k \neq 1 \, , \ \ \ \ \ \
E^{(n,p)}_{0,z_j} = (e_{j1}^{(n,p)})^t + \sum_{i=p+2}^{j-1} (e_{ij}^{(n,p)})^t \, .
\label{defenp}
\eeq
The choice of fibration basis is adapted to the arrangement (\ref{expa.1}) of the unintegrated 
punctures $z_{p+2},\ldots,z_{n-2}$ on the real line and amounts to integrating the
KZ form $\Omega^{(n,p)}$ in $\dd F^{(n,p)} = \Omega^{(n,p)} F^{(n,p)}$
along the path
\begin{align}
(0,0,\ldots,0) &\rightarrow (0,\ldots,0,z_{n-2}) \rightarrow (0,\ldots,0,z_{n-3},z_{n-2}) \rightarrow
\ldots
\label{integrationOrderDefault}
\\  
&\rightarrow \ldots \rightarrow (0,z_{p+3},\ldots,z_{n-2}) \rightarrow (z_{p+2},z_{p+3},\ldots,z_{n-2}) \, .
 \notag 
\end{align}
The series (\ref{apexp.26}) in polylogarithms act by right-multiplication on the $z_j$-independent matrices 
$ \mathbb P^{(n,p)},\mathbb M^{(n,p)}$ in
(\ref{apexp.35}) that are claimed to carry the MZVs according to (\ref{expa.15}).
As exemplified by (\ref{expa.21}), (\ref{expa.29}) and (\ref{apexp.40}) for $p=1$ 
and appendix \ref{app:A.4} for $(n,p) =(6,2)$, the entries of $ \mathbb P^{(n,p)},\mathbb M^{(n,p)}$ are expected to be expressible in terms of the disk integrals $F^{(k+3,k)}$ in string amplitudes with $k\leq p$. Their compositions can be determined via monodromy relations from the initial values $z_{p+2},\ldots,z_{n-2}\rightarrow 0$ in a basis of contours where these limits for the punctures commute with $\ap$-expansions.

\section{Coaction properties of $F^{(n,p)}_{ab}$ and their building blocks}
\label{sec:4}

The goal of this section is to investigate the coaction formula (\ref{conv7})
of the $F^{(n,p)}_{ab}$ at the level of their factorized $\alpha'$-expansion 
(\ref{apexp.35}). We will identify conjectural coaction properties of 
the building blocks $\mathbb G^{(n,p)}_{\{0,1,z_{j+1},z_{j+2},\ldots,z_{n-2}\}}(z_j)$
in (\ref{apexp.26}) which imply (\ref{conv7}) and mix different braid matrices and the matrices
$M^{(n,p)}_{2k+1}$ accompanying the MZVs. The subsequent expressions for
$\Delta \mathbb G^{(n,p)}$ are generating functions for coactions of polylogarithms: Each contribution
is already cast into a fibration basis, and they drastically simplify order-by-order tests
of (\ref{conv7}).

\subsection{Coaction of multiple polylogarithms}
\label{sec:4.0}

The structures to be described in this section originate from the  coproduct in the Hopf algebra of multiple polylogarithms  taken modulo their branch cuts, or equivalently modulo $i\pi$ \cite{Goncharov:2001iea, Goncharov:2005sla},
\begin{align}\label{goncharov-coproduct}
& \Delta I(a_0;a_1,\ldots,a_n;a_{n+1}) \\
& = \sum_{0=i_0<i_1<\cdots<i_k<i_{k+1}=n+1}
I(a_0;a_{i_1},\ldots,a_{i_k};a_{n+1}) \otimes
\prod_{p=0}^k
I(a_{i_p};a_{i_p+1},\ldots,a_{i_{p+1}-1};a_{i_{p+1}})\,, \notag
\end{align}
where the iterated integrals $I$ are defined as
\beq
I(a_0;a_1,\ldots,a_n;a_{n+1}) = \int_{a_0}^{a_{n+1}}
\frac{\dd t}{t-a_n}I(a_0;a_1,\ldots,a_{n-1};t)\,,
\eeq
and are thus related to the multiple polylogarithms defined in (\ref{expa.3}) by a shift of base point,
\beq\label{defIG}
I(0;a_1,\ldots,a_n;a_{n+1}) = G(a_n,\ldots,a_1;a_{n+1})\,.
\eeq
It is thus possible to convert any integral $I$ with general arguments into combinations of the integrals $G$ (see \cite{Duhr:2012fh} for examples), but the coproduct is more neatly expressed in terms of the former, as seen in (\ref{goncharov-coproduct}).

The coproduct can be lifted to a coaction \cite{Brown:2011ik,Duhr:2012fh} that reincorporates $i\pi$ with the additional definition
\beq\label{coacipi}
\Delta(i\pi) = i\pi \otimes 1\,,
\eeq
which implies
\beq\label{zetaEven}
\Delta(\zeta_n) = \zeta_n\otimes 1\,,\qquad  \textrm{for $n$ even}\,,
\eeq
for even zeta values, in addition to the straightforward operation on odd zeta values,
\beq\label{zetaOdd}
\Delta(\zeta_n) = \zeta_n\otimes 1+ 1\otimes \zeta_n\,,\qquad \textrm{for $n$ odd}\,.
\eeq
Strictly speaking, the coaction is only defined for motivic MZVs $\zeta^{\mathfrak m}_{n_1,\ldots,n_r}$, and we informally omit their superscripts in (\ref{zetaEven}), (\ref{zetaOdd}) and similar equations below. Moreover, the second entries of the coaction feature de Rham periods associated with the respective motivic MZVs. See for instance \cite{Francislecture} for their distinction which is implicit in our notation. The absence of $1 \otimes \zeta_n$ in (\ref{zetaEven}) can be understood from the vanishing of the de Rham version of $\zeta_2$. 

As a consequence of (\ref{goncharov-coproduct}) and (\ref{defIG}), the coaction always includes a particularly simple collection of terms
\beq
\Delta G(u_1,u_2,\ldots,u_w;z) = \sum_{j=0}^w G(u_{j+1},u_{j+2},\ldots,u_w;z) \otimes G(u_1,u_2,\ldots,u_j;z) + \cdots
\label{coprop.2}
\eeq
that arise from deconcatenations of the labels $\vec{u}=(u_1,u_2,\ldots,u_w)$. The terms in the ellipsis in turn still involve polylogarithms
of the form $G(\ldots;z)$ in the first entry, but the second entry carries at least one unit of transcendental weight
via polylogarithms $G(\ldots;u_j)$ that do not depend on $z$ and may reduce to MZVs. 
In other words, the deconcatenation terms in (\ref{coprop.2}) make all terms 
contributing to $\Delta G(\vec{u};z) $ explicit that take the form $G(\ldots;z)\otimes G(\ldots;z)$ with the same original argument $z$ in both entries. This property is perhaps most easily understood from the representation of the terms of the coproduct (\ref{goncharov-coproduct}) as polygons inscribed in a semicircle \cite{Goncharov:2005sla,Duhr:2012fh}.

For generating series of the form in (\ref{apexp.26}), the deconcatenation terms in (\ref{coprop.2}) translate into matrix products: We shall illustrate this in the one-variable case with an abstract version of (\ref{apexp.24})
\beq
\mathbb G_{\{0,1\} }(z) = \sum_{\vec{u} \in \{0,1\}^{\times}} G(\vec{u}^t;z) E_{u_1} E_{u_2}\ldots E_{u_w}\, ,
\label{coprop.3}
\eeq
where $E_0, E_1$ are unspecified matrices without any relations prescribed among their products. 
Here and below, $\vec{u}^t=(u_w,\ldots,u_2,u_1)$ denotes the reversal of $\vec{u}=(u_1,u_2,\ldots,u_w)$,
and we write $\vec{u} \in \{0,1,z,\ldots\}^{\times}$ when all 
words $(u_1,u_2,\ldots,u_w)$ of arbitrary length $w=0,1,2,\ldots$ 
in the alphabet $u_i \in \{0,1,z,\ldots\}$ are summed over.
With row and column indices $a,b,\ldots$ for $E_0$ and $E_1$ as well as
Einstein summation for repeated indices, we have
\begin{align}
\Delta  \mathbb G_{\{0,1\} }(z)_{ab} &= \mathbb G_{\{0,1\} }(z)_{ac} \otimes \mathbb G_{\{0,1\} }(z)_{cb}
\label{coprop.4} \\
& +  \sum_{\vec{u} \in \{0,1\}^{\times}} G(\vec{u};z) \otimes \sum_{\vec{k} \in (2\mathbb N{+}1)^\times}
\phi^{-1}(f_{k_1}f_{k_2}\ldots f_{k_\ell}) W(\vec{u}|\vec{k})_{ac} \mathbb G_{\{0,1\} }(z)_{cb} \, , \notag
\end{align}
where the MZVs arising from the terms in the ellipsis of (\ref{coprop.2}) have been translated 
into the $f$-alphabet (the second entry of the coaction does not admit any $f_2$). The
objects $W(\vec{u}|\vec{k})$ are products of $E_0, E_1$ with rational coefficients whose composition is determined
by (\ref{goncharov-coproduct}). Finally, the right-multiplicative generating series $\mathbb G_{\{0,1\} }(z)_{cb}$ in 
the second entry of (\ref{coprop.4}) ensures the property that the $z$-derivatives operate in the second entry \cite{Duhr:2012fh}, 
\beq
\Delta \partial_z\mathbb  G_{\{0,1\} }(z) = ({\rm id} \otimes  \partial_z) \Delta\mathbb  G_{\{0,1\} }(z) \, .
\label{coprop.5}
\eeq
The fact that each term in the ellipsis of (\ref{coprop.2}) carries at least one unit of weight in
polylogarithms independent on $z$ translates into $W(\vec{u}|\emptyset)=0$ in (\ref{coprop.4}),
i.e.\ each term in the second line involves MZVs with at least one letter $f_{k_i}$.

As a simple example of non-vanishing $W(\vec{u}|\vec{k})_{ac}$ in (\ref{coprop.4}), we rewrite 
\begin{align}
\Delta   G(0,0,1,1;z) &=  1 \otimes G(0,0,1,1;z) + G(1;z) \otimes G(0,0,1;z)  
+ G(1,1;z) \otimes G(0,0;z)     \notag \\
&\ \ \ \
+ G(0,1,1;z) \otimes G(0;z)+ G(0,0,1,1;z) \otimes 1 + G(1;z) \otimes \zeta_3  
\label{coprop.6}
\end{align}
and similar weight-four coactions in generating-function form. Since $1\otimes \zeta_3$ is always accompanied
by $G(1;z)\otimes 1$ rather than $G(0;z)\otimes 1$ in any $\Delta   G(u_1,u_2,u_3,u_4;z) $ with $u_i \in \{0,1\}$,
we have $W(0|3)=0$ and
\begin{align}
W(1|3) &= 
{-}E_0E_0E_1E_1 + 2 E_0E_1E_0E_1 - 2 E_1E_0E_1E_0 + E_1E_1E_0E_0  \notag \\
& \ \ \ \ +  E_0E_1E_1E_1 -  3 E_1E_0E_1E_1  +  3 E_1E_1E_0E_1 - E_1E_1E_1E_0 \label{C01.7}  \\
&=
 [[[ E_0,E_1],E_0],E_1]
+ [[[ E_0,E_1],E_1],E_1]\, . \notag
\end{align}
In the remainder of this section, we specialize the abstract $E_0, E_1$ to the braid matrices 
of various $F^{(n,p)}$ as for instance in (\ref{apexp.LE}) and find relations involving commutators 
of matrices and $M_k^{(n,p)}$.

\subsection{Coaction of $F^{(n,p)}$ with $p=n{-}4$}
\label{sec:4.1}

In this section, we explore the consequences of the coaction property at $p=n{-}4$,
i.e.\ for functions in factorized form (\ref{apexp.35}) that depend on one puncture $z=z_{n-2}$
\beq
F^{(n,n-4)}(z)_{ad} = \mathbb P^{(n,n-4)}_{ab}  \mathbb M^{(n,n-4)}_{bc} \mathbb G^{(n,n-4)}_{\{0,1\}}(z)_{cd} \, ,
\label{coprop.8}
\eeq
see (\ref{expa.14}) and (\ref{expa.15}) for the structure of $\mathbb P^{(n,n-4)}$ and $\mathbb M^{(n,n-4)}$.
We will find recursive relations among the coefficients $W(\vec{u}|\vec{k})$ of the coaction in the second line of (\ref{coprop.4}), and their solution can be resummed in terms of repeated adjoint actions in the generating functions in (\ref{coprop.8}).

The conjectural coaction property for the full disk integrals is
\beq
\Delta F^{(n,n-4)}(z)_{ac} = 
F^{(n,n-4)}(z)_{ab} 
\otimes F^{(n,n-4)}(z)_{bc}  \, ,
\label{coprop.9}
\eeq
and we start by investigating the regularized $z \rightarrow 0$ limit that sets 
$ \mathbb G^{(n,n-4)}_{\{0,1\}}(z) \rightarrow 1$ and relates the contributions 
involving MZVs via
\beq
\Delta ( \mathbb P^{(n,n-4)}_{ab}  \mathbb M^{(n,n-4)}_{bd}  ) = 
\mathbb P^{(n,n-4)}_{ab}  \mathbb M^{(n,n-4)}_{bc} \otimes  \mathbb M^{(n,n-4)}_{cd}  \, .
\label{coprop.10}
\eeq
This $z \rightarrow 0$ limit of (\ref{coprop.9}) is implied by the assumptions (\ref{expa.16}) on $\mathbb P^{(n,n-4)}$ and 
$\mathbb M^{(n,n-4)}$ which in turn follow from the expansion (\ref{expa.14}) and (\ref{expa.15}) in terms of matrices
$P^{(n,n-4)}_w$, $M^{(n,n-4)}_w$ of fixed polynomial degree $w$ in $s_{ij}$. In order for (\ref{coprop.9})
to hold at nonzero $z$, the series $ \mathbb M^{(n,n-4)}$ and $\mathbb G^{(n,n-4)}_{\{0,1\}}(z)$
need to be interrelated through the coaction,
\beq
\Delta \big( \mathbb M^{(n,n-4)}_{ac} \mathbb G^{(n,n-4)}_{\{0,1\}}(z)_{ce} \big) = 
\mathbb M^{(n,n-4)}_{ab} \mathbb G^{(n,n-4)}_{\{0,1\}}(z)_{bc}  \otimes
\mathbb M^{(n,n-4)}_{cd} \mathbb G^{(n,n-4)}_{\{0,1\}}(z)_{de}   \, .
\label{coprop.11}
\eeq
With the property $\Delta  \mathbb M^{(n,n-4)}_{ac}= \mathbb M^{(n,n-4)}_{ab} \otimes \mathbb M^{(n,n-4)}_{bc}$
assumed in (\ref{expa.16}) and the ansatz (\ref{coprop.4}) for the coaction of $\mathbb G^{(n,n-4)}_{\{0,1\}}(z)$,
the desired property (\ref{coprop.11}) implies
\begin{align}
&\mathbb G^{(n,n-4)}_{\{0,1\}}(z)_{ab}  \otimes \mathbb M^{(n,n-4)}_{bc}  - \mathbb G^{(n,n-4)}_{\{0,1\}}(z)_{bc}  \otimes \mathbb M^{(n,n-4)}_{ab} \label{coprop.12} \\
&\ \ \ \ = 
 \sum_{\vec{u} \in \{0,1\}^{\times}} G(\vec{u};z) \otimes \mathbb M^{(n,n-4)}_{ab} \shuffle \sum_{\vec{k} \in (2\mathbb N{+}1)^\times}
\phi^{-1}(f_{k_1}f_{k_2}\ldots f_{k_\ell}) W(\vec{u}|\vec{k})_{bc} 
\notag
\end{align}
upon left- and right-multiplication with the inverses of $ \mathbb M^{(n,n-4)}$ and $\mathbb G_{\{0,1\} }(z)$. 
The shuffle symbol in the second entry acts on the combinations of $f_{k}$ that are explicit in the second 
line of (\ref{coprop.12}) and those in the expansion of $\mathbb M^{(n,n-4)}$. The row- and column indices
$a,b,\ldots$ are spelt out since the order of matrix multiplication does not always line up with the sequence of entries in
the coaction as for instance for the term $ \mathbb G^{(n,n-4)}_{\{0,1\}}(z)_{bc}  \otimes \mathbb M^{(n,n-4)}_{ab}$ on the left-hand side.

By isolating the coefficients of various $G(\vec{u};z) \otimes f_{k_1}f_{k_2}\ldots f_{k_\ell}$ in (\ref{coprop.12}),
one obtains a recursion that relates $W(\vec{u}| k_1,k_2,\ldots ,k_\ell)$ associated with different 
numbers $\ell$ of letters $f_{k}$. With the shorthand notation 
\begin{align}
E(\vec{u}) = E^{(n,n-4)}_{u_w}\ldots 
E^{(n,n-4)}_{u_2} E^{(n,n-4)}_{u_1}
\label{shrtE}
\end{align}
for the matrix product accompanying $G(u_1,u_2,\ldots,u_w;z)$
in (\ref{coprop.3}) and suppressing the superscripts of $M^{(n,n-4)}_{k}$, the coefficient equations at $\ell=0,1,2$ read 
\begin{align}
E(\vec{u})_{ab}  \mathds{1}_{bc} - E(\vec{u})_{bc}   \mathds{1}_{ab} &= W(\vec{u}|\emptyset)_{ac} \notag \\
E(\vec{u})_{ab}   (M_{k_1})_{bc} - E(\vec{u})_{bc}   (M_{k_1})_{ab} &= W(\vec{u}| k_1 )_{ac} + (M_{k_1})_{ab} W(\vec{u}|\emptyset)_{bc} \, ,\label{coprop.13} \\
E(\vec{u})_{ab}   (M_{k_1} M_{k_2})_{bc} - E(\vec{u})_{bc}   (M_{k_1} M_{k_2})_{ab} &= W(\vec{u}| k_1,k_2 )_{ac}
+ (M_{k_1})_{ab} W(\vec{u}|k_2)_{bc} \, , \notag \\
&\ \ \ \
+ (M_{k_2})_{ab} W(\vec{u}|k_1)_{bc}
+ (M_{k_1} M_{k_2})_{ab} W(\vec{u}|\emptyset)_{bc}\, .
\notag
\end{align}
It is easy to see from (\ref{coprop.12}) that the
generalization to coefficients of $G(\vec{u};z) \otimes f_{k_1}f_{k_2}\ldots f_{k_\ell}$ at arbitrary $\ell$
is captured by the deshuffle $\sum_{\vec{p}\shuffle \vec{q} = \vec{k}}$ on the right-hand side. The latter instructs to sum over all pairs $\vec{p}=(p_1,p_2,\ldots,p_i)$ and  $\vec{q}=(q_1,q_2,\ldots,q_j)$ of ordered sets such that a given $\vec{k}=
(k_1,k_2,\ldots,k_\ell)$ with $\ell=i{+}j$ occurs in their shuffle product:
\begin{align}
\big[ E(\vec{u}) \, , \, M_{k_1} M_{k_2} \ldots M_{k_\ell} \big]_{ac} 
&= \sum_{\vec{p}\shuffle \vec{q} = \vec{k}}
 (M_{p_1} M_{p_2} \ldots M_{p_i})_{ab}
 W(\vec{u}|\vec{q})_{bc}\, . \label{coprop.14}
\end{align}
The recursion for the $W(\vec{u}|k_1,k_2,\ldots,k_\ell)$ in (\ref{coprop.13}) and (\ref{coprop.14})
can be straightforwardly solved in terms of nested matrix commutators such as
\begin{align}
W(\vec{u}|\emptyset)&= 0 \ , \notag \\
W(\vec{u}|k_1)&= [ E(\vec{u}),  M_{k_1}] \ , \label{coprop.15} \\
W(\vec{u}|k_1,k_2)&= [ [ E(\vec{u}),  M_{k_1}], M_{k_2} ] \ , \notag
\end{align}
and more generally
\beq
W(\vec{u}|k_1,k_2,\ldots ,k_\ell)= [[ \ldots [ [ E(\vec{u}),  M_{k_1}], M_{k_2} ],\ldots ,M_{k_{\ell-1}}], M_{k_\ell}]\, .
\label{coprop.16}
\eeq
For words $\vec{u}$ of length one, (\ref{coprop.15}) relates the commutators
$[E_0,M_{2k+1}]$ and $[E_1,M_{2k+1}]$ to products of braid matrices. One can for instance find
\begin{align}
[E_0,M_3]&= 0\, , \ \ \ \ \ \
[E_0,M_5]=0  \notag  \\
[E_1,M_3]&=  [[[ E_0,E_1],E_0],E_1]
+ [[[ E_0,E_1],E_1],E_1]
\notag \\
[E_1,M_5]&=
[[[[[  E_0,E_1] , E_0], E_0], E_0], E_1 ]
+ \frac{3}{2}
[[[[[  E_0,E_1] , E_0], E_0], E_1], E_1 ]
  \label{coprop.16a} \\
&+ \frac{1}{2}
[[[[[  E_0,E_1] , E_0], E_1], E_0], E_1 ] 
+ \frac{1}{2}[[[[[  E_0,E_1] , E_1], E_0], E_1], E_1 ]
\notag  \\
&+ \frac{3}{2}
[[[[[  E_0,E_1] , E_1], E_1], E_0], E_1 ]
+
[[[[[  E_0,E_1] , E_1], E_1], E_1], E_1 ] \notag 
\, ,
\end{align}
based on $W(0|3)=W(0|5)=0$, and $W(1|3)$ in (\ref{C01.7}) (with a similar expression for $W(1|5)$). Up to the outermost bracket with $E_1$, the right-hand sides of (\ref{coprop.16a}) match the coefficients of $\zeta_3$ and $\zeta_5$ in the Drinfeld associator $\Phi(E_0,E_1)$ (when reducing the MZVs to the standard conjectural $\QQ$-bases), see (\ref{drineqs.3}) below. Multiples of these expressions also feature as the nested brackets that define the elements $D_{f_3}$ and $D_{f_5}$ in the stable derivation algebra \cite{MR1248930, 2003695}.\footnote{For any element $f(x,y)$ in a free Lie algebra with generators $x,y$, the derivation $D_f$ is defined by proposition 2 of \cite{MR1248930}. The cases of $f(x,y)$ relevant to (\ref{coprop.16a}) are \cite{2003695}
\begin{align*}
f_3 &= -[[x,y],x] - [[x,y],y]
\\
f_5 &= -2 [[[[ x,y],x], x], x] - 3 [[[[ x,y] , x],x],y] - [ [ [ [ x,y],x],y],x] \\
& \ \ \ \
- [[[[ x,y], y], x], y] - 3 [[[[ x,y], y], y], x] - 2 [[[[x,y], y], y], y]
\end{align*}
with $x \rightarrow E_0$ and $y \rightarrow E_1$ which are not to be confused with the $f$-alphabet description of MZVs.}

Given that each $W(\vec{u}|\vec{k})$ in (\ref{coprop.16}) involves the matrix product
$E(\vec{u})$ in (\ref{shrtE}), the sum over
$\sum_{\vec{u} \in \{0,1\}^\times} G(\vec{u};z) W(\vec{u}|\vec{k})$ in (\ref{coprop.4}) 
is expressible in terms of the generating series $\mathbb G_{ \{0,1\}}(z)$ in (\ref{coprop.3}):
The coaction property (\ref{coprop.11}) along with the ansatz (\ref{coprop.4}) are 
equivalent to
\begin{align}
&\Delta \mathbb G^{(n,p)}_{\{0,1\}}(z) = \mathbb G^{(n,p)}_{\{0,1\}}(z) \otimes  \mathbb G^{(n,p)}_{\{0,1\}}(z) 
+ \sum_{k_1 \in 2\mathbb N+1} \big[  \mathbb G^{(n,p)}_{\{0,1\}}(z) \, , \, M^{(n,p)}_{k_1} \big] \otimes \phi^{-1}( f_{k_1} ) \mathbb G^{(n,p)}_{\{0,1\}}(z)  \notag \\
&\ \ \ \ + \sum_{k_1,k_2 \in 2\mathbb N+1} \big[ \, \big[   \mathbb G^{(n,p)}_{\{0,1\}}(z) \, , \, M^{(n,p)}_{k_1} \big] \, , \, M^{(n,p)}_{k_2} \big] \otimes \phi^{-1}( f_{k_1} f_{k_2} )\mathbb G^{(n,p)}_{\{0,1\}}(z) 
\label{C01.25} \\
& \ \ \ \ + \! \! \! \sum_{k_1,k_2,k_3 \in 2\mathbb N+1} \! \! \! \big[ \, \big[ \, \big[  \mathbb G^{(n,p)}_{\{0,1\}}(z) \, , \, M^{(n,p)}_{k_1} \big] \, , \, M^{(n,p)}_{k_2} \big] \, , \, M^{(n,p)}_{k_3} \big] \otimes \phi^{-1}( f_{k_1} f_{k_2} f_{k_3} ) \mathbb G^{(n,p)}_{\{0,1\}}(z)  + \ldots 
\notag \\
&= \! \!\! \sum_{ \vec{k} \in (2\mathbb N+1)^\times}\! \! \! \! \!
\big[  \big[ \! \ldots \! \big[  \big[  \mathbb G^{(n,p)}_{\{0,1\}}(z)  ,  M^{(n,p)}_{k_1} \big]  ,  M^{(n,p)}_{k_2} \big] \! \ldots  ,  M^{(n,p)}_{k_{\ell-1}} \big]   ,  M^{(n,p)}_{k_\ell} \big] \otimes \phi^{-1}( f_{k_1} \ldots f_{k_\ell} ) \mathbb G^{(n,p)}_{\{0,1\}}(z)  \notag
\end{align}
with terms involving four or more $f_{k}$ in the ellipsis in the third line.
In fact, this derivation of (\ref{C01.25}) not only applies to $p=n{-}4$ but also to general values of $p$:
One imposes the coaction properties of $F^{(n,p)}$ to hold for the matrix
product (\ref{apexp.35}) at generic $z=z_{n-2}$ and vanishing $z_{n-3},z_{n-4},\ldots,z_{p+2}$.
Note that the pattern of $f_{k_i}$ and $M_{k_i}$ in (\ref{C01.25}) amounts to translating the matrix products in the
expansion (\ref{expa.15}) of $ \mathbb M ^{(n,p)}$ to the adjoint representation: By
introducing the formal operation
\begin{align}
&X \otimes \textrm{ad}_\textrm{L}  \big( \phi^{-1}(f_{k_1} f_{k_2} \ldots f_{k_r}) M_{k_1} M_{k_2} \ldots M_{k_r}\big) Y 
\label{C01.30}
\\
&= \left[ \left[ \ldots \left[\left[X,M_{k_1}\right],M_{k_2}\right],\ldots ,M_{k_{r-1}} \right], M_{k_r}\right] \otimes 
\phi^{-1}(f_{k_1} f_{k_2} \ldots f_{k_r}) Y \notag
\end{align}
that converts matrix products to nested commutators in the appropriate order
and acts linearly $\textrm{ad}_\textrm{L} (P+Q) =\textrm{ad}_\textrm{L}(P)+\textrm{ad}_\textrm{L}(Q)$,
one can compactly rewrite (\ref{C01.25}) as
\beq
\Delta \mathbb G^{(n,p)}_{\{0,1\}}(z) =  \mathbb G^{(n,p)}_{\{0,1\}}(z) \otimes \textrm{ad}_\textrm{L} \left( \mathbb{M}^{(n,p)} \right) \mathbb G^{(n,p)}_{\{0,1\}}(z) \, .
\label{C01.32}
\eeq
We emphasize that (\ref{C01.25}) is still conjectural and can be thought of as an economic
reformulation of the coaction conjecture (\ref{conv7}) for $F^{(n,p)}$: We have
started to decompose the coaction relation involving all the contributions $G(\ldots;z_{n-2}),\ldots,G(\ldots;z_{p+2})$ and MZVs to the $\alpha'$-expansion of
$F^{(n,p)}$ into simpler coaction formulae for the building blocks in (\ref{apexp.35}).
In the next section, this decomposition will be extended to polylogarithms in several variables.

We have tested (\ref{C01.25}) and (\ref{C01.32}) order by order in the $\alpha'$-expansion, namely up to and including $\alpha'^{11}$ for $(n,p)=(5,1)$ and $\alpha'^{10}$ for $(n,p)=(6,2)$. The relevant braid matrices and $M^{(n,p)}_{2k+1}$ can be found in (\ref{apexp.LE}) and (\ref{expa.21}) for $(n,p)=(5,1)$ as well as appendix \ref{app:A.4} for $(n,p)=(6,2)$. Note in particular that the $\alpha'^{\geq 9}$-orders at $(n,p)=(6,2)$ are sensitive to the commutator structure of $ \big[ [  \mathbb G^{(n,p)}_{\{0,1\}}(z)  ,  M^{(n,p)}_{k_1} ] , M^{(n,p)}_{k_2} \big] \otimes f_{k_1} f_{k_2}$ along with $\zeta_{3,5}$; see appendix \ref{appA4det} for details. These checks go beyond the reach of $(n,p)=(5,1)$ since $[M^{(5,1)}_{3} , M^{(5,1)}_{5}]=0$ and therefore $\big[ [  \mathbb G^{(5,1)}_{\{0,1\}}(z)  ,  M^{(5,1)}_{3} ] , M^{(5,1)}_{5} \big]=\big[ [  \mathbb G^{(5,1)}_{\{0,1\}}(z)  ,  M^{(5,1)}_{5} ] , M^{(5,1)}_{3} \big]$.

\subsection{The general case}
\label{sec:4.2}

In preparation for the multivariable generalization of the expression
(\ref{C01.32}) for $\Delta \mathbb G^{(n,p)}_{\{0,1\}}(z) $, we briefly repeat the
analysis of the previous section in the two-variable case $p=n{-}5$ with $z=z_{n-3}$ and $y=z_{n-2}$,
\beq
F^{(n,n-5)}(z,y)_{ae} = \mathbb P^{(n,n-5)}_{ab}  \mathbb M^{(n,n-5)}_{bc} \mathbb G^{(n,n-5)}_{\{0,1\}}(y)_{cd}
\mathbb G^{(n,n-5)}_{\{0,1,y\}}(z)_{de} 
\label{coprop.41}
\eeq
and study the coaction of $\mathbb G^{(n,n-5)}_{\{0,1,y\}}(z)_{de}$. We will arrive at a compact
form of the generating function for coactions (\ref{coprop.2}) of polylogarithms $G(\vec{u};z)$ with
labels $u_i$ in the three-letter alphabet $\{0,1,y\}$. Again, the general coaction formula (\ref{goncharov-coproduct})
leads to the simple class of terms from deconcatenation of $\vec{u}$ that are explicit in (\ref{coprop.2}),
and we will elaborate on the additional terms in the ellipsis with some $G(\ldots; u_i), \ u_i\in \{0,1,y\}$ in their second entry.
In terms of generating functions
\beq
\mathbb G_{\{0,1,y \} }(z) = \sum_{\vec{u} \in \{0,1,y \}^{\times}} G(\vec{u}^t;z) E_{u_1,z} E_{u_2,z}\ldots E_{u_w,z}
\label{coprop.42}
\eeq
with unspecified matrices $E_{0,z},E_{1,z},E_{y,z}$, it remains to determine the 
$W(\vec{u}|\vec{k}|\vec{m})$ comprising products of matrices with rational coefficients in
\begin{align}
\Delta  \mathbb G_{\{0,1,y\} }(z)_{ab} &= \mathbb G_{\{0,1,y\} }(z)_{ac} \otimes \mathbb G_{\{0,1,y\} }(z)_{cb}+  \sum_{\vec{u} \in \{0,1,y\}^{\times}} G(\vec{u};z)
\label{coprop.43} \\
& \! \! \! \! \! \! \! \! \! \! \! \!  \otimes \sum_{\vec{k} \in (2\mathbb N{+}1)^\times}
\phi^{-1}(f_{k_1}f_{k_2}\ldots f_{k_\ell}) \sum_{\vec{m} \in\{0,1\}^\times}G(\vec{m};y) W(\vec{u}|\vec{k}|\vec{m})_{ac} \mathbb G_{\{0,1,y\} }(z)_{cb} \, . \notag
\end{align}
From coactions at weight two with $G(\ldots;y)$ in their second entry such as
\begin{align}
\Delta G(1,y;z) &= 1 \otimes G(1,y;z) + G(y;z) \otimes G(1;z) + G(1,y;z) \otimes 1
\label{coprop.44}
\\
& \ \ \ \ - G(1;z) \otimes G(0;y) + G(1;z) \otimes G(1;y) - G(y;z) \otimes G(1;y)\, , 
\notag
\end{align}
one can for instance read off
\begin{align}
W(0| \emptyset |0) &= W(0| \emptyset |1) =0 \notag \\
W(y| \emptyset |0) &= [E_{0,z}, E_{y,z}]  
\label{coprop.45}\\
W(1| \emptyset |0) &= - W(1| \emptyset |1) = W(y| \emptyset |1) = [E_{1,z},E_{y,z}] \, .\notag
\end{align}
The matrix products $W(\vec{u}|\vec{k}|\vec{m})$ in the coaction can again be
determined by imposing (\ref{coprop.41}) and furthermore assuming that
$\Delta \mathbb P^{(n,n-5)} = \mathbb P^{(n,n-5)} \otimes 1$
and that (\ref{coprop.11}) holds for $\mathbb M^{(n,n-5)}$ and $\mathbb G^{(n,n-5)}_{\{0,1\}}$.
In this setting, the ansatz (\ref{coprop.43}) for the coaction of interest has to satisfy
\begin{align}
&\mathbb G^{(n,n-5)}_{\{0,1,y\}}(z)_{ab}  \otimes \mathbb M^{(n,n-5)}_{bc} \mathbb G^{(n,n-5)}_{\{0,1\}}(y)_{cd}   - 
\mathbb G^{(n,n-5)}_{\{0,1,y\}}(z)_{cd}  \otimes \mathbb M^{(n,n-5)}_{ab} \mathbb G^{(n,n-5)}_{\{0,1\}}(y)_{bc} \label{coprop.47} \\
&\ = \! \! \!  \!
 \sum_{\vec{u} \in \{0,1,y\}^{\times}} \! \! \! \! G(\vec{u};z) \otimes \mathbb M^{(n,n-5)}_{ab}  \mathbb G^{(n,n-5)}_{\{0,1\}}(y)_{bc} 
  \shuffle \! \! \! \! \! \sum_{\vec{k} \in (2\mathbb N{+}1)^\times
 \atop{ 
\vec{m} \in \{0,1\}^\times  
 } }  \! \! \! \! \!
\phi^{-1}(f_{k_1}f_{k_2}\ldots f_{k_\ell}) G(\vec{m};y) W(\vec{u}|\vec{k}|\vec{m})_{cd} \, .
\notag
\end{align}
By isolating the coefficients of $G(\vec{u};z) \otimes f_{k_1} \ldots f_{k_\ell} G(m_1,\ldots,m_j;y)$,
we obtain a recursion for  $W(\vec{u}|\vec{k}|\vec{m})$ in the total number of letters in $\vec{k}$ and $\vec{m}$.
With the shorthand notation 
\begin{align}
E_z(\vec{u}) = E_{u_w,z} \ldots E_{u_2,z} E_{u_1,z}
\label{shrtEE}
\end{align}
and an expansion 
of $\mathbb G^{(n,n-5)}_{\{0,1\}}(y)$ in terms of $G(\vec{m};y)E_{m_j,y} \ldots E_{m_2,y}E_{m_1,y}$, 
the simplest examples are
\begin{align}
[ E_z(\vec{u}), \mathds{1}]_{ac}  &= W(\vec{u}|\emptyset|\emptyset)_{ac} \notag \\
[ E_z(\vec{u}), M_{k_1}]_{ac}&= W(\vec{u}| k_1|\emptyset )_{ac} + (M_{k_1})_{ab} W(\vec{u}|\emptyset |\emptyset)_{bc} \notag \\
[ E_z(\vec{u}), E_{m_1,y}]_{ac}&= W(\vec{u}| \emptyset |m_1)_{ac} + (E_{m_1,y})_{ab} W(\vec{u}|\emptyset |\emptyset)_{bc}
\notag \\
[E_z(\vec{u}), M_{k_1} M_{k_2}]_{ac} &= W(\vec{u}| k_1,k_2| \emptyset  )_{ac}
+ (M_{k_1})_{ab} W(\vec{u}| k_2 | \emptyset)_{bc} \notag \\
&\ \ \ \
+ (M_{k_2})_{ab} W(\vec{u}|k_1 | \emptyset )_{bc}
+ (M_{k_1} M_{k_2})_{ab} W(\vec{u}| \emptyset | \emptyset)_{bc} \label{coprop.48} \\
[E_z(\vec{u}), E_{m_2} E_{m_1}]_{ac} &= W(\vec{u}| \emptyset  | m_1,m_2 )_{ac}
+ (E_{m_1,y})_{ab} W(\vec{u}|\emptyset |m_2)_{bc} \notag \\
&\ \ \ \
+ (E_{m_2,y})_{ab} W(\vec{u} | \emptyset  |m_1)_{bc}
+ (E_{m_2,y} E_{m_1,y})_{ab} W(\vec{u}|\emptyset | \emptyset )_{bc} \notag \\
[E_z(\vec{u}), M_{k_1} E_{m_1}]_{ac} &= W(\vec{u}| k_1  | m_1 )_{ac}
+ (E_{m_1,y})_{ab} W(\vec{u}| k_1 | \emptyset)_{bc} \notag \\
&\ \ \ \
+ (M_{k_1})_{ab} W(\vec{u} | \emptyset  |m_1)_{bc}
+ (M_{k_1} E_{m_1,y})_{ab} W(\vec{u}|\emptyset | \emptyset )_{bc} \, . \notag
\end{align}
The general formula can again be written in terms of deshuffles similar to (\ref{coprop.14}).
Note that the extraction of these identities from (\ref{coprop.47}) hinges on the fact that all
polylogarithms are already in a fibration basis.

Similar to (\ref{coprop.15}) and (\ref{coprop.16}),
the solution to the recursion furnished by (\ref{coprop.48}) and higher-weight generalizations
features nested commutators, starting with
\begin{align}
W(\vec{u}| \emptyset  | \emptyset )&=0 \, , 
&W(\vec{u}| k_1,k_2  | \emptyset ) &= [[E_z(\vec{u}), M_{k_1}] , M_{k_2} ] \notag  \\
W(\vec{u}| k_1  | \emptyset ) &= [E_z(\vec{u}), M_{k_1}]\, ,
&W(\vec{u}| \emptyset  | m_1 ,m_2) &= [[E_z(\vec{u}), E_{m_2,y}], E_{m_1,y}] \label{coprop.49} \\
W(\vec{u}| \emptyset  | m_1 ) &= [E_z(\vec{u}), E_{m_1,y}] \, ,
&W(\vec{u}| k_1  | m_1 ) &= [[E_z(\vec{u}), M_{k_1}] , E_{m_1,y} ]
\notag
\end{align}
and more generally (note the reversal of the commutators of the $E_{m_i,y}$)
\begin{align}
&W(\vec{u}| k_1,k_2,\ldots,k_\ell  | m_1,m_2,\ldots,m_j) \label{coprop.50} \\
&= [[ \ldots [[ \ldots [[ E_z(\vec{u}), M_{k_1}], M_{k_2}],\ldots , M_{k_\ell}] , E_{m_j,y} ],\ldots ,E_{m_2,y}], E_{m_1,y}] \, .
\notag
\end{align}
As in the transition from (\ref{coprop.15}) to (\ref{C01.25}), we recover the generating
series $\mathbb G_{\{0,1,y\}}(z)$ by summing the combinations of $E_z(\vec{u})$ (defined in (\ref{shrtEE}) and coming from $W(\vec{u}|\vec{k}|\vec{m})$)
and $G(\vec{u};z)$ over $\vec{u} \in \{0,1,y\}^\times$. In the context of the $F^{(n,p)}$ with $y=z_{n-2}$ and $z=z_{n-3}$, this yields
\begin{align}
&\Delta \mathbb G^{(n,p)}_{\{0,1,y\}}(z) =  \mathbb G^{(n,p)}_{\{0,1,y\}}(z) \otimes  \mathbb G^{(n,p)}_{\{0,1,y\}}(z) \notag \\
&\ \ \ \ 
+  \sum_{k_1 \in 2\mathbb N+1} 
\big[  \mathbb G^{(n,p)}_{\{0,1,y\}}(z) \, , \, M^{(n,p)}_{k_1} \big] \otimes \phi^{-1}( f_{k_1})  \mathbb G^{(n,p)}_{\{0,1,y\}}(z) \notag \\
&\ \ \ \ 
+ \sum_{m_1\in \{0,1\}} \big[  \mathbb G^{(n,p)}_{\{0,1,y\}}(z) \, , \, E^{(n,p)}_{m_1,y} \big] \otimes G(m_1;y)  \mathbb G^{(n,p)}_{\{0,1,y\}}(z)  \notag \\
&\ \ \ \ + \sum_{k_1,k_2 \in 2\mathbb N+1} \big[ \, \big[   \mathbb G^{(n,p)}_{\{0,1,y\}}(z) \, , \, M^{(n,p)}_{k_1} \big] \, , \, M^{(n,p)}_{k_2} \big] \otimes \phi^{-1}( f_{k_1} f_{k_2} )\mathbb G^{(n,p)}_{\{0,1,y\}}(z) 
\label{coprop.51} \\
&\ \ \ \ + \sum_{m_1,m_2 \in  \{0,1\}} \big[ \, \big[   \mathbb G^{(n,p)}_{\{0,1,y\}}(z) \, , \, E^{(n,p)}_{m_2,y} \big] \, , \, E^{(n,p)}_{m_1,y} \big] \otimes G(m_1,m_2;y) \mathbb G^{(n,p)}_{\{0,1,y\}}(z) 
\notag \\
&\ \ \ \ + \sum_{m_1\in  \{0,1\} \atop{k_1 \in 2\mathbb N+1}} \big[ \, \big[   \mathbb G^{(n,p)}_{\{0,1,y\}}(z) \, , \, M^{(n,p)}_{k_1} \big] \, , \, E^{(n,p)}_{m_1,y} \big] \otimes \phi^{-1}( f_{k_1} )G(m_1;y) \mathbb G^{(n,p)}_{\{0,1,y\}}(z) +\ldots
\notag \\
&= \! \! \sum_{  \vec{m} \in \{0,1\}^\times \atop{ \vec{k} \in (2\mathbb N+1)^\times }} \! \! \! \!
\big[  \big[  \!\ldots  \! \big[  \big[  \! \ldots  \! \big[  \big[  \mathbb G^{(n,p)}_{\{0,1,y\}}(z)  ,  M^{(n,p)}_{k_1} \big]  ,  M^{(n,p)}_{k_2} \big] \ldots  ,  M^{(n,p)}_{k_{\ell}} \big]   ,  E^{(n,p)}_{m_j,y} \big] \ldots  ,E^{(n,p)}_{m_2,y} \big], E^{(n,p)}_{m_1,y} \big] \notag \\
&\ \ \ \ \ \ \ \ \otimes \phi^{-1}( f_{k_1} f_{k_2}\ldots f_{k_\ell} ) G(m_1,m_2,\ldots,m_j;y) \mathbb G^{(n,p)}_{\{0,1,y\}}(z)  \, .\notag
\end{align}
The sum over $\vec{m}$ and $\vec{k}$ in the last line can be conveniently
absorbed into a generalization of the notation (\ref{C01.30}) to 
\begin{align}
&X \otimes \textrm{ad}_\textrm{L}  \left( \phi^{-1}(f_{k_1} f_{k_2} \ldots f_{k_r}) M_{k_1} M_{k_2} \ldots M_{k_r} G(m_1,m_2,\ldots,m_j;y) E_{m_j,y} \ldots E_{m_2,y} E_{m_1,y}\right) Y  \notag \\
& \ \ = \left[ \left[\ldots\left[ \left[\ldots \left[\left[X,M_{k_1}\right],M_{k_2}\right], \ldots ,M_{k_r} \right], E_{m_j,y}\right]  , \ldots \, E_{m_2,y} \right] , E_{m_1,y} \right] 
\label{coprop.55}
\\
&\ \ \ \ \ \ \ \  \otimes \phi^{-1}( f_{k_1} f_{k_2} \ldots f_{k_r} ) G(m_1,m_2,\ldots,m_j;y) Y \, , \notag
\end{align}
namely
\beq
\Delta \mathbb G^{(n,p)}_{\{0,1,y\}}(z)  =  \mathbb G^{(n,p)}_{\{0,1,y\}}(z)  \otimes 
 \textrm{ad}_\textrm{L}  \big( \mathbb M^{(n,p)} \mathbb G^{(n,p)}_{\{0,1\}}(y) \big)
\mathbb G^{(n,p)}_{\{0,1,y\}}(z) \, .
\label{coprop.56} 
\eeq
We have explicitly verified this to be the case order by order in the $\alpha'$-expansion, namely up to and including $\alpha'^{6}$ for both $(n,p)=(6,1)$ and $(n,p)=(7,1)$.

The strategy of this section to obtain a conjectural coaction formula for the series
$\mathbb G^{(n,p)}$ of polylogarithms in the $F^{(n,p)}$ can be inductively extended
to any number of unintegrated punctures. With the obvious generalization of (\ref{coprop.55})
to several species of braid matrices $E^{(n,p)}_{z_k,z_j}$ in (\ref{defenp}) and polylogarithms in the fibration bases specified below, our conjecture
for the coaction properties of the constituents of (\ref{apexp.35}) is
\begin{align}
&\Delta \mathbb G^{(n,p)}_{\{0,1,z_{j+1},z_{j+2}, \ldots ,z_{n-2}\}}(z_{j})  =   \mathbb G^{(n,p)}_{\{0,1,z_{j+1},z_{j+2}, \ldots ,z_{n-2}\}}(z_{j}) \label{coprop.57}   \\
&\ \ \otimes \textrm{ad}_\textrm{L} \Big( \mathbb M ^{(n,p)} \mathbb G^{(n,p)}_{\{0,1\}}(z_{n-2}) \mathbb G^{(n,p)}_{\{0,1,z_{n-2}\}}(z_{n-3}) \ldots
 \notag  \\
& \hspace{3.5cm} 
\ldots\mathbb G^{(n,p)}_{\{0,1,z_{j+2},z_{j+3}, \ldots ,z_{n-2}\}}(z_{j+1})            \Big)  \mathbb G^{(n,p)}_{\{0,1,z_{j+1},z_{j+2}, \ldots ,z_{n-2}\}}(z_{j})  \, .
\notag
\end{align}
These coaction formulae at $j=3,4, \ldots ,n{-}2$ and (\ref{expa.16}) are necessary and sufficient conditions for
the factorized $\alpha'$-expansion (\ref{apexp.35}) to obey the coaction formula (\ref{conv7}) of the $F^{(n,p)}$.
In an order-by-order check of the coaction properties of the $\alpha'$-expansion, the individual cases 
of (\ref{coprop.57}) are considerably simpler to verify than dealing with the complete expressions for
$F^{(n,p)}$ at once. The simplest examples of (\ref{coprop.57}) with $j=n{-}2$ and $j=n{-}3$ can be found in
(\ref{C01.32}) and (\ref{coprop.56}), respectively. We have performed the order-by-order checks for the cases with
$(n,p)=(5,1),(6,2),(6,1)$ and $(7,1)$ to the orders of $\ap^{11},\ap^{10},\ap^{6}$ and $\ap^{6}$, respectively.

\section{Analytic continuation}
\label{sec:5}

In this section we study the analytic continuation of the functions $F^{(n,p)}_{ab}(z_{p+2},\ldots,z_{n-2})$ while keeping the orthonormal bases of forms and cycles fixed. Previously, we have defined this family of functions  with a specific branch choice in mind: the branch consistent with
\beq
0=z_1<z_{p+2}<z_{p+3}< \ldots <z_{n-2}<z_{n-1}=1
\eeq
when all the punctures sit on the real line.\footnote{This is the usual branch choice for the polylogarithms appearing in $F^{(n,p)}(z_{p+2},\ldots,z_{n-2})$.} 
This branch choice is implicit in our selection of cycles, ${\gamma^{(n,p)}_a}$; the regularized initial values for these functions,  $\mathbb P^{(n,p)} 
\mathbb M^{(n,p)}$; and explicit in the order of path-ordered integration from these initial values -- schematically shown in (\ref{integrationOrderDefault}) -- which induces a fibration basis on the multiple polylogarithms appearing in the $\ap$-expansion of $F^{(n,p)}(z_{p+2},\ldots,z_{n-2})$.  

The key nontrivial example to keep in mind for this section is $F^{(6,1)}(z_{p+2},z_{n-2})=F^{(6,1)}(z_{3},z_{4})$, where we have assumed
$0<z_3<z_4<1$ as discussed in section \ref{sec:3.3}. The analytic continuation of this function into the branch $\{z_4<z_3\}$ has to be seen \textit{not} as a permutation of $z_3$ and $z_4$ but rather as a \textit{braiding} of these punctures. 
Fortunately, the theory of the KZ equations provides a representation of the braid group acting on certain solutions to these equations \cite{KasselQuantumGroupsBook}. In what follows and in appendix \ref{app:B}, we spell out how this representation furnishes a  \textit{group action} on the solution space in which our functions $F^{(n,p)}(z_{p+2},\ldots,z_{n-2})$ live.

\subsection{Warm-up example: Monodromies of $F^{(5,1)}(z_3)$}

A monodromy is, of course, an  example of analytic continuation. Because all the $F^{(n,p)}(z_{p+2},\ldots,z_{n-2})$ are themselves defined to be holomorphic functions, the solutions of our KZ equations have certain \textit{prescribed monodromies}.\footnote{One can build solutions to the KZ equations with no monodromy, by using certain non-holomorphic initial values, see the discussion of sphere integrals and single-valued polylogarithms in section \ref{sec:6} and for instance \cite{Brown:2013gia, DelDuca:2016lad}.}

In the case of $F^{(5,1)}(z_3) = \mathbb P^{(5,1)}  \mathbb M^{(5,1)} \mathbb G^{(5,1)}_{\{0,1\}}(z_3)$, the monodromy  is determined solely from the generating series of  polylogarithms $\mathbb G^{(5,1)}_{\{0,1\}}(z_3)$ in (\ref{apexp.24}). For example, the monodromies for $z_3$ going anticlockwise around $0$ and $1$ are given by
\begin{align}
\mathcal{M}_{0,z_3} \mathbb G^{(5,1)}_{\{0,1\}}(z_3) &=
\exp(2 \pi i E^{(5,1)}_{31})
\mathbb G^{(5,1)}_{\{0,1\}}(z_3)
\label{eqn5p1monodromy} \\
\mathcal{M}_{1,z_3} \mathbb G^{(5,1)}_{\{0,1\}}(z_3) &=
\Phi(E_{31}^{(5,1)},E_{34}^{(5,1)})
\exp(2 \pi i E^{(5,1)}_{34})
\Phi(E_{31}^{(5,1)},E_{34}^{(5,1)})^{-1}
 \mathbb G^{(5,1)}_{\{0,1\}}(z_3)\, ,
\notag
\end{align}
where the exponentials generalize the weight-one identities 
\begin{align}
\mathcal{M}_{0,z_3} G(0;z_3) =
G(0;z_3)  +2 \pi i
\, , \ \ \ \ \ \ 
\mathcal{M}_{1,z_3} G(1;z_3) =
G(1;z_3)  + 2 \pi i \, .
\label{weionemono}
\end{align}
Throughout this section, we shall use the shorthand
\beq
E_{ij}^{(n,p)} = (e_{ij}^{(n,p)})^t
\label{drineqs.1}
\eeq
for transposed braid matrices, not to be confused with the special combinations $E_{i,z_j}^{(n,p)}$ or
$E_{z_i,z_j}^{(n,p)}$ in (\ref{apexp.28})
with $z$-variables appearing in the subscript. In the second line of (\ref{eqn5p1monodromy}), the expression
\beq
\Phi(E_{31}^{(5,1)},E_{34}^{(5,1)}) = 
 \mathbb G^{(5,1)}_{\{0,1\}}(z_3{=}1)
 \label{drineqs.2}
\eeq
is a special case of the Drinfeld associator whose expansion in terms of MZVs and arbitrary non-commutative indeterminates $E_0,E_1$ is given by \cite{LeMura} 
\begin{align}
\Phi(E_{0},E_{1}) &= \sum_{r=0}^{\infty}  \sum_{a_1,a_2,\ldots \atop{\ldots,a_r \in \{0,1\}} } G(a_r,\ldots,a_2,a_1;1)
E_{a_1}E_{a_2} \ldots E_{a_r} \notag \\
&=1 + \zeta_2 [E_0,E_1] - \zeta_3[E_0{+}E_1,[E_0,E_1]]+ \ldots \, ,
\label{drineqs.3}
\end{align}
in lines with (\ref{apexp.24}).
Its inverse can be written in two different ways:
\begin{align}
\Phi(E_{0},E_{1})^{-1} &= \sum_{r=0}^{\infty} (-1)^r \sum_{a_1,a_2,\ldots \atop{\ldots,a_r \in \{0,1\}} } G(a_1,a_2,\ldots,a_r;1)
E_{a_1}E_{a_2} \ldots E_{a_r} \notag \\
&= \Phi(E_{1},E_{0}) \, .
\label{drineqs.4}
\end{align}
Thus, the monodromy of $F^{(5,1)}(z_3)$ is given by \cite{Brown:2009qja} 
\begin{align}
\mathcal{M}_{0,z_3}   \mathbb P^{(5,1)}  \mathbb M^{(5,1)}  \mathbb G^{(5,1)}_{\{0,1\}}(z_3) =  \mathbb P^{(5,1)}  \mathbb M^{(5,1)}  
\exp(2 \pi i E^{(5,1)}_{31})
\mathbb G^{(5,1)}_{\{0,1\}}(z_3)\,,
\label{monodromyOfFn5p1}
\end{align}
with a similar expression for $\mathcal{M}_{1,z_3}   \mathbb P^{(5,1)}  \mathbb M^{(5,1)}  \mathbb G^{(5,1)}_{\{0,1\}}(z_3)$.
Also for $F^{(n,p)}(z_{p+2},\ldots,z_{n-2})$ at more general $n,p$, the
monodromies are clearly determined by the  generating series of polylogarithms, i.e.\ the $\mathbb G^{(n,p)}$ in (\ref{apexp.26}). The monodromies of generating functions of multiple polylogarithms, as studied in this work, have already been spelled out in detail in \cite{Brown:2009qja}. From now on we will focus on the analytic continuation of the functions $F^{(n,p)}(z_{p+2},\ldots,z_{n-2})$ from $z_i<z_{i+1}$ to branches with
$z_{i+1}<z_i$, which are not monodromies.

\subsection{Warm-up example: Analytic continuation of $F^{(6,1)}(z_3,z_4)$}
\label{sec:warmupp}

We shall now study the analytic continuation of $F^{(n,p)}(z_{p+2},\ldots ,z_{n-2})$ from $0<z_{p+2}<z_{p+3}<\ldots <z_{n-2}<1$ to different arrangements of the unintegrated punctures $z_j$ with $j=p{+}2,\ldots,n{-}2$ in the unit interval. These analytic continuations are implemented via braid-group generators $\sigma_{j,j+1}$ involving unintegrated punctures $z_j,z_{j+1}$ which have not been ${\rm SL}_2$-fixed to $(0,1,\infty)$. More details on braid groups and examples involving $(0,1,\infty)$ can be found in appendix \ref{app:B}.

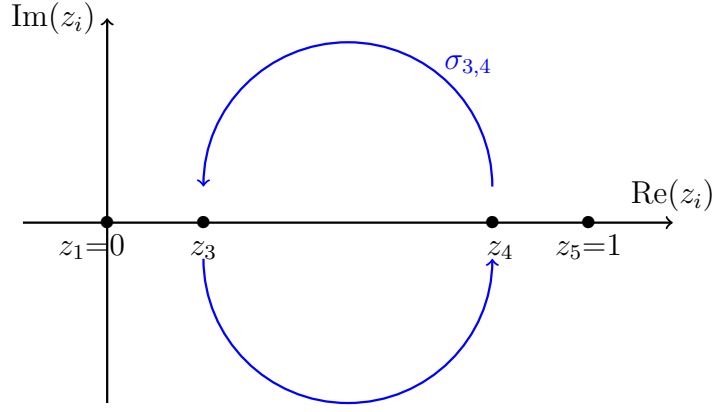
\begin{figure}
	\begin{center}
		\begin{tikzpicture}[scale = 1.6,line width=0.30mm]
			\draw[->](-2.7,0) -- (2.7,0)node[above]{$\Re(z_i)$};
			\draw[->](-2,-1.5) -- (-2,1.7)node[left]{$\Im(z_i)$};
			\draw(-2,0.00)node{$\bullet$}node[below]{$z_1{=}0\ \ \ $};
			\draw(-1.2,0.00)node{$\bullet$}node[below]{$\phantom{0}z_3\phantom{0}$};
			\draw(1.2,0.00)node{$\bullet$}node[below]{$\phantom{0} z_4$};
			\draw(2.0,0.00)node{$\bullet$}node[below]{$z_5{=}1$};
			\draw[blue,->] (1.2,0.3) arc (0:180:1.2cm);
			\draw[blue] (1,1.3)node{$\sigma_{3,4}$};
			\draw[blue,->] (-1.2,-0.3) arc (-180:0:1.2cm);
		\end{tikzpicture}
	\end{center}
	\caption{The elementary braid operation $\sigma_{3,4}$ braids puncture $z_4$ counterclockwise around $z_3$. }
	\label{fig:braidsigma34}
\end{figure}

The formula for the monodromy of $F^{(5,1)}(z_3)$ in (\ref{monodromyOfFn5p1}) suggests that to understand the analytic continuation of $F^{(6,1)}(z_3,z_4)$ into $\{z_4<z_3\}$, we need to focus on the analytic continuation of its generating series of polylogarithms, $\mathbb G^{(6,1)}_{\{0,1\}}(z_4)\mathbb G^{(6,1)}_{\{0,1,z_{4}\}}(z_3) $.  
There is only one element of the braid group we will consider, which is $\sigma_{3,4}$, which braids punctures $z_3$ and $z_4$ around each other, with $z_4$ going around $z_3$ counterclockwise. This choice of orientation of the braiding is depicted in figure \ref{fig:braidsigma34} and determines the phase in
\beq
\log(z_3{-}z_4) = \log \big(e^{i \pi}(z_4{-}z_3)\big) = \log(z_4{-}z_{3})+ i \pi \,
\label{angez3z4}
\eeq
or equivalently in 
\beq
G(z_3;z_4) = G(z_4;z_3) +G(0;z_4)- G(0;z_3) +  i \pi \, .
\label{changez3z4}
\eeq
Similarly, our choice of $\sigma_{3,4}$ fixes the prescription to perform a change of fibration basis from arbitrary $G(a_j \in \{0,1,z_4\};z_3)$ to $G(a_j \in \{0,1,z_3\};z_4)$. One can view
(\ref{changez3z4}) as the braiding analogue of the weight-one monodromies (\ref{weionemono}). In the same way as the latter have a compact uplift to generating functions in (\ref{eqn5p1monodromy}), the generalizations of (\ref{changez3z4}) to higher weight are most conveniently given at the level of the generating function
$\mathbb G^{(6,1)}_{\{0,1,z_{4}\}}(z_3) $, see (\ref{eqBraidn6p1GGeqXGG}) below. More precisely, we will study the $\sigma_{3,4}$ action on the combination
\beq
\mathcal G^{(6,1)}(z_3,z_4) \vcentcolon =  \mathbb G^{(6,1)}_{\{0,1\}}(z_4)\mathbb G^{(6,1)}_{\{0,1,z_{4}\}}(z_3)
\label{calG34}
\eeq
entering the $\alpha'$-expansion of $F^{(6,1)}(z_3,z_4) = \mathbb P^{(6,1)} \mathbb M^{(6,1)} \mathcal G^{(6,1)}(z_3,z_4)$ in (\ref{apexp.27}).
As discussed in section \ref{sec:3.3}, the matrix product $\mathcal G^{(6,1)}(z_3,z_4)$ is a solution of the KZ equations (\ref{61.4}) and (\ref{61.4two}) obtained from integrating the form $\Omega^{(6,1)}$ in $\dd F^{(6,1)} = \Omega^{(6,1)} F^{(6,1)}$ along the path
$(0,0) \rightarrow (0,z_4) \rightarrow (z_3,z_4) $. The braid-group generator $\sigma_{3,4}$ maps (\ref{calG34}) to another solution $\tilde{\mathcal G}^{(6,1)}(z_4,z_3)$ of the same KZ equation where the form $\Omega^{(6,1)}$ is now integrated along the alternative path
\beq
(0,0) \rightarrow (z_3,0) \rightarrow (z_3,z_4) 
\label{integrationOrderAltn6p1}
\eeq
adapted to the branch choice after braiding, i.e.
\beq
0=z_1<z_4<z_3<z_5=1 \, .
\label{branchChoiceAltn6p1}
\eeq
By the arguments in section \ref{sec:3.3}, the solution  $\tilde{\mathcal G}^{(6,1)}(z_4,z_3)$ due to (\ref{integrationOrderAltn6p1}) is composed of
\beq
\tilde{\mathcal G}^{(6,1)}(z_4,z_3)   =  \tilde{\mathbb G}^{(6,1)}_{\{0,1\}}(z_3) \tilde{\mathbb G}^{(6,1)}_{\{0,1,z_{3}\}}(z_4)\, ,
\label{calG43}
\eeq
where the form of $\tilde{\mathbb G}^{(6,1)}_{\{0,1\}}(z_3)$ and $\tilde{\mathbb G}^{(6,1)}_{\{0,1,z_{3}\}}(z_4)$ follows from the way we perform the path-ordered integration, namely
\begin{align}
\tilde{\mathbb G}^{(6,1)}_{\{0,1\}}(z_3) &= \sum_{r=0}^{\infty}  \sum_{a_1,a_2,\ldots,a_r\atop{\in \{0,1\}} } G(a_r,\ldots,a_2,a_1;z_3)
\tilde{E}^{(6,1)}_{a_1,z_3}\tilde E^{(6,1)}_{a_2,z_3} \ldots \tilde E^{(6,1)}_{a_r,z_3} \, ,
\\
\tilde{\mathbb G}^{(6,1)}_{\{0,1,z_{3}\}}(z_4) 
&= \sum_{r=0}^{\infty}  \sum_{a_1,a_2,\ldots ,a_r\atop{ \in \{0,1,z_{3}\}} } G(a_r,\ldots,a_2,a_1;z_4)
\tilde{E}^{(6,1)}_{a_1,z_4}\tilde E^{(6,1)}_{a_2,z_4} 
\ldots \tilde E^{(6,1)}_{a_r,z_4} \, .
\end{align}
The $\tilde E^{(6,1)}_{a_r,z_j}$ matrices are also determined by the integration order in (\ref{integrationOrderAltn6p1}),
\begin{align}
	\tilde E^{(6,1)}_{0,z_3}&=(e^{(6,1)}_{31}{+}e_{34}^{(6,1)})^t
	= \cccb 
	s_{12}{+}s_{13}& {-}s_{24}& 0\\{-}s_{12}& s_{23} {+} s_{24}& 0\\0& {-} s_{23}& 0
	\ccce \, ,  \label{nwapexp.28} \\
	\tilde E^{(6,1)}_{1,z_3}&=(e_{35}^{(6,1)})^t
	= \cccb
	0& {-}s_{25}& {-}s_{25}\\0& s_{25}& s_{25}\\0& s_{23}& s_{23}
	\ccce  \, , \ \ \ \ \ \ 
	\tilde E^{(6,1)}_{0,z_4}=(e_{41}^{(6,1)})^t
	= \cccb 
	s_{24} & s_{24}& 0\\ s_{12}& s_{12}& 0\\{-}s_{12}& {-}s_{12}& 0 \ccce \, ,
	\notag \\
	\tilde E^{(6,1)}_{z_3,z_4}&=(e_{34}^{(6,1)})^t
	= \cccb 
	0& {-}s_{24}& 0\\0& s_{23} {+} s_{24}& 0\\0& {-}s_{23}& 0
	\ccce   \, , \ \ \ \ \ \ 
	\tilde E^{(6,1)}_{1,z_4}=(e_{45}^{(6,1)})^t
	= \cccb
	0& 0& 0\\0& 0& -s_{25}\\0& 0& s_{24} {+} s_{25}
	\ccce  \,  .\notag
\end{align}
See (\ref{apexp.28}) for the analogous braid matrices in $\mathbb G^{(6,1)}_{\{0,1\}}(z_4)\mathbb G^{(6,1)}_{\{0,1,z_{4}\}}(z_3)$ that arise from the earlier choice of integration path $(0,0) \rightarrow (0,z_4) \rightarrow (z_3,z_4) $.

Since $\mathcal G^{(6,1)}(z_3,z_4)$ and $\tilde{\mathcal G}^{(6,1)}(z_4,z_3)$ solve the same KZ equations, they must be related by a left-multiplicative constant series $ \mathbb X^{(6,1)}$,
\beq
\tilde{\mathcal G}^{(6,1)}(z_4,z_3)= \sigma_{3,4 } \mathcal G^{(6,1)} ( z_3,z_4 ) =  \mathbb X^{(6,1)}\left(\sigma_{3,4}\right) \mathcal G^{(6,1)}\left(z_3,z_4  \right) \,. 
\label{alsocooleq}
\eeq
Comparison of (\ref{calG34}) and (\ref{calG43}) with the
phase of (\ref{changez3z4}) in changing fibration basis completely determines the series $\mathbb X^{(6,1)}$ to be
\beq
\mathbb X^{(6,1)}\left(\sigma_{3,4}\right) =  \Phi(E^{(6,1)}_{41},E^{(6,1)}_{34})  \exp (i \pi  E^{(6,1)}_{34}) \Phi(E^{(6,1)}_{34},E^{(6,1)}_{31}) \, .
\label{cooleq}
\eeq
The composition of Drinfeld associators (\ref{drineqs.3}) with the exponential of a braid matrix resembles the structure of the ${\cal M}_{1,z_3}$ monodromy (\ref{eqn5p1monodromy}). However, the phase
of $\exp (i \pi  E^{(6,1)}_{34})$ in the braiding relation
\begin{align}
\label{eqBraidn6p1GGeqXGG}
\tilde{\mathbb G}^{(6,1)}_{\{0,1\}}(z_3) \tilde{\mathbb G}^{(6,1)}_{\{0,1,z_{3}\}}(z_4)   &=  
 \Phi(E^{(6,1)}_{41}\!\!\!,E^{(6,1)}_{34})  \exp (i \pi E^{(6,1)}_{34}) \Phi(E^{(6,1)}_{34} \! \!\!,E^{(6,1)}_{31})
\mathbb G^{(6,1)}_{\{0,1\}}(z_4)\mathbb G^{(6,1)}_{\{0,1,z_{4}\}}(z_3)  
\end{align}
is half of the phase in the monodromies (\ref{eqn5p1monodromy}). We have used \texttt{PolyLogTools} \cite{Duhr2019PolylogTools} to perform the changes of fibration basis to verify (\ref{eqBraidn6p1GGeqXGG}) order by order in the $E_{ij}^{(6,1)}$ or Mandelstam invariants. 

Note that the signs of the $i \pi$-terms in (\ref{angez3z4}) and (\ref{changez3z4}) as well as the phases in the generating-function identities (\ref{cooleq}) and (\ref{eqBraidn6p1GGeqXGG}) are reversed when changing the orientation of the braiding $\sigma_{3,4}$. The analogous signs of $i \pi$ in the fibration-basis formulas returned by computer packages are controlled by 
the sign of imaginary part of $z_3$ in case of \texttt{HyperInt} \cite{Panzer2015HyperInt} and by the sign of $\arg(z_3)-\arg(z_4)$ in case of
\texttt{PolyLogTools} \cite{Duhr2019PolylogTools}, respectively.\footnote{In particular, (\ref{eqBraidn6p1GGeqXGG}) is consistent with the numerics of \texttt{PolyLogTools} if $\arg(z_3)>\arg(z_4)$.}

Going back to our original question about analytic continuation, we can now pinpoint the behavior of our solution $F^{(6,1)}(z_3,z_4) =\mathbb P^{(6,1)} \mathbb M^{(6,1)} \mathcal G^{(6,1)} (z_3,z_4)$ in passing from $z_3<z_4$ to $z_4<z_3$ for real $z_3,z_4$. Instead of analytically continuing each  polylogarithm in the $\alpha'$-expansion of $F^{(6,1)}(z_3,z_4)$, we have performed this analytic continuation at the level of the generating function. With the constant matrix $\mathbb{X}^{(6,1)} (\sigma_{3,4})$ in (\ref{cooleq}) and the composition of the polylogarithmic series $\tilde{ \mathcal G}^{(6,1)}(z_4,z_3)$ in (\ref{calG43}), we have
\beq
F^{(6,1)}(z_3,z_4) = \mathbb P^{(6,1)} \mathbb M^{(6,1)} \left[ \mathbb{X}^{(6,1)} (\sigma_{3,4})\right]^{-1} \tilde{ \mathcal G}^{(6,1)}(z_4,z_3) \, ,
\label{61ancont.1}
\eeq
in terms of functions naturally defined on the branch (\ref{integrationOrderAltn6p1}), or equivalently
\beq
\sigma_{3,4}F^{(6,1)}(z_3,z_4) = \mathbb P^{(6,1)} \mathbb M^{(6,1)} 
\tilde{ \mathcal G}^{(6,1)}(z_4,z_3) \, .
\label{61ancont.2}
\eeq
While (\ref{61ancont.1}) is simply a rewriting of $F^{(6,1)}(z_3,z_4) =\mathbb P^{(6,1)} \mathbb M^{(6,1)} \mathcal G^{(6,1)} (z_3,z_4)$, the image (\ref{61ancont.2}) under the braid group  furnishes the analytic continuation of $F^{(6,1)}(z_3,z_4)$ into the branch with $0<z_4<z_3<1$. Further examples of analytic continuations to one of $z_3,z_4$ being $<0$ or $>1$ follow the lines of the $(n,p)=(5,1)$ example in appendix \ref{appbtwo}.

\subsection{Initial values of $F^{(6,1)}$ and their coaction}
\label{initbraid}

As a side effect of (\ref{61ancont.1}), it determines a formal initial value for $F^{(6,1)}$ adapted to
path-ordered integration in the order shown in (\ref{integrationOrderAltn6p1}). On top of the
$\mathbb Q$-linear combinations of MZVs seen in the $\ap$-expansion (\ref{expa.14}), (\ref{expa.15}) of the initial values $\mathbb P^{(6,1)} \mathbb M^{(6,1)}$, the $\ap$-expansion of  the initial value in (\ref{61ancont.1}) involves powers of $i\pi$. Hence, we reorganize
\beq
\mathbb P^{(6,1)} \mathbb M^{(6,1)} \left[ \mathbb{X}^{(6,1)} (\sigma_{3,4})\right]^{-1}
=  \tilde {\mathbb P}^{(6,1)} \tilde {\mathbb M}^{(6,1)} 
\label{newini.1}
\eeq
with an expansion of $ \tilde {\mathbb P}^{(6,1)}$, $ \tilde {\mathbb M}^{(6,1)} $ in terms of alternative $3\times 3$ matrices $\tilde P^{(6,1)}_w,\tilde M^{(6,1)}_w$
whose entries are still degree-$w$ polynomials in $s_{ij}$ with rational coefficients:
\begin{align}
\tilde {\mathbb P}^{(6,1)} &= 
\mathds{1}+i \pi \tilde P^{(6,1)}_{1} + \zeta_2 \tilde P^{(6,1)}_{2} + i \pi \zeta_2 \tilde P^{(6,1)}_{3}+ \zeta_2^2 \tilde P^{(6,1)}_{4} + i\pi \zeta_2^2 \tilde P^{(6,1)}_{5} +\zeta_2^3 \tilde P^{(6,1)}_{6}+ {\cal O}(s_{ij}^7)
\notag \\
&=\mathds{1}+\sum_{k=1}^\infty ( i \pi \zeta_2^{k-1} \tilde P^{(6,1)}_{2k-1}
 + \zeta_2^{k} \tilde P^{(6,1)}_{2k})\,, \label{newini.2}
\\
\tilde {\mathbb M}^{(6,1)} &= \phi^{-1}\sum_{r=0}^{\infty} \sum_{k_1,k_2,\ldots, k_r =1}^{\infty} f_{2k_1+1} f_{2k_2+1} \ldots f_{2k_r+1}  \tilde M^{(6,1)}_{2k_1+1}   \tilde M^{(6,1)}_{2k_2+1}\ldots  \tilde M^{(6,1)}_{2k_r+1}\,.  \notag
\end{align}
The $\tilde P^{(6,1)}_{w}$-matrices associated with odd $w$ do not have any counterparts
in the expansion of $\mathbb P^{(6,1)}$. Moreover, the $ \tilde P^{(6,1)}_{2k} $ and $ \tilde M^{(6,1)}_{2k+1} $ resulting
from (\ref{newini.1}) differ from the $P^{(6,1)}_{2k} $ and $M^{(6,1)}_{2k+1} $ determined by (\ref{general.29}) and (\ref{expa.29}) as exemplified in appendix \ref{tildepms}.

Still, the coefficients of the MZVs and their products with $i\pi$ in (\ref{newini.1}) are expected to be compatible with the coaction principle in the sense that
\beq
\Delta( \tilde {\mathbb P}^{(6,1)}_{ab} \tilde {\mathbb M}^{(6,1)}_{bd}) = 
\tilde {\mathbb P}^{(6,1)}_{ab} \tilde {\mathbb M}^{(6,1)}_{bc} \otimes  \tilde {\mathbb M}^{(6,1)}_{cd} \,,
\label{coprop.90}
\eeq
which we have tested up to and including the order of $\ap^8$. At the level of the MZVs that solely arise from words in $f_{2k+1}$, the coaction (\ref{coprop.90}) is again equivalent to an expansion
\begin{align}
 \tilde {\mathbb M}^{(6,1)} &=
\mathds{1} + \zeta_3  \tilde M^{(6,1)}_3 + \zeta_5  \tilde M^{(6,1)}_5+\frac{1}{2} \zeta_3^2  \tilde M^{(6,1)}_3  \tilde M^{(6,1)}_3
+ \zeta_7  \tilde M^{(6,1)}_7 \notag \\
&\ \ \ \ + \zeta_3\zeta_5  \tilde  M^{(6,1)}_5  \tilde  M^{(6,1)}_3
+ \frac{1}{5} \zeta_{3,5}[  \tilde M^{(6,1)}_5,  \tilde  M^{(6,1)}_3]
+{\cal O}(s_{ij}^{9})
\label{coprop.99}
\end{align}
as in (\ref{expa.7}), where the commutator $[  \tilde M^{(6,1)}_5,  \tilde  M^{(6,1)}_3]$ vanishes just like $[   M^{(6,1)}_5,    M^{(6,1)}_3]=0$. In other words, $\zeta_{3,5}$ drops out from $\tilde {\mathbb M}^{(6,1)}$ in the same way as it does from $\mathbb M^{(6,1)}$. In fact, we have checked that all irreducible MZVs of depth $\geq 2$ at weight $\leq 11$ already cancel from the individual Drinfeld associators in (\ref{cooleq}).

Moreover, already the matrix-multiplicative structure on the right-hand side of (\ref{newini.1}) is
not manifest on its left-hand side. Hence, the fact that the coefficients of $i\pi \zeta_3, i \pi \zeta_5$ and $i \pi \zeta_2 \zeta_3$ in (\ref{newini.1}) are given by matrix products $\tilde P^{(6,1)}_1\tilde M^{(6,1)}_3, \tilde P^{(6,1)}_1\tilde M^{(6,1)}_5$ and $\tilde P^{(6,1)}_3\tilde M^{(6,1)}_3$, respectively, can be viewed as non-trivial checks of the coaction principle.

\subsection{Analytic continuation  of $F^{(n,p)}$}

The examples in section \ref{sec:warmupp} have set the stage to describe the analytic continuation of $F^{(n,p)}(z_{p+2},z_{p+3},\ldots,z_{n-2})$. The simplest analytic continuation of these functions was  described  in   (\ref{alsocooleq}) and (\ref{cooleq}) as a group action of certain generators $\sigma_{3,4}$. The group in question is $B_\mathrm{N}$, the braid group of $\mathrm{N}$ strands, acting on the $\mathrm{N}=n{-}p$ unintegrated\footnote{Doing a complete turn around $\mathrm{SL}(2,\mathbb{C})$-fixed punctures performs a monodromy as for instance in (\ref{eqn5p1monodromy}). These operations can also be described as part of a braid group. See Appendix \ref{app:B} for more details.} punctures. The braid group $B_\mathrm{N}$ can be defined as the non-commutative group with generators $\sigma_i \vcentcolon= \sigma_{i,i+1}$, where $1\leq i \leq \mathrm{N}{-}1$, that satisfy the relations \cite{kohno2002conformal}
\begin{align}\label{eqPresentationBraidGroup}
\sigma_i \sigma_j &= \sigma_j \sigma_i 
&&\hspace{-2cm}\textrm{for} \qquad |i{-}j| \geq 2   \, , \\
\sigma_i \sigma_{i+1} \sigma_i &= \sigma_{i+1} \sigma_i \sigma_{i+1}  \notag
&&\hspace{-2cm}\textrm{for} \qquad 1\leq i \leq \mathrm{N}{-}2 \, .
\end{align}
For convenience, we will label the generators according to the punctures, i.e.\ $\sigma_{i,i+1}$ denotes the generator that interchanges punctures $z_i$ and $z_{i+1}$ via braiding, with $z_{i+1}$ going around $z_i$ counterclockwise.

We will now describe the group action of a generator of the braid group $\sigma_{i,i+1}$ on $F^{(n,p)}(z_{p+2},z_{p+3},\ldots,z_{n-2})$. This corresponds to performing a change of branch from the branch consistent with

\beq
0=z_{1}<z_{p+2}<z_{p+3}<\cdots < z_i < z_{i+1}    <  \cdots< z_{n-2}<z_{n-1}=1
\eeq
when all the punctures lie on the real line, into a branch consistent with
\beq
0=z_{1}<z_{p+2}<z_{p+3}<\cdots < z_{i+1}< z_{i}    <  \cdots< z_{n-2}<z_{n-1}=1  \, .
\eeq
Now, the analytic continuation of $F^{(n,p)} = \mathbb{P}^{(n,p)} \mathbb M ^{(n,p)} \mathcal G^{(n,p)}(z_{p+2},\ldots,z_i,z_{i+1},\ldots, z_{n-2})$ with $\mathcal G^{(n,p)}(\ldots)$ comprising all the factors of $\mathbb G^{(n,p)}$ in (\ref{apexp.35}) is given by a matrix acting on the generating series of polylogarithms\footnote{While this is a known formula in the literature, it is not usually written down explicitly. An explicit version of it can be found in Proposition 5.1 of \cite{CEEuniversalKZB} for the genus 1 case, which apparently has the same formula.},

\begin{align}
  \sigma_{i,i+1 } F^{(n,p)}  &=  \mathbb{P}^{(n,p)} \mathbb M ^{(n,p)}  \mathbb X^{(n,p)} (\sigma_{i,i+1}) \mathcal G^{(n,p)}(z_{p+2},\ldots,z_{i},z_{i+1},\ldots, z_{n-2}) 
  \label{braidGeneralNP}
  \\
  \notag
   &=\vcentcolon   \mathbb{P}^{(n,p)} \mathbb M ^{(n,p)} \tilde{ \mathcal G}^{(n,p)}(z_{p+2},\ldots,z_{i+1},z_{i},\ldots, z_{n-2})  \, ,
\end{align}
where $ \mathbb X^{(n,p)} (\sigma_{i,i+1}) $ is given as follows in terms
of transposed braid matrices 
(\ref{drineqs.1})
\begin{align}
\mathbb X^{(n,p)}\left(\sigma_{i,i+1}\right) &= \Phi \left(E^{(n,p)}_{1,i+1}+ \! \! \sum_{j=p+2}^{i-1} \! E^{(n,p)}_{j,i+1} , E^{(n,p)}_{i,i+1}\right)  
\exp \left(i\pi E^{(n,p)}_{i,i+1}\right) \notag \\
& \ \ \ \ \ \ \ \ \ \ \ \times
\Phi \left(E^{(n,p)}_{i,i+1},E^{(n,p)}_{1,i}+\! \! \sum_{j=p+2}^{i-1} \! E^{(n,p)}_{j,i} \right) \, .
  \label{braidGeneralNP2}
\end{align}
The $(n,p)=(6,1)$ cases of these expressions for $ \sigma_{i,i+1 } F^{(n,p)} $ and $\mathbb X^{(n,p)}\left(\sigma_{i,i+1}\right)$ can be found in (\ref{61ancont.2}) and (\ref{cooleq}), respectively.
Before the analytic continuation in (\ref{braidGeneralNP}), one can translate the rewriting $F^{(n,p)}=\mathbb{P}^{(n,p)} \mathbb M ^{(n,p)} [\mathbb X ^{(n,p)}( \sigma_{i,i+1 })]^{-1} \tilde{ \mathcal G}^{(n,p)}$ into a modified initial value 
\beq
\tilde{\mathbb{P}}^{(n,p)} \tilde{\mathbb M} ^{(n,p)}=\mathbb{P}^{(n,p)} \mathbb M ^{(n,p)} [\mathbb X ^{(n,p)}( \sigma_{i,i+1 })]^{-1}
\eeq
as done in section \ref{initbraid} at $(n,p)=(6,1)$. We expect $\tilde{\mathbb{P}}^{(n,p)} \tilde{\mathbb M} ^{(n,p)}$ to inherit the coaction properties of $F^{(n,p)}$, i.e.\ to generalize (\ref{coprop.90}) to arbitrary $n$ and $p$. Accordingly, the $\ap$-expansion of $ \tilde{\mathbb M} ^{(n,p)}$ will share the structure of the leading-order terms in (\ref{coprop.99}), and the $\ap$-expansion of $ \tilde{\mathbb P} ^{(n,p)}$ will involve odd powers of $i\pi$ as in (\ref{newini.2}).

The image $\tilde{ \mathcal G}^{(n,p)}(z_{p+2},\ldots,z_{i+1},z_{i},\ldots, z_{n-2}) $ under $\sigma_{i,i+1}$ describes the path-ordered integration of the KZ form $\Omega^{(n,p)}$ in $\dd F^{(n,p)} = \Omega^{(n,p)} F^{(n,p)}$, with an initial value equal to the identity, and along the path where $z_i$ is moved to nonzero values before $z_{i+1}$,
\begin{align}
(0, 0, \ldots,0) & \rightarrow (0, \ldots, 0, z_{n-2}) \rightarrow (0, \ldots, 0, z_{n-3}, z_{n-2}) 
\notag  
\\
&\rightarrow  \ldots \rightarrow (0 , \ldots,0, 0,z_{i+2}, \ldots , z_{n-2})  
\notag 
\\
&\rightarrow (0 , \ldots,0,  z_{i} , 0,z_{i+2}, \ldots , z_{n-2}) 
\label{integrationOrderAltNPbraid}
\\
& \rightarrow(0 , \ldots,0,  z_{i} , z_{i+1},z_{i+2}, \ldots , z_{n-2})  
\notag
\\
&\rightarrow(0 , \ldots,0, z_{i-1,}  z_{i} , z_{i+1},z_{i+2}, \ldots , z_{n-2}) 
\notag 
\\
& \rightarrow \ldots \rightarrow (z_{p+2},z_{p+1}, \ldots z_{n-2} )  \, .
\notag
\end{align}
Both the matrices that enter the definition of $\tilde{ \mathcal G}^{(n,p)}(z_{p+2},\ldots,z_{i+1},z_{i},\ldots, z_{n-2}) $ and the fibration basis of its component polylogarithms respect this integration order above. Equivalently, we can define $\tilde{ \mathcal G}^{(n,p)}(z_{p+2},\ldots,z_{i+1},z_{i},\ldots, z_{n-2}) $ to be given by
\beq
\tilde{ \mathcal G}^{(n,p)}(z_{p+2},\ldots,z_{i+1},z_{i},\ldots, z_{n-2})  =\mathcal G^{(n,p)}(z_{p+2},\ldots,z_{i},z_{i+1},\ldots, z_{n-2})  \big|_{i \leftrightarrow i+1} \, ,
\eeq
where $i \leftrightarrow i{+}1$ instructs to interchange $z_i$ with $z_{i+1}$ and  $E_{i,j}$ with $E_{i+1,j}$ everywhere, but without modifying the Mandelstam variables in their entries. In the case of $(n,p)=(6,1)$, this procedure converts the series (\ref{calG34}) to (\ref{calG43}) and the braid matrices in (\ref{apexp.28}) to those in (\ref{nwapexp.28}).

We have explicitly verified the $(n,p)=(7,1)$ cases of (\ref{braidGeneralNP}) and (\ref{braidGeneralNP2}) for $\sigma_{3,4}$ and $\sigma_{4,5}$ up to and including $\ap^5$. For these explicit checks, changes of fibration bases were performed via \texttt{PolyLogTools} \cite{Duhr2019PolylogTools}, with the sign of $i \pi$ as in the last term of (\ref{changez3z4}) and the analogous identity with $(z_3,z_4) \rightarrow  (z_4,z_5)$ to take the orientation of the braiding into account.  

In conclusion, the key achievement in this section is to spell out the action of the  elementary braid $\sigma_{i,i+1}$ involving neighboring punctures on $F^{(n,p)}$. Since the braid group of $\mathrm{N}$ strands, $B_\mathrm{N}$, is generated by these $\sigma_{i,i+1}$, the results of this section determine the analytic continuation due to arbitrary braiding of the punctures. Further examples of analytic continuation can be found in Appendix \ref{app:B}.

\section{Sphere integrals}
\label{sec:6}

This section is dedicated to sphere integrals over the forms $\omega_a^{(n,p)}$
of section \ref{sec:2} and their complex conjugates. When interpreting the $F_{ab}^{(n,p)}$ as open-string
integrals with a subset of the vertex-operator insertions integrated out, the
sphere integrals in this section can be viewed are their closed-string counterparts. Moreover, they are directly applicable to computations of correlation functions in two-dimensional conformal field theories as will be further elaborated on in section~\ref{sec:7}.

We will express the $\ap$-expansion of the sphere integrals in this section both as
single-valued maps of the $F_{ab}^{(n,p)}$ and as Kawai--Lewellen--Tye (KLT) formulae
involving products of open-string integrals and their complex conjugates. We will propose two prescriptions for computing the entries of the KLT matrix and its inverse. The latter will be given in terms of combinatorial rules describing adjacency properties of Stasheff polytopes associated to each integration cycle, while the former will be an explicit expression in terms of polynomials of trigonometric functions.

\subsection{General formulae}
\label{sec:6.1}

The sphere integrals of interest in this section take the form
\beq
\int_{{\cal C}^{(n,p)}} \Big( \prod_{k=2}^{p+1} \dd^2 z_k \Big)\, | {\rm KN}^{(n,p)} |^2 \, \overline{ \hat{\omega}_a^{(n,p)}}
\,  \hat{\omega}_b^{(n,p)}\, ,
\label{sph.1}
\eeq
where $\dd^2 z =\frac{i}{2} \dd z \wedge \dd \bar z$ and $a,b =1,2,\ldots,d^{(n,p)}$ independently run over the bases of forms $\omega_a^{(n,p)} = \hat{\omega}_a^{(n,p)} \prod_{j=2}^{p+1} \dd z_j$
specified in section \ref{sec:2.2}. For $p=n{-}3$, we recover the sphere integrals of closed-string amplitudes
which are known to be single-valued maps of open-string integrals $F_{ab}^{(n,n-3)}$ if $ \overline{ \hat{\omega}_a^{(n,p)}}$
are replaced by suitably chosen Parke--Taylor forms \cite{Schlotterer:2012ny, Stieberger:2013wea, Stieberger:2014hba, Schlotterer:2018zce, Brown:2018omk, Vanhove:2018elu, Brown:2019wna}:
\begin{align}
{\rm sv} \, F_{ab}^{(n,n-3)} &=  \frac{1}{\pi^{n-3}} \int_{{\cal C}^{(n,n-3)}} \Big( \prod_{k=2}^{n-2} \dd^2 z_k \Big) \, | {\rm KN}^{(n,n-3)} |^2 \, \overline{ \hat{\nu}_a^{(n,n-3)}}
\,  \hat{\omega}_b^{(n,n-3)}, 
\label{sph.2} 
\end{align}
with
\begin{align}
\overline{ \hat{\nu}_a^{(n,n-3)}} &=  \frac{(-1)^{n-3}\, \overline{z_{1,n-1}}  }{\rho_a(  \overline{z_{1,2}}  \,\overline{z_{2,3}} \ldots  \overline{z_{n-3,n-2}} \,  \overline{z_{n-2,n-1}}  ) }\, .
\label{sph.3}
\end{align}
The $\overline{ \hat{\nu}_a^{(n,n-3)}}$ are SL$(2,\mathbb{C})$-fixed antiholomorphic Parke--Taylor factors, furnish the Betti--de Rham duals \cite{betti1,betti2, Brown:2018omk} 
to disk orderings of the $F_{ab}^{(n,n-3)}$ and are indexed by permutations
$\rho_a \in S_{n-3}$ of the labels $\{2,3,\ldots,n{-}2\}$ in lexicographic ordering.

One of the goals of this section
is to extend (\ref{sph.2}) to generic $p$, i.e., to spell out the forms $\overline{ \hat{\nu}_a^{(n,p)}}$ that
generalize (\ref{sph.3}) to the Betti--de Rham dual of the cycles $\gamma_a^{(n,p)}$ with an arbitrary
number of integrated and unintegrated punctures. For each collection of adjacent integrated punctures
$z_{i_1},z_{i_2},\ldots,z_{i_k}$ located between unintegrated ones $z_b$, $z_c$, the forms $\overline{ \hat{\nu}_a^{(n,p)}}$
pick up a factor as on the right-hand side of (\ref{sph.3}), i.e.,
\beq
\{z_b < z_{i_1}<z_{i_2}<\ldots< z_{i_k}< z_c\} \ \ \ \leftrightarrow \ \ \ 
\frac{(-1)^k \, \overline{z_{b,c}}  }{  \overline{z_{b,i_1}}  \,\overline{z_{i_1,i_2}}
  \,\overline{z_{i_2,i_3}} \ldots  \overline{z_{i_{k-1},i_k}} \,  \overline{z_{i_k,c}}  }
\, .
\label{sph.4}
\eeq
After combining the contributions from all integrated and unintegrated punctures, one obtains the basis of $\overline{ \nu_a^{(n,p)}}$ given in \eqref{nu-forms} which
reduces to \eqref{sph.3} in the special case of $p=n{-}3$.
This will be further illustrated through the examples at various $(n,p)$ in the next subsections.

Given the $d^{(n,p)}$-element basis of forms $\overline{ \nu_a^{(n,p)}}$ defined in this way, we claim that a basis
of sphere integrals (\ref{sph.1}) can be computed from the single-valued map acting on the MZVs
and polylogarithms in the $\ap$-expansion of $F_{ab}^{(n,p)}$:\footnote{The notation $\langle \overline{\nu} | \omega \rangle$ with an antiholomorphic form $\overline{\nu}$ will always refer to \eqref{sph.5} as opposed to the right-hand side of \eqref{conv6b}, which takes two holomorphic forms instead.}
\beq
{\rm sv} \, F_{ab}^{(n,p)} = \frac{1}{\pi^p} \int_{{\cal C}^{(n,p)}} \Big( \prod_{k=2}^{p+1} \dd^2 z_k \Big) \, | {\rm KN}^{(n,p)} |^2 \, \overline{ \hat{\nu}_a^{(n,p)}}
\,  \hat{\omega}_b^{(n,p)} = \langle \overline{ \nu_a^{(n,p)}} | \omega_b^{(n,p)} \rangle
\, .
\label{sph.5}
\eeq
The single-valued map is compatible with the product of MZVs and polylogarithms and can be evaluated separately for each factor in the $\ap$-expansion of $F_{ab}^{(n,p)}$ in \eqref{apexp.35}.

For example, single-valued MZVs relevant for ${\rm sv}\, \mathbb P^{(n,p)}= 1$ and ${\rm sv} \, \mathbb M^{(n,p)}$ have been introduced in \cite{Schnetz:2013hqa, Brown:2013gia}: Their simplest cases include\footnote{Strictly speaking, the single-valued map is only well-defined in the setting of motivic MZVs. We will informally drop the superscript of $\zeta^{\mathfrak m}_{n_1,n_2,\ldots,n_r}$ in (\ref{exmzvs}) and use the same notation sv for the single-valued map of motivic MZVs and their images in the $f$-alphabet in (\ref{svinfs}).}
\beq
{\rm sv}\, \zeta_{2k+1} = 2\zeta_{2k+1} \, , \ \ \ \ \ \
{\rm sv}\, \zeta_{2k} = 0 \, ,
\ \ \ \ \ \
{\rm sv}\, \zeta_{3,5} = - 10 \zeta_3 \zeta_5 \, ,
\label{exmzvs}
\eeq
and the $f$-alphabet admits the closed formula (with $i_1,\ldots,i_r \in 2\mathbb N{+}1$)\footnote{The conventions of this work differ from those of \cite{Schnetz:2013hqa, Brown:2013gia} by $A \otimes B \to B \otimes A$ and therefore by a reversal $f_{2k_1+1}f_{2k_2+1} \ldots f_{2k_r+1} \mapsto f_{2k_r+1} \ldots f_{2k_2+1} f_{2k_1+1}$. Accordingly, (\ref{svinfs}) features a reversal in the first part $f_{i_j}  \ldots f_{i_2} f_{i_1}$ of the deconcatenated word $f_{i_1}f_{i_2}\ldots f_{i_r}$ on the right-hand side and not in the second part $f_{i_{j+1}} f_{i_{j+2}} \ldots  f_{i_r}$ as seen in the references.}
\beq
{\rm sv}\, f_2^N f_{i_1}f_{i_2}\ldots f_{i_r} = \delta_{N,0} \sum_{j=0}^r
f_{i_j}  \ldots f_{i_2} f_{i_1} \shuffle  f_{i_{j+1}} f_{i_{j+2}} \ldots  f_{i_r}\, .
\label{svinfs}
\eeq
The expansion coefficients of ${\rm sv}  \, \mathbb G^{(n,p)}_{\{0,1\}}(z_{n-2})$ are
single-valued polylogarithms in one variable \cite{svpolylog} that include 
\begin{align}\label{svG1}
{\rm sv} \, G(0;z) &= {\rm sv}\, \log z = \log |z|^2 \ , \\
 {\rm sv} \, G(1;z)&={\rm sv}\, \log (1{-}z) = \log |1{-}z|^2 \, , \notag
\end{align}
as well as
\begin{align}
{\rm sv} \, G(0,1;z) &= - {\rm sv} \, {\rm Li}_2(z) \notag \\
&= G(0,1;z) + G(0;z) \overline{G(1;z)} + \overline{G(1,0;z)}  \notag \\
&= - {\rm Li}_2(z) + {\rm Li}_2(\bar z)
+ \log(1{-}\bar z) \log|z|^2 \label{svG2}
 \\
{\rm sv} \, G(1,0;z)&=
{\rm sv} \, G(0;z) \, {\rm sv} \, G(1;z)
- {\rm sv} \, G(0,1;z) \notag \\
{\rm sv} \, G(a,a;z)&= \frac{1}{2} {\rm sv} \, G(a;z)^2
\, , \notag
\end{align}
and
\begin{align}
 {\rm sv} \, G(0,0,1,1;z) &=
  G(0,0,1,1;z)
  + G(0,0,1;z)\overline{G(1;z)}
  + G(0,0;z)\overline{G(1,1;z)} \label{exsvpoly}\\
  & \ \ \ \
  + G(0;z)\overline{G(1,1,0;z)}
  +\overline{G(1,1,0,0;z)}
  +2\zeta_3 \overline{G(1;z)} \, .
  \notag
\end{align}
Single-valued polylogarithms in multiple variables that enter the remaining ${\rm sv}  \, \mathbb G^{(n,p)}_{\ldots}$ can be found in \cite{Broedel:2016kls, DelDuca:2016lad}.

It should be possible to derive \eqref{sph.5} from the inductive techniques of \cite[section 3.3]{Schlotterer:2018zce}. However, this is not a fully rigorous proof since the use of the single-valued map relies on transcendentality conjectures on MZVs. The techniques of Brown and Dupont \cite{Brown:2018omk, Brown:2019wna} in turn should allow for a proof without any such assumptions.

\subsection{First look at KLT formulae}

\begin{figure}[h]
    \centering
    \includegraphics{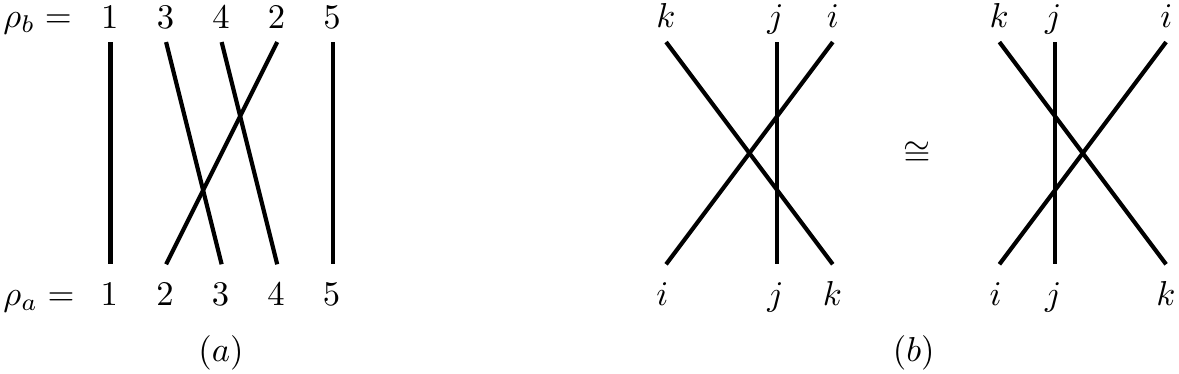}
    \caption{$(a)$ Example graphical computation of the phase $\phi_{ab} = s_{23} + s_{24}$, where the two Mandelstam invariants contribute because of the crossing of lines associated to labels $(2,3)$ and $(2,4)$. The final label, here $n=5$, is always held fixed. $(b)$ Illustration of the independence of the phase (in this case $s_{ij} + s_{jk} + s_{ik}$) on the way of drawing straight lines, as long as they only intersect pairwise.}
    \label{fig:phase}
\end{figure}

An second goal of this section is to write the sphere integrals (\ref{sph.5}) as bilinears in the open-string integrals
$F^{(n,p)}_{ab}$ and their complex conjugates, following the Kawai--Lewellen--Tye (KLT) formula for the
case $p=n{-}3$ \cite{Kawai:1985xq} and its generalization to $p=n{-}4$ \cite{Vanhove:2018elu}. Since the integrand of \eqref{sph.5} is already holomorphically factorized, it can be easily written down as a double sum over pairs of \emph{all} $(n{-}1)!/(n{-}p{-}1)!$ real cycles in ${\cal C}^{(n,p)}$, including those outside of the $d^{(n,p)}$ basis. To be precise, introducing
\beq
\tildeF_{ab}^{(n,p)} = \int_{\gamma_a^{(n,p)}} \Big( \prod_{k=2}^{p+1} \dd z_k \Big) \,  {\rm KN}^{(n,p)}  \,  \hat{\nu}_b^{(n,p)},
\label{sph.6b}
\eeq
we have
\beq
{\rm sv}\, F_{ab}^{(n,p)} = \medmath{\left(\frac{-1}{2\pi i }\right)^{\!p}} 
\sum_{c,d=1}^{\frac{(n-1)!}{(n-p-1)!}} e^{i \pi \phi_{cd}}\, \overline{\tildeF_{da}^{(n,p)}}\,
F_{cb}^{(n,p)}.
\label{sph.7b}
\eeq
Both of $c,d$ run over the $\frac{(n-1)!}{(n-p-1)!}$ cycles $\gamma^{(n,p)}_c$ that impose the ordering (\ref{order-unintegrated}) of the unintegrated punctures $z_1,z_{p+2},\ldots,z_{n-1}$
but allow the integrated ones $z_2,z_3,\ldots,z_{p+1}$
to be in $(-\infty,0)$ or $(1,+\infty)$ besides the standard interval $(0,1)$ of the $d^{(n,p)}$-element basis.
The only subtlety in (\ref{sph.7b}) comes from the fact that each integral on the right-hand side comes with a specific phase of the Koba--Nielsen factor prescribed in \eqref{conv2}. This is corrected by the explicit phase factor $e^{i \pi \phi_{cd}}$, where 
\be
\phi_{cd} = \!\!\! \sum_{1\leq i<j  \leq n-1}\!\!\! \theta_{cd}^{ij} \qquad\text{with}\qquad \theta_{cd}^{ij} = \begin{dcases}
s_{ij} \quad &\text{if}\quad (\rho_{c}^{-1}(i) {-} \rho_{c}^{-1}(j))(\rho_{d}^{-1}(i) {-} \rho_{d}^{-1}(j)) < 0,\\
0 \quad &\text{otherwise}.
\label{phi-cd}
\end{dcases}
\ee
Here $\rho_{c}^{-1}(i)$ denotes the position of the label $i$ in $\rho_c$. In other words, $\phi_{cd}$ is the sum of all Mandelstam invariants $s_{ij}$ for which $i$ and $j$ appear in reversed order in $\rho_c$ than in $\rho_d$ (recall that we always fix $z_n = \infty$). Thus the phase can be computed easily using the graphical rules illustrated in figure~\ref{fig:phase}. For $p=n{-}3$ this formula was given in \cite{Kawai:1985xq,10.1093/qmath/38.4.385}, where the sum in \eqref{sph.7b} is over $[(n{-}1)!/2]^2$ terms.

For practical purposes it is beneficial to eliminate redundant terms from \eqref{sph.7b} to involve only a sum over the minimal $d^{(n,p)}$ basis. To this end we construct dual cycles $\beta_a^{(n,p)}$
such that
\beq
I_{ab}^{(n,p)} = \int_{\beta_a^{(n,p)}} \Big( \prod_{k=2}^{p+1} \dd z_k \Big) \,  {\rm KN}^{(n,p)}  \,  \hat{\nu}_b^{(n,p)}
\label{sph.6}
\eeq
reduce to $\delta_{ab}$ in the $\ap \rightarrow 0$ limit. One can always expand the $I_{ab}^{(n,p)}$ in a basis
of $\tilde\tildeF_{cb}^{(n,p)}$, which are the integrals from \eqref{sph.6b} but with a shifted basis of cycles $\tilde\gamma_a^{(n,p)}$ (defined below in \eqref{allklt.4}) instead of $\gamma_a^{(n,p)}$:
\be
I_{ab}^{(n,p)} = \sum_{c=1}^{d^{(n,p)}}\,
S_{\alpha'}(\rho_a|\rho_c)\,
\tilde\tildeF_{cb}^{(n,p)},
\label{J-to-I}
\ee
for example by the use of monodromy relations.
We will describe two distinct prescriptions for deriving the coefficients $S_{\alpha'}(\rho_a|\rho_c)$, which we will refer to as the generalized KLT kernel.\footnote{Our terminology is not to be confused with the generalized KLT kernel in \cite{Frost:2020eoa}. This reference generalizes the field-theory version of the 
KLT kernel at $p=n{-}3$ to a $(n{-}2)! \times (n{-}2)!$ matrix (instead of the conventional $(n{-}3)! \times (n{-}3)!$ format) and generates its entries from a Lie-bracket based
on the S-map. Our generalization of the KLT kernel concerns the cases with
$p \neq n{-}3$, and it would be interesting to also derive the recursion relations \eqref{allklt.5} for its entries
from the S-bracket of \cite{Frost:2020eoa}. We would like to thank Carlos Mafra for discussions on this point.} As is known from the $p=n{-}3$, the inverse of the matrix $S_{\alpha'}(\rho_a|\rho_c)$ are intersection numbers of cycles $\gamma_a^{(n,p)}$ \cite{Mizera:2016jhj,Mizera:2017cqs,Matsubara-Heo:2020lqa}. In fact, this is a general feature of complex integrals (see \cite{cho1995} and \cite[Section 6]{Matsubara-Heo:2020lzo}), which allows us to extend this prescription to all other values of $p$. Intersection numbers are given by combinatorial rules describing how the cycles $\gamma_a^{(n,p)}$ intersect one another in the moduli space.
Based on this computation and direct manipulations using monodromy relations, we propose an explicit recursive expression for the KLT matrix $S_{\alpha'}(\rho_a|\rho_c)$ and verify its correctness up to $n=8$ with any $p$.

Putting everything together, the resulting expression is the second major claim of this section:
\begin{align}
{\rm sv}\, F_{ab}^{(n,p)} &=  \frac{1}{\pi^p}
\sum_{c=1}^{d^{(n,p)} } \overline{I_{ca}^{(n,p)}}
F_{cb}^{(n,p)}  \label{sph.7}\\
&= \frac{1}{\pi^p}\sum_{c,d=1}^{d^{(n,p)} } \overline{\tilde \tildeF_{da}^{(n,p)}} S_{\alpha'}(\rho_c|\rho_d)
F_{cb}^{(n,p)},
\notag
\end{align}
which is the generalization of the KLT formula to arbitrary $(n,p)$.

\subsection{\label{sec:6.2}Intersection numbers of Stasheff polytopes}

In this subsection we describe combinatorial rules for computing the intersection numbers of twisted cycles
\be
H^{(n,p)}_{ab} = \langle \gamma_{a}^{(n,p)} | \gamma_{b}^{(n,p)} \rangle
\label{intmatr}
\ee
in terms of adjacency properties of Stasheff polytopes (or associahedra) \cite{10.2307/1993608} tiling the real slice of the configuration space $\Re {\cal C}^{(n,p)}$.
In fact, it will prove rewarding to construct the $d^{(n,p)} \times d^{(n,p)}$ matrix 
\be
\tilde H^{(n,p)}_{ab} = \langle \gamma_{a}^{(n,p)} | \tilde \gamma_{b}^{(n,p)} \rangle
\ee
with an alternative basis of cycles $\tilde \gamma_{b}^{(n,p)}$ in the second entry, where
some of the punctures $\underline{j_i}$
are integrated over subsets of $(-\infty,0)$,
\begin{align}
 \tilde\gamma^{(n,p)}_{\vec{B},\vec{j}} \quad\leftrightarrow\quad \rho_{\vec{B},\vec{j}} = (B_1,\underline{j_1},B_2,\underline{j_2},\ldots,B_p,\underline{j_p},B_{p+1},n{-}2,n{-}1,n) \, .
 \label{allklt.4}
\end{align}
In this setup, the KLT matrix in (\ref{sph.7}) is given by
\be
S_{\alpha'}(\rho_a | \rho_b) = (\tilde H^{(n,p)})^{-1}_{ba}.
\label{inverse-H}
\ee

\begin{figure}[h]
    \centering
    \includegraphics{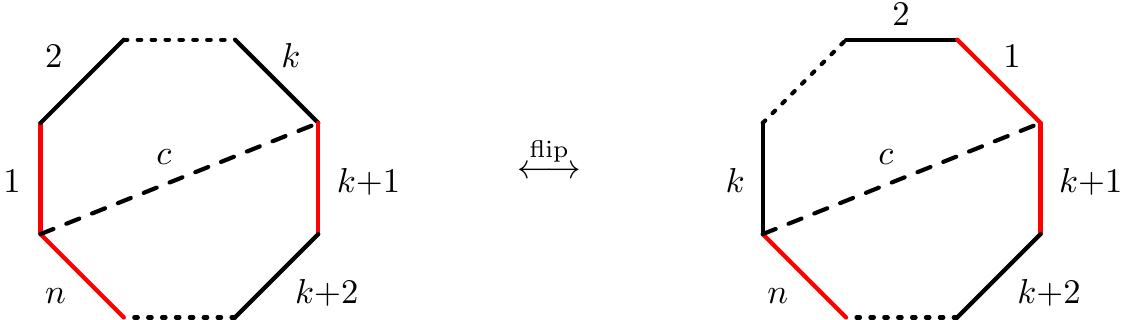}
    \caption{Example of an admissible flip by a chord $c$ between $n$-gons labelled by $(1,\underline{2},\ldots,\underline{k},k{+}1,\underline{k{+}2},\ldots,n)$ and $(\underline{k},\ldots,\underline{2},1,k{+}1,\underline{k{+}2},\ldots,n)$. Edges corresponding to unintegrated punctures are indicated in red. Flipped side of the $n$-gon contains only one red edge, which makes the flip admissible.}
    \label{fig:flip}
\end{figure}

For a given $(n,p)$, the cycles $\gamma_a^{(n,p)}$ are in bijection to $n$-gons with edges labelled according to the given ordering $\rho_a$ of labels $\{1,2,\ldots,n\}$. We will consider all possible permutations $\rho_a$ where the unintegrated (fixed) labels $(1, p{+}2, p{+}3,\ldots, n)$ always appear in this specific order. By an extension of the combinatorics describing the moduli space ${\cal M}_{0,n}$ \cite{Devadoss98tessellationsof,Brown:2009qja}, adjacency properties on $\Re {\cal C}^{(n,p)}$ can be described by drawing tessellations of decorated $n$-gons. A flip move corresponds to drawing a single chord $c$ and reflecting one side of the chord as illustrated in figure~\ref{fig:flip}.
A given chord is admissible only if the result of the flip leaves the fixed labels $(1, p{+2}, p{+}3,\ldots,n)$ in the same order. In other words, the side of the $n$-gon we flip has to have exactly $0$ or $1$ edges corresponding to fixed punctures. (In particular, for $p=n{-}3$ all chords are admissible.) To every chord $c$ we associate the Mandelstam variable
\be
s_c = \alpha' \left( \sum_{i \in F_c} k_i \right)^2,
\ee
where $F_c$ is the set of edges being flipped ($2 \leq |F_c| \leq n{-}2$). In the example of figure~\ref{fig:flip} we have $s_c = s_{12\ldots k}$. If a chord is admissible, it labels an element of the boundary\footnote{More precisely, here and in the following, whenever we talk about boundaries of cycles, we mean $\partial\overline{\pi^{-1}((\gamma_{a}^{(n,p)})^{\mathrm{o}})}$, the boundary of the closure of the interior of $\gamma_a^{(n,p)}$ after the resolution of exceptional divisors of the configuration space ${\cal C}^{(n,p)}$ by a blowup map $\pi^{-1}$ (see \cite{Fulton:1994hh}).} of $\gamma_a^{(n,p)}$, and the whole boundary structure is governed by how these chords fit into tessellations. The cycles $\gamma_a^{(n,p)}$ are combinatorially isomorphic to Stasheff polytopes and their direct products.

\subsubsection{Self-intersection numbers}

A given tessellation $T_a$ associated to $\gamma_a^{(n,p)}$ is admissible if it includes only admissible chords $\{c_\ell\}_{\ell=1,2,\ldots,|T_a|} \in T_a$, where $|T_a| \in \{0,1,\ldots,p\}$ is the number of chords used, and one imposes that these chords do not cross. Following \cite{MANA:MANA19941660122,doi:10.1002/mana.19941680111} we find the following formula for self-intersection numbers:
\be
\langle \gamma_a^{(n,p)} | \gamma_a^{(n,p)} \rangle = (2i)^p \sum_{T_a} \prod_{c_\ell \in T_a} \frac{1}{e^{2\pi i s_{c_\ell}}-1}, 
\label{self-int}
\ee
where the sum goes over all admissible tessellations (for $|T_a|=0$ the set of chords is empty and the term contributing to the sum is $1$) and we introduce the following shorthand
\beq
t_{ij\ldots} = e^{2\pi i s_{ij\ldots}} - 1 \, .
\eeq
The geometric understanding of this formula is that a given tessellation $T_a$ with $|T_a|$ chords labels the codimension-$|T_a|$ boundary of $\gamma_a^{(n,p)}$. For example, the terms with maximum number of chords, $\max |T_a|$, label its vertices, and those with a single chord label its facets. Tessellations describe the combinatorics of how these elements of the boundary fit together. 

For example, at $p=1$:
\be
\langle \gamma_{1\underline{2}3\cdots n}^{(n,1)} | \gamma_{1\underline{2}3\cdots n}^{(n,1)} \rangle = 2i \left(1 + \frac{1}{t_{12}} + \frac{1}{t_{23}}\right)
= \frac{\sin(\pi(s_{12}{+}s_{23})) }{ \sin(\pi s_{12}) \sin(\pi  s_{23})}
\, .
\label{n1-self-int}
\ee
Here and below, to make the connection with $n$-gons easier to follow, we label the cycles $\gamma_a^{(n,p)}$ directly by their permutation $\rho_a$ and underline the integrated (unfixed) labels. For $p=2$ the answer depends on the number of fixed punctures separating the two unfixed ones:
\begin{align}
\langle \gamma_{1\underline{2}\underline{3}4\cdots n}^{(n,2)} | \gamma_{1\underline{2}\underline{3}4\cdots n}^{(n,2)} \rangle = -4\bigg(&\,1 + \frac{1}{t_{12}} + \frac{1}{t_{23}} + \frac{1}{t_{34}} + \frac{1}{t_{123}} + \frac{1}{t_{234}} \\
& + \frac{1}{t_{12} t_{34}} + \frac{1}{t_{12}t_{123}} + \frac{1}{t_{23}t_{123}} + \frac{1}{t_{23}t_{234}}
+ \frac{1}{t_{34}t_{234}}\bigg) \, ,\notag
\end{align}
\begin{align}
\langle \gamma_{1\underline{2}4\underline{3}5\cdots n}^{(n,2)} | \gamma_{1\underline{2}4\underline{3}5\cdots n}^{(n,2)} \rangle = -4\bigg(&\,1 + \frac{1}{t_{12}} + \frac{1}{t_{24}} + \frac{1}{t_{34}} + \frac{1}{t_{35}} + \frac{1}{t_{234}} \\
& + \frac{1}{t_{12}t_{34}} + \frac{1}{t_{12}t_{35}} + \frac{1}{t_{24}t_{35}} + \frac{1}{t_{24}t_{234}}
+ \frac{1}{t_{34}t_{234}} \bigg) \, ,\notag
\end{align}
and
\be
\langle \gamma_{1\underline{2}4\ldots k\underline{3} k+1\cdots n}^{(n,2)} | \gamma_{1\underline{2}45\underline{3}6\cdots n}^{(n,2)} \rangle = -4 \left( 1 + \frac{1}{t_{12}} + \frac{1}{t_{24}} \right)\left( 1 + \frac{1}{t_{3k}} + \frac{1}{t_{3,k+1}} \right)
\ee
for $k\geq 5$.
Factorization of the final example reflects the fact that the corresponding chamber is combinatorially a square (a product of two one-dimensional Stasheff polytopes), while the first two were two-dimensional Stasheff polytopes, combinatorially pentagons.

\subsubsection{Generic intersection numbers}

\begin{figure}[h]
    \centering
    \includegraphics{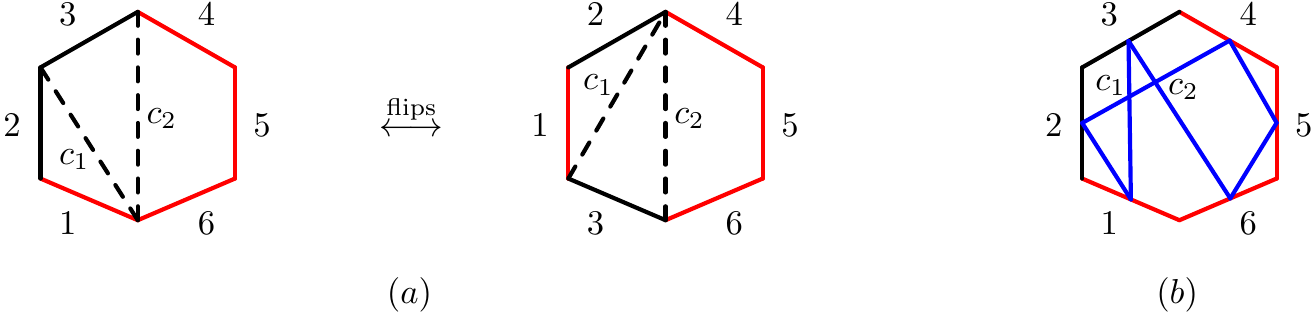}
    \caption{$(a)$ Example of the tessellation $T_{ab}$ by admissible chords $c_1, c_2$ for $\gamma_{1\underline{23}456}^{(6,2)}$ and $\gamma_{\underline{3}1\underline{2}456}^{(6,2)}$. We have $s_{c_1} = s_{12}$ and $s_{c_2} = s_{123}$. $(b)$ The set of chords can be determined, following \cite{Cachazo:2013iea,Mizera:2016jhj}, by embedding the second $n$-gon (blue) inside the first one by connecting midpoints of its edges in the order $\rho_b$. Provided this can be done without self-overlaps, the chords are determined by places where the second $n$-gon folds over (admissibility criteria need to be checked separately).}
    \label{fig:tess}
\end{figure}

A more interesting case is the intersection number of distinct cycles, which geometrically describes the boundary of their intersection in the moduli space.
If two $n$-gons cannot be transformed into one another with a series of admissible flips, their intersection number is zero. Otherwise, associated to $\gamma_a^{(n,p)}$ and $\gamma_b^{(n,p)}$, there exists a unique set of chords $T_{ab}$ that flips one into another in the minimal number of steps, as illustrated in figure~\ref{fig:tess}.
The resulting $n$-gon is tessellated into a number of smaller polygons $P_{ab}$. For each $P_{ab}$ we can define the set of admissible tessellations $T_{P_{ab}}$ that have chords only within $P_{ab}$ (admissibility is determined with respect to the original $n$-gon). The formula for the intersection number becomes
\be
\langle \gamma_a^{(n,p)} | \gamma_b^{(n,p)} \rangle = (-1)^{w(\rho_a|\rho_b)+1}\left( \prod_{c_\ell \in T_{ab}}
\frac{1}{\sin (\pi s_{c_\ell})}
\right) \prod_{P_{ab}} (2i)^{\max |T_{P_{ab}}|} \sum_{T_{P_{ab}}} \prod_{c_\ell \in T_{P_{ab}}} \frac{1}{e^{2\pi i s_{c_\ell}}-1} \, ,
\label{gamma-int-number}
\ee
where $w(\rho_a|\rho_b)$ is the relative winding number of the two permutations as defined in \cite[Appendix~A]{Mizera:2016jhj}. The proof of this formula is analogous to those in \cite{MANA:MANA19941660122,Mizera:2017cqs}.
The definition collapses to \eqref{self-int} when $a=b$ because $T_{ab}$ is the empty set and $P_{ab}$ is simply the original $n$-gon, so $T_{P_{ab}} = T_{ab}$ and $\max |T_{P_{ab}}| = p$.

\begin{figure}[h]
    \centering
    \includegraphics{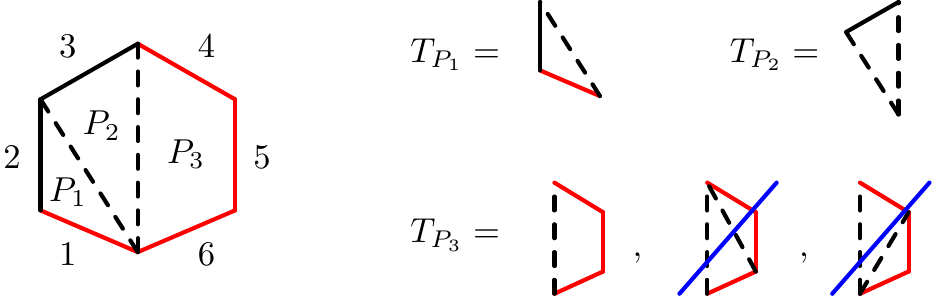}
    \caption{Example polygon decomposition needed for the computation of the intersection number in \eqref{123456-example}. Out of the three polygons, $P_1$ and $P_2$ are triangles and hence admit only one admissible tessellation. The last one, $P_3$, has three possible tessellation, but the last two are not admissible as they separate two red edges on either side of the additional chord.}
    \label{fig:example}
\end{figure}

For example, let us compute the intersection number of $\gamma_{1\underline{23}456}^{(6,2)}$ and $\gamma_{\underline{3}1\underline{2}456}^{(6,2)}$, which we already know is non-zero from figure~\ref{fig:tess}. We found two chords defining $T_{ab}$, which dissects the original $6$-gon into three polygons we will call $\{P_1, P_2, P_3\} \ni P_{ab}$ below. For each polygon we find that there is exactly one admissible tessellation, see figure~\ref{fig:example}. 
In this case the winding number is $w(123456|312456)=2$. This leaves us with the final answer:
\be
\langle \gamma_{1\underline{23}456}^{(6,2)} | \gamma_{\underline{3}1\underline{2}456}^{(6,2)} \rangle = \frac{(-1)^{2+1}}{\sin (\pi s_{12}) \sin (\pi s_{123})} \times 1 \times 1 \times 1 \, .
\label{123456-example}
\ee
As another example, we can consider the intersection of $\gamma_{1\underline{234}56}^{(6,3)}$ and $\gamma_{\underline{3}1\underline{24}56}^{(6,3)}$, which only differs from (\ref{123456-example}) by the fact that the label $4$ is now integrated. Hence, all the computations are identical, except for the fact that the final two tessellations in $T_{P_3}$ of figure~\ref{fig:example} are now admissible. We therefore find
\begin{align}
\langle \gamma_{1\underline{234}56}^{(6,3)} | \gamma_{\underline{3}1\underline{24}56}^{(6,3)} \rangle &= \frac{(-1)^{2+1}}{\sin (\pi s_{12}) \sin (\pi s_{123})} \times 1 \times 1 \times 2i \left( 1 + \frac{1}{t_{45}} + \frac{1}{t_{56}} \right) \notag \\
&= - \frac{ \sin (\pi (s_{45}{+}s_{56}))}{ \sin (\pi s_{12}) \sin (\pi s_{123})
\sin (\pi s_{45}) \sin (\pi s_{56})} \, .
\label{123456-example2}
\end{align}
Another way of stating this result is that the intersection of the two cycles is a one-dimensional Stasheff polytope, while in \eqref{123456-example} it was a zero-dimensional one (a point).

We will provide more examples in the following subsections. Alternative prescriptions for computing intersection numbers $\langle \gamma_a^{(n,p)} | \gamma_b^{(n,p)} \rangle$ were given in \cite{Mimachi2003,mimachi2004}. The advantage of our approach is that it provides combinatorial insight in terms of tessellations of $n$-gons (or equivalently planar trees).

\subsection{Case $p=1$}

Let us start with an instructive case of $p=1$, which will inspire the choice of bases of cycles for the $p>1$ cases as well. Recall that $d^{(n,1)} = n{-}3$
and the canonical basis of cycles we use is 
\begin{align}
\gamma_{1}^{(n,1)} &= \{ z_2 \in \mathbb{R} \,|\, z_1 < z_2 < z_{3} < \cdots < z_n \} \, ,\\
\gamma_{a}^{(n,1)} &= \{ z_2 \in \mathbb{R} \,|\, z_1 < \cdots < z_{a+1} < z_2 < z_{a+2} < \cdots < z_n \} \quad\mathrm{for}\quad 2 \leq a \leq n{-}3 \, ,\notag
\end{align}
or in the notation introduced above
\be
\gamma_{a}^{(n,1)} = \left( \gamma_{1\underline{2}34\ldots n}^{(n,1)},\; \gamma_{13\underline{2}4\ldots n}^{(n,1)},\;
\ldots,\; \gamma_{134\ldots n-2,\underline{2},n-1,n}^{(n,1)} \right)_a \, .
\label{n1-basis}
\ee

\subsubsection{Symmetric bases}

Let us compute the intersection matrix $H_{ab}^{(n,1)}$
in (\ref{intmatr}) explicitly. For the $n$-gon associated to $\gamma_{\ldots j\underline{2}k\ldots}^{(n,1)}$, only two chords $c_{s_{j2}}, c_{s_{2k}}$ are admissible: precisely those corresponding to the Mandelstam invariants $s_{j2}, s_{2k}$. Hence, we conclude that elements in the basis \eqref{n1-basis} which are more than one element apart have no common chords and thereby zero intersection number.

It remains to consider the other two cases. For the self-intersection number, we already computed the answer in \eqref{n1-self-int}, which after relabeling and expressing in terms of trigonometric functions gives
\be
\langle \gamma_{\ldots j\underline{2} k\ldots}^{(n,1)} | \gamma_{\ldots j\underline{2} k\ldots}^{(n,1)} \rangle =
\frac{\sin(\pi(s_{j2}{+}s_{2k})) }{ \sin(\pi s_{j2}) \sin(\pi  s_{2k})}=\cot(\pi s_{j2}) + \cot(\pi s_{2k}) \, .
\ee
Two adjacent cycles $\gamma_{\ldots j\underline{2} k,k+1,\ldots}^{(n,1)}$ and $\gamma_{\ldots j k \underline{2},k+1,\ldots}^{(n,1)}$ share a single chord $c_{s_{2k}}$, which decomposes the $n$-gon into a triangle $P_1$ and an $(n{-}1)$-gon $P_2$. Both of these have only one admissible empty tessellation, which is given by the polytope itself, $T_{P_i} = P_i$. Together with the fact that the relative winding number of the two permutations is $2$, we have
\be
\langle \gamma_{\ldots j\underline{2} k,k+1,\ldots}^{(n,1)} | \gamma_{\ldots j k \underline{2},k+1,\ldots}^{(n,1)} \rangle = - \frac{1}{\sin(\pi s_{2k})}  = - \csc(\pi s_{2k})
\ee
and the same result for $\langle \gamma_{\ldots j k \underline{2},k+1,\ldots}^{(n,1)} | \gamma_{\ldots j\underline{2} k,k+1,\ldots}^{(n,1)} \rangle$ by hermitian symmetry of the intersection product.
Organizing these results into an $(n{-}3)\times(n{-}3)$ symmetric tridiagonal matrix we obtain
\be\label{n-1-int}
H^{(n,1)}  = \ccccb
\cot(\pi s_{12}) {+} \cot(\pi s_{23}) &
- \csc(\pi s_{23}) &
0 &
\cdots \\
- \csc(\pi s_{23}) &
\cot(\pi s_{23}) {+} \cot(\pi s_{24}) &
- \csc(\pi s_{24}) 
&
\cdots \\
0 &
- \csc(\pi s_{24}) &
\cot(\pi s_{24}) {+} \cot(\pi s_{25}) &
\cdots \\
\vdots &
\vdots &
\vdots &
\ddots
\cccce \, .
\ee
We find that these matrices have the inverse with entries (recall that $s_{ii}=0$):
\be
(H^{(n,1)})^{-1}_{ab} = \frac{\sin (\pi \sum_{i=1}^{\min(a,b)+1} s_{2i}) \sin (\pi \sum_{i=\max(a,b)+2}^{n-1} s_{2i})}{\sin (\pi \sum_{i=1}^{n-1} s_{2i})} \, .
\label{H-inv-ab}
\ee

\subsubsection{Alternative bases}

In spite of the appeal of a symmetric basis choice for the entries of $\tilde{H}^{(n,p)}_{ab}$, we found a more convenient choice of bases that simplifies the entries of the KLT matrix. Let us denote the corresponding intersection matrix by
\be
\tilde{H}^{(n,1)}_{ab} = \langle \gamma_{a}^{(n,1)} | \tilde{\gamma}_{b}^{(n,1)} \rangle \, ,
\label{altHbasis}
\ee
where the right basis is now taken to be
\be
\tilde\gamma_{a}^{(n,1)} = \left( \gamma_{\underline{2}134\ldots n}^{(n,1)},\; \gamma_{1\underline{2}34\ldots n}^{(n,1)},\;
\ldots,\; \gamma_{134\ldots n-3,\underline{2},n-2,n-1,n}^{(n,1)} \right)_a \, .
\label{n1-basis-tilde}
\ee
We simply ``shifted'' the position of $2$ by one slot to the left compared to \eqref{n1-basis} which effectively moves the diagonals of the intersection matrix and leads to the new form
\be
\tilde{H}^{(n,1)} = \ccccb
- \csc(\pi s_{12})& \cot(\pi s_{12}) {+} \cot(\pi s_{23}) &
- \csc(\pi s_{23}) &
\cdots \\
0 &
- \csc(\pi s_{23}) &
\cot(\pi s_{23}) {+} \cot(\pi s_{24}) &
\cdots \\
0 &
0 &
- \csc(\pi s_{24}) &
\cdots \\
\vdots &
\vdots &
\vdots &
\ddots
\cccce \, .
\ee
This fact is crucial in simplifying the computation of the inverse, which can easily be seen to take the upper-triangular form
\be
(\tilde{H}^{(n,1)})^{-1}  = - \ccccb
\sin(\pi s_{12})& \sin(\pi (s_{12}{+}s_{23})) &
\sin(\pi (s_{12}{+}s_{23}{+}s_{24})) &
\cdots \\
0 &
\sin(\pi s_{23}) &
\sin(\pi (s_{23}{+}s_{24})) &
\cdots \\
0 &
0 &
\sin(\pi s_{24}) &
\cdots \\
\vdots &
\vdots &
\vdots &
\ddots
\cccce \, ,
\ee
or more explicitly
\be
(\tilde{H}^{(n,1)})_{ab}^{-1} = - \sin \bigg(\pi \sum_{i=a+1-\delta_{a1}}^{b+1} s_{2i} \bigg) \, .
\label{H-tilde-inv-ab}
\ee
The entries vanish for $b<a$ and are polynomial in terms of the sines, i.e.\ do not have any analogue of the denominator in (\ref{H-inv-ab}). This choice of bases will inform the choices for general $(n,p)$.

The matrix entries in \eqref{H-tilde-inv-ab} can be used to compute $\beta_a^{(n,1)}$ cycles in terms of the basis $\tilde{\gamma}_a^{(n,1)}$ needed in the definition of the integrals $I_{ab}^{(n,1)}$ given in \eqref{sph.6}. Using \eqref{J-to-I} and \eqref{inverse-H} we have, for instance,
\be
\beta^{(4,1)}_1 = - \sin (\pi s_{12}) \gamma_{\underline{2}134}^{(4,1)}\, ,
\ee
as well as
\begin{align}
\beta^{(5,1)}_1 &= - \sin (\pi s_{12}) \gamma_{\underline{2}1345}^{(5,1)}\, , \notag \\
\beta^{(5,1)}_2 &= - \sin (\pi (s_{12}{+}s_{23})) \gamma_{\underline{2}1345}^{(5,1)} - \sin (\pi s_{23}) \gamma_{1\underline{2}345}^{(5,1)}\, ,
\end{align}
and
\begin{align}
\beta^{(6,1)}_1 &= - \sin (\pi s_{12}) \gamma_{\underline{2}13456}^{(6,1)}\, , \notag \\
\beta^{(6,1)}_2 &= - \sin (\pi (s_{12}{+}s_{23})) \gamma_{\underline{2}13456}^{(6,1)} - \sin (\pi s_{23}) \gamma_{1\underline{2}3456}^{(6,1)}\, ,\\
\beta^{(6,1)}_3 &= - \sin (\pi (s_{12}{+}s_{23}{+}s_{24})) \gamma_{\underline{2}13456}^{(6,1)} - \sin (\pi (s_{23}{+}s_{24})) \gamma_{1\underline{2}3456}^{(6,1)} - \sin (\pi s_{24}) \gamma_{13\underline{2}456}^{(6,1)} \, . \notag
\end{align}

\subsubsection{Overcomplete form of KLT relations}

Before looking at $p=2$ examples, let us see how the same results could have been obtained from the overcomplete (but extremely simple) form of the KLT relations from \eqref{sph.7b}. We can write them with an $(n{-}1)\times(n{-}1)$ kernel matrix $\Phi$ with entries $\Phi_{ab} = \tfrac{i}{2} e^{i\pi \phi_{ab}}$ given by \eqref{phi-cd}. More explicitly, we have
\be
\Phi = \frac{i}{2} \cccccb
1 & e^{i \pi s_{12}} & e^{i \pi (s_{12}+s_{23})} & \cdots & e^{i \pi \sum_{i=1}^{n-1} s_{2i}}\\
e^{i \pi s_{12}} & 1 & e^{i\pi s_{23}} & \cdots & e^{i \pi \sum_{i=3}^{n-1} s_{2i}} \\
e^{i \pi (s_{12}+s_{23})} & e^{i\pi s_{23}} & 1 & \cdots & e^{i \pi \sum_{i=4}^{n-1} s_{2i}} \\
\vdots & \vdots & \vdots & \ddots & \vdots \\
e^{i \pi \sum_{i=1}^{n-1} s_{2i}} & e^{i \pi \sum_{i=3}^{n-1} s_{2i}} & e^{i \pi \sum_{i=4}^{n-1} s_{2i}} & \cdots & 1
\ccccce,
\ee
where the columns and rows are labelled by \emph{all} cycles in $\mathbb{R} \setminus \{ z_2 = z_j\}$,
\be
\Gamma = \left( \gamma_{\underline{2}134\ldots n}^{(n,1)},\; \gamma_{1\underline{2}34\ldots n}^{(n,1)},\;
\ldots,\; \gamma_{134\ldots n-1,\underline{2},n}^{(n,1)} \right)^t.
\label{Gamma}
\ee
In order to reduce this to the $(n{-}3)\times(n{-}3)$ form, one makes use of the fact that only $n{-}3$ cycles in \eqref{Gamma} are linearly dependent. They satisfy a pair of monodromy relations \cite{Plahte:1970wy, BjerrumBohr:2009rd, Stieberger:2009hq}:
\be
\cccccb
1 & e^{i\pi s_{12}} & e^{i\pi (s_{12}+s_{23})} & \cdots & e^{i\pi
\sum_{i=1}^{n-1} s_{2i}} \\
1 & e^{-i\pi s_{12}} & e^{-i\pi (s_{12}+s_{23})} & \cdots & e^{-i\pi 
\sum_{i=1}^{n-1} s_{2i}}
\ccccce 
\Gamma = \begin{pmatrix} 0 \\ 0 \end{pmatrix},
\ee
where $s_{22}=0$ and the first (second) row comes from considering a contour right above (below) the real $z_2$-axis and deforming it to a point in the upper-half (lower-half) plane, also see appendix \ref{app:A.1}. Let us invert these relations to construct projectors onto the two bases \eqref{n1-basis}, \eqref{n1-basis-tilde} we considered in this subsection. We can write $(\Gamma_1, \Gamma_{n-1}) P = \Gamma$ with
\begin{align}
P &= - \ccb 1 & e^{i\pi 
\sum_{i=1}^{n-1} s_{2i}}\\
1 & e^{-i\pi 
\sum_{i=1}^{n-1} s_{2i}} \cce^{-1}
\ccccb
e^{i\pi s_{12}} & e^{i\pi (s_{12}+s_{23})} & \cdots & e^{i\pi 
\sum_{i=1}^{n-2} s_{2i}} \\
e^{-i\pi s_{12}} & e^{-i\pi (s_{12}+s_{23})} & \cdots & e^{-i\pi 
\sum_{i=1}^{n-2} s_{2i}}
\cccce \notag
\\
&= \frac{1}{\sin(\pi s_{2n})}\ccccb
\sin(\pi \sum_{i=3}^{n-1} s_{2i}) & \sin(\pi \sum_{i=4}^{n-1} s_{2i}) & \cdots & \sin(\pi s_{2,n-1}) \\
\sin(\pi s_{12}) & \sin(\pi (s_{12}{+}s_{23})) & \cdots & \sin(\pi \sum_{i=1}^{n-2} s_{2i})
\cccce ,
\end{align}
which expresses the first and last elements of $\Gamma$ in terms of the basis $\gamma^{(n,1)}_a$. Similarly, eliminating the second-last and last elements we have a projector onto the $\tilde{\gamma}^{(n,1)}_a$ basis:
\begin{align}
\tilde{P} &= - \ccb e^{i\pi 
\sum_{i=1}^{n-2} s_{2i}} & e^{i\pi 
\sum_{i=1}^{n-1} s_{2i}}\\
e^{-i\pi 
\sum_{i=1}^{n-2} s_{2i}} & e^{-i\pi
\sum_{i=1}^{n-1} s_{2i}} \cce^{-1}
\cccccb
1 & e^{i\pi s_{12}} & e^{i\pi (s_{12}+s_{23})} & \cdots & e^{i\pi 
\sum_{i=1}^{n-3} s_{2i}} \\
1 & e^{-i\pi s_{12}} & e^{-i\pi (s_{12}+s_{23})} & \cdots & e^{-i\pi
\sum_{i=1}^{n-3} s_{2i}}
\ccccce 
\\
&= \frac{1}{\sin(\pi s_{2,n-1})}\ccccb
{-}\sin(\pi \sum_{i=1}^{n-1} s_{2i}) & {-}\sin(\pi \sum_{i=3}^{n-1} s_{2i}) & \cdots & {-}\sin(\pi (s_{2,n-2}{+}s_{2,n-1})) \\
\sin(\pi \sum_{i=1}^{n-2} s_{2i}) & \sin(\pi \sum_{i=3}^{n-2} s_{2i}) & \cdots & \sin(\pi s_{2,n-2})
\cccce \, . \notag
\end{align}
With these computations in place, we can simply apply the projectors to the relevant columns and rows of the overcomplete KLT matrix to obtain:
\be
\begin{pmatrix} P_1 \\ \mathds{1}_{n-3} \\ P_2 \end{pmatrix}^{\!\!t}
\Phi
\begin{pmatrix} P_1 \\ \mathds{1}_{n-3} \\ P_2 \end{pmatrix} = (H^{(n,1)})^{-1} \, , \qquad
\begin{pmatrix} \mathds{1}_{n-3} \\ \tilde{P}_1 \\ \tilde{P}_2 \end{pmatrix}^{\!\!t}
\Phi
\begin{pmatrix} P_1 \\ \mathds{1}_{n-3} \\ P_2 \end{pmatrix} = (\tilde{H}^{(n,1)})^{-1} \, ,
\ee
reproducing the results from \eqref{H-inv-ab} and \eqref{H-tilde-inv-ab}.

\subsection{Case $p=2$}

Recall that for $p=2$ the basis $\gamma_a^{(n,2)}$ consists of all cycles where $z_1 < z_4 < z_5 < \cdots < z_n = \infty$ are fixed in this order and $z_2, z_3$ are placed between $z_1 = 0$ and $z_{n-1} = 1$, for the total of $d^{(n,2)} = (n{-}3)(n{-}4)$ elements in the basis.

Motivated by the simplicity of the results for $p=1$, we will also introduce a second basis $\tilde{\gamma}_a^{(n,2)}$ where $z_2, z_3$ are placed between $z_n = -\infty$ and $z_{n-2}$ (or $z_1$ for $p=n{-}3$). To make the notation a bit more clear and recognize a pattern, let us compute the KLT kernel for the examples $n=5,6$.

\subsubsection{Example $(n,p) = (5,2)$}

The two bases are given by
\be
\gamma_{a}^{(5,2)} = \left( \gamma_{1\underline{23}45}^{(5,2)},\; \gamma_{1\underline{32}45}^{(5,2)} \right)_a,
\qquad
\tilde{\gamma}_{b}^{(5,2)} = \left( \gamma_{\underline{23}145}^{(5,2)},\; \gamma_{\underline{32}145}^{(5,2)} \right)_b.
\ee
Computing the intersection matrix according to the above combinatorial rules we find
\be
\tilde H^{(5,2)} = \csc(\pi s_{123}) \ccb
- \csc(\pi s_{23})  & \cot(\pi s_{12}) + \cot(\pi s_{23}) \\
\cot(\pi s_{13}) + \cot(\pi s_{23}) & - \csc(\pi s_{23})
\cce \, ,
\ee
where the tilde refers to the asymmetric basis choice $\tilde H^{(5,2)}_{ab} = \langle \gamma_{a}^{(5,2)}|
\tilde{\gamma}_{b}^{(5,2)} \rangle$ analogous to (\ref{altHbasis}).
Inverting the matrix we obtain the well-known result
for the local KLT matrix
\be
(\tilde H^{(5,2)})^{-1} = \ccb
\sin(\pi s_{12}) \sin(\pi s_{13})  & \sin(\pi s_{13}) \sin(\pi (s_{12}{+}s_{23}))  \\
\sin(\pi s_{12}) \sin(\pi (s_{13}{+}s_{23})) & \sin(\pi s_{12}) \sin(\pi s_{13})
\cce \, ,
\ee
in agreement with momentum-kernel techniques \cite{momentumKernel}. With this result the basis cycles $\beta_{a}^{(5,2)}$
in (\ref{sph.6}) and (\ref{sph.7}) read as follows,
\begin{align}
\beta_1^{(5,2)} = \sin(\pi s_{12}) \left( \sin(\pi s_{13}) \gamma_{\underline{23}145}^{(5,2)} + \sin(\pi (s_{13}{+}s_{23})) \gamma_{\underline{32}145}^{(5,2)} \right)\, , \notag \\
\beta_2^{(5,2)} = \sin(\pi s_{13}) \left(  \sin(\pi (s_{12}{+}s_{23})) \gamma_{\underline{23}145}^{(5,2)} + \sin(\pi s_{12}) \gamma_{\underline{32}145}^{(5,2)} \right)\, .
\end{align}

\subsubsection{Example $(n,p) = (6,2)$}

In this case the two bases are given by
\begin{gather}
\gamma_{a}^{(6,2)} = \left(
\gamma_{1\underline{23}456}^{(6,2)},\;
\gamma_{1\underline{2}4\underline{3}56}^{(6,2)},\;
\gamma_{1\underline{32}456}^{(6,2)},\;
\gamma_{1\underline{3}4\underline{2}56}^{(6,2)},\;
\gamma_{14\underline{23}56}^{(6,2)},\;
\gamma_{14\underline{32}56}^{(6,2)}
\right)_a,\\
\tilde{\gamma}_{b}^{(6,2)} = \left(
\gamma_{\underline{23}1456}^{(6,2)},\;
\gamma_{\underline{2}1\underline{3}456}^{(6,2)},\;
\gamma_{1\underline{23}456}^{(6,2)},\;
\gamma_{\underline{32}1456}^{(6,2)},\;
\gamma_{\underline{3}1\underline{2}456}^{(6,2)},\;
\gamma_{1\underline{32}456}^{(6,2)}
\right)_b \, .
\end{gather}
Among the intersection numbers $\tilde H^{(6,2)}_{ab} = \langle \gamma_{a}^{(6,2)}|
\tilde{\gamma}_{b}^{(6,2)} \rangle$ in the asymmetric
basis choice, we already computed one example in the entry $\tilde H_{15}^{(6,2)}$ in \eqref{123456-example}. Due to space limitations we do not present the full intersection matrix here. Its $6 \times 6$ inverse, however, takes a relatively compact form:
\be
(\tilde H^{(6,2)})^{-1} = 
\left (\begin {array} {ccc}
                     \sin (\pi  s_ {12}) \sin (\pi  s_ {13}) & \sin (\pi 
                       s_ {12}) \sin (\pi  s_ {14, 
                    3}) & \sin (\pi  s_ {13})
                       \sin (\pi  s_ {13, 
                    2})  \\
                     
                    0 & \sin (\pi  s_ {12}) \sin (\pi  s_ {34}) & 0  \\
                     
                    0 & 0 & 0  \\
                     \sin (\pi  s_ {12}) \sin (\pi  s_ {12, 
                    3}) & \sin (\pi 
                       s_ {12}) \sin (\pi  s_ {124, 
                    3}) & \sin (\pi  s_ {12})
                       \sin (\pi  s_ {13})  \\
                     
                    0 & 0 & 0  \\
                0 & 0 & 0  \\
    \end {array}
   \right.
 \ee
 \be\qquad\qquad
   \left. \begin {array} {ccc}
                     \sin (\pi  s_ {13}) \sin (\pi 
                       s_ {134, 2}) & \sin (\pi  s_ {14, 2}) \sin (\pi 
                       s_ {14, 3}) & \sin (\pi  s_ {14, 
                    3}) \sin (\pi 
                       s_ {134, 2}) \\
 0 & \sin
                       (\pi  s_ {34}) \sin (\pi  s_ {14, 
                    2}) & \sin (\pi 
                       s_ {34}) \sin (\pi  s_ {134, 2}) \\
              0 & \sin (\pi  s_ {24}) \sin (\pi  s_{34}) & \sin
                       (\pi  s_ {34}) \sin (\pi  s_ {34, 2}) \\  \sin (\pi  s_ {13}) \sin
(\pi 
                       s_ {14, 2}) & \sin (\pi  s_ {14, 
                    2}) \sin (\pi 
                       s_ {124, 3}) & \sin (\pi  s_ {14, 2}) \sin (\pi 
                       s_ {14, 3}) \\ \sin (\pi  s_ {13}) \sin (\pi  s_{24}) & \sin
                     (\pi  s_ {24}) \sin (\pi  s_ {124, 
                    3}) & \sin (\pi 
                     s_ {24}) \sin (\pi  s_ {14, 3}) \\ 
                 0 & \sin (\pi  s_ {24}) \sin (\pi  s_ {24,
          3}) &
       \sin (\pi  s_ {24}) \sin (\pi  s_ {34}) \\
    \end {array}
   \right) \, ,\notag
\ee
where we use the notation $s_{i_1 i_2\ldots,j} = s_{i_1 j} + s_{i_2 j} + \cdots$. In terms of the  $\beta_a^{(6,2)}$ cycles from (\ref{sph.6}), this translates to
\begin{align}
\beta_1^{(6,2)} &= \sin (\pi  s_{12}) \left(\sin (\pi  s_{13})
   \gamma^{(6,2)}_{\underline{23}1456} + \sin (\pi  s_{12,3})
   \gamma^{(6,2)}_{\underline{32}1456}\right) \, , \notag \\
 \beta_2^{(6,2)} &= \sin (\pi  s_{12}) \left(\sin (\pi s_{14,3})
   \gamma^{(6,2)}_{\underline{23}1456} + \sin (\pi  s_{34})
   \gamma^{(6,2)}_{\underline{2}1\underline{3}456} + \sin (\pi 
   s_{124,3}) \gamma^{(6,2)}_{\underline{32}1456}\right)\, , \notag \\
 \beta_3^{(6,2)} &=  \sin (\pi  s_{13}) \left(\sin (\pi s_{13,2})
   \gamma^{(6,2)}_{\underline{23}1456} +
 \sin (\pi  s_{12})
   \gamma^{(6,2)}_{\underline{32}1456}\right)\, , \\
   \beta_4^{(6,2)} &=  \sin (\pi  s_{13}) \left(
   \sin (\pi 
   s_{134,2}) \gamma^{(6,2)}_{\underline{23}1456}
   + \sin (\pi s_{14,2})
   \gamma^{(6,2)}_{\underline{32}1456}
   + \sin (\pi  s_{24})
   \gamma^{(6,2)}_{\underline{3}1\underline{2}456} \right)\, , \notag \\
 \beta_5^{(6,2)} &=
 \sin(\pi s_{14,2}) \left(
 \sin(\pi s_{14,3})\gamma_{\underline{23}1456}^{(6,2)} +
\sin(\pi s_{34})\gamma_{\underline{2}1\underline{3}456}^{(6,2)} +
\sin(\pi s_{124,3}) \gamma_{\underline{32}1456}^{(6,2)} \right) \notag\\
& \quad + \sin(\pi s_{24}) \left(
\sin(\pi s_{34}) \gamma_{1\underline{23}456}^{(6,2)} +
\sin(\pi s_{124,3}) \gamma_{\underline{3}1\underline{2}456}^{(6,2)} +
\sin(\pi s_{24,3}) \gamma_{1\underline{32}456}^{(6,2)}
\right) \, , \notag \\
 \beta_6^{(6,2)} &= \sin(\pi s_{34}) \left( 
\sin(\pi s_{134,2})\gamma_{\underline{2}1\underline{3}456}^{(6,2)}+
\sin(\pi s_{34,2}) \gamma_{1\underline{23}456}^{(6,2)} +
\sin(\pi s_{24}) \gamma_{1\underline{32}456}^{(6,2)}
\right)  \notag \\
&\quad + \sin(\pi s_{14,3}) \left(
\sin(\pi s_{134,2}) \gamma_{\underline{23}1456}^{(6,2)} +
\sin(\pi s_{14,2}) \gamma_{\underline{32}1456}^{(6,2)} +
\sin(\pi s_{24}) \gamma_{\underline{3}1\underline{2}456}^{(6,2)}
\right)\, .
 \notag
\end{align}
This example already illustrates the general rule: For each integrated puncture $i \in \{2,3\}$ we have a sine factor in the generalized KLT kernel $S_{\alpha'}$ in (\ref{inverse-H}). The arguments of the sine functions are given by the overlap between labels to the left of $i$ in $\gamma_{a}^{(6,2)}$ which are also to the right of $i$ in $\tilde{\gamma}_{b}^{(6,2)}$. We make this observation more concrete in the following.

\subsection{\label{sec:6.5}Recursion for general $(n,p)$}

The goal of this subsection is to find the explicit form of the cycles $\beta^{(n,p)}$
in (\ref{sph.6}) which need to be field-theory orthonormal with respect to the forms $\nu^{(n,p)}$ that are the Betti--de Rham 
duals of the integration cycles
\beq
\gamma^{(n,p)}_{\vec{A},\vec{i}} \quad\leftrightarrow\quad \rho_{\vec{A},\vec{i}} = (1,A_1,\underline{i_1},A_2,\underline{i_2},\ldots, A_p,\underline{i_p},A_{p+1},n{-}1,n)
\label{allklt.1}
\eeq
in order to avoid inconsistency in the $\alpha' \rightarrow 0$ limit of (\ref{sph.7}).

As in section~\ref{sec:2.2}, we gather the unintegrated punctures different from $(z_1,z_{n-1},z_n)=(0,1,\infty)$
in a vector $\vec{A}$ of words $A_1,A_2,\ldots$ each of which gathers (possibly zero) adjacent 
unintegrated punctures. The $i_1,i_2,\ldots,i_p$ in turn are a permutation of the $p$ integrated punctures
$z_2,z_3,\ldots,z_{p+1}$.

The examples of the $\beta^{(n,p)}$ in the earlier subsections 
have orthonormal intersection numbers
with the $\gamma^{(n,p)}$ in (\ref{allklt.1}) in the sense that
\beq
\langle \gamma^{(n,p)}_{\vec{A},\vec{i}}  | \beta^{(n,p)}_{\vec{B},\vec{j}}  \rangle = \delta_{\vec{A},\vec{B}} \delta_{\vec{i},\vec{j}} \, .
\label{allklt.2}
\eeq
At general $n$ and $p$, the $\beta^{(n,p)}$ with this property are conjecturally given by
\begin{align}
 \beta^{(n,p)}_{\vec{A},\vec{i}} = \sum_{\vec{B},\vec{j}} S_{\alpha'}(1,\vec{A},\vec{i}  \ | \vec{B},\vec{j},n{-}2)\,
  \tilde\gamma^{(n,p)}_{\vec{B},\vec{j}}\, ,
  \label{allklt.3}
\end{align}
where we employ the alternative basis (\ref{allklt.4}) of $d^{(n,p)}$ cycles $\tilde\gamma^{(n,p)}_{\vec{B},\vec{j}}$ instead of (\ref{allklt.1}), in order to obtain a local expression for the generalized KLT kernel $S_{\alpha'}$. Since the final two labels are always the same, we suppress them in (\ref{allklt.3}) and below for clarity, i.e., $S_{\alpha'}(X|Y) = S_{\alpha'}(X,n{-}1,n |Y,n{-}1,n)$. The latter 
is claimed to obey the following recursion in the number of integrated punctures $i_k$
\begin{align}
&S_{\alpha'}(1,A_1,\underline{i_1},A_2,\ldots,A_p,\underline{i_p},A_{p+1} | X, \underline{i_p}, Y)
\label{allklt.5} \\
&=  - \sin(2 \pi \alpha' k_{i_p} {\cdot} \textstyle\sum_{\ell \in Y \cap (1,A_1,i_1,\ldots,i_{p-1},A_p) } k_\ell)\, S_{\alpha'}(1,A_1,\underline{i_1},A_2,\ldots,\underline{i_{p-1}},A_p,A_{p+1} | X, Y) \, . \notag
\end{align}
This step may only be applied to remove the rightmost integrated puncture $i_p$ in the
first entry, and the recursion terminates with 
\beq
S_{\alpha'}(1,\vec{A} \ |\vec{B},n{-}2)= \delta_{(1,\vec{A}),(\vec{B},n{-}2)}
\label{allklt.6}
\eeq
when there are no more integrated punctures left.
This has been verified up to and including $n=8$ for any value of $p\leq n{-}4$ by checking that $S_{\alpha'}$ is indeed the inverse of $\tilde H^{(n,p)}$ obtained with combinatorial rules of the previous subsections.\footnote{Throughout this subsection we have assumed that $p\neq n{-}3$: otherwise, the 
basis (\ref{allklt.4}) of $\tilde \gamma^{(n,p)}$ would be of the form
$(\ldots,1,n{-}1,n)$ rather than $(\ldots,n{-}2,n{-}1,n)$ when maintaining the recursion
for $S_{\alpha'}$. Since local representations of the KLT formula for $p= n{-}3$
are well explored in the literature \cite{Bern:1998sv, momentumKernel}, there is no loss of generality in demanding $p<n{-}3$ here.}
More generally, the recursion (\ref{allklt.5}) can be rewritten as 
\beq
S_{\alpha'}(P,\underline{i},Q|X,\underline{i},Y) = - \sin(2\pi \alpha' k_i\cdot k_{P \cap Y})\, S_{\alpha'}(P,Q|X,Y) \, ,
\label{allklt.7}
\eeq
where $Q$ has no integrated punctures,
i.e., the momenta in the sine functions are determined
by the punctures that appear on opposite sides of $i$ in the two entries of $S_{\alpha'}(\cdot | \cdot)$. The structure
of the recursion (\ref{allklt.7}) resonates with the
momentum-kernel formalism \cite{momentumKernel} and its generalization to the KLT formulae for $p=n{-}4$ \cite{Vanhove:2018elu}.

With the expansion (\ref{allklt.3}), the desired orthonormality property (\ref{allklt.2}) takes
the form (with collective indices $a,b,c$ taking the role of $\vec{A},\vec{i}$),
\beq
\langle \gamma^{(n,p)}_a | \beta_b^{(n,p)} \rangle = \sum_{c=1}^{d^{(n,p)} } \langle \gamma_a^{(n,p)} | \tilde{\gamma}^{(n,p)}_c \rangle (S_{\alpha'})_{bc} = \delta_{ab} \,,
\label{allklt.8}
\eeq
In order to deduce the desired orthonormality of $ \beta_a^{(n,p)}$ and $ \nu_b^{(n,p)}$ in the 
$\alpha' \rightarrow 0 $ limit, we insert complete sets of cycles $\gamma^{(n,p)}_c$ and 
cocycles $\overline{\omega_d^{(n,p)}}$,
\begin{align}
\lim_{\alpha' \rightarrow 0} \langle \beta_a^{(n,p)} | \nu^{(n,p)}_b \rangle 
&= \lim_{\alpha' \rightarrow 0} \sum_{c,d=1}^{d^{(n,p)}}  \langle \beta_a^{(n,p)} | \gamma_c^{(n,p)} \rangle
\, (\overline{F^{(n,p)}})^{-1}_{cd} \,
\langle \overline{\omega_d ^{(n,p)}}  | \nu^{(n,p)}_b \rangle 
\notag \\
&=  \sum_{c,d=1}^{d^{(n,p)}} \delta_{ac} \delta_{cd} \delta_{db} = \delta_{ab}\, ,
\label{allklt.9}
\end{align}
where $\overline{F^{(n,p)}_{cd}} = \langle \overline{\omega_d^{(n,p)}} | \gamma_c^{(n,p)} \rangle
=
\overline{\langle \gamma_c^{(n,p)} | \omega_d^{(n,p)} \rangle}$ as in \eqref{conv5}. 
The final two Kronecker deltas in passing to the last line stem from the fact that 
the $\overline{\omega_d ^{(n,p)}}$ are engineered to be field-theory orthonormal to both
$ \gamma_c^{(n,p)} $ and $\nu^{(n,p)}_b $. The first Kronecker delta 
arises from the conjectural orthonormality (\ref{allklt.8}).\footnote{In general, intersection numbers satisfy $\langle \gamma | \tilde{\gamma} \rangle = \overline{\langle \tilde{\gamma} | \gamma \rangle}$, but in our normalizations they are purely real, which is why the equality \eqref{allklt.8} also implies $\langle \beta_a^{(n,p)} | \gamma^{(n,p)}_c \rangle = \delta_{ac}$.} It would be interesting to
find a rigorous all-multiplicity proof that (\ref{allklt.3}) together with the recursion (\ref{allklt.5}) indeed
leads to orthonormal intersection numbers.

\section{\label{sec:7}Implications for minimal models}

In this section we give an interpretation of our results in terms of correlation functions in two-dimensional conformal field theories (CFTs). We will focus on the family of theories known as the \emph{minimal models} whose spectrum can been completely classified and solved in terms of irreducible representations of the Virasoro algebra.\footnote{Saying that minimal models are solved by no means implies that their correlation functions have been computed, or are easy to compute, in general.}
We start with a lightning review of these models, where we focus only on the parts necessary to make connections with the rest of this paper. For more comprehensive expositions we refer the reader to \cite{dotsenko1988,DiFrancesco:1997nk,Kapec:2020xaj}.

\subsection{Lightning review of the Coulomb gas formalism}

Our starting point is the action of a free boson $\phi(x)$ coupled linearly to the scalar curvature $R$ of the genus-zero surface:
\be
S_{\mathfrak{p},\mathfrak{p}'} = \int_{\CP^1} \dd^2 x \sqrt{g} \left( \frac{1}{2} \partial_\mu \phi \partial^\mu \phi + \frac{i}{\sqrt{2}} Q_{\mathfrak{p},\mathfrak{p}'} \phi R \right).
\ee
Here the strength of the coupling is given by the \emph{background charge} $Q_{\mathfrak{p},\mathfrak{p}'}$, which makes the $\mathrm{U}(1)$ symmetry anomalous. Since the action is complex, it does not automatically give rise to a unitary theory. In fact, families of unitary models written in this way are heavily constrained and can be classified by a pair of co-prime integers $(\mathfrak{p},\mathfrak{p}')$, in terms of which
\be
Q_{\pp,\pp'} = \frac{\pp - \pp'}{\sqrt{\pp \pp'}} \, .
\ee
The central charge is $c_{\pp,\pp'}=1- 6 Q_{\pp,\pp'}^2$, and we take $\pp > \pp'$ by convention. These are the minimal models. For example
$(\pp,\pp')=(4,3)$ gives the critical Ising model with $Q_{4,3} = \frac{1}{2\sqrt{3}}$ and $c_{4,3} =\frac{ 1}{2}$,
while $(\pp,\pp')=(5,2)$ is the Yang--Lee edge singularity with $Q_{5,2} = \frac{3}{\sqrt{10}}$ and $c_{5,2} = -\frac{22}{5}$.

Conformal primary operators ${\cal O}_{q_{(r,s)}}$ in the $(\mathfrak{p},\mathfrak{p}')$ minimal model are classified by two integers $(r,s)$ such that
\be
1 \leq r \leq \pp'{-}1, \qquad 1 \leq s \leq \pp{-}1  \, .
\ee
Charges $q_{(r,s)}$ and conformal dimensions $h_{(r,s)}$ of these operators are given by
\be\label{q-r-s}
q_{(r,s)} = \frac{\pp(1{-}r) - \pp'(1{-}s)}{2 \sqrt{\pp \pp'}} \,, \qquad h_{(r,s)} = \frac{(r\pp {-} s\pp')^2 - (\pp {-} \pp')^2}{4 \pp \pp'} \, .
\ee
Notice that operators ${\cal O}_q$ and ${\cal O}_{Q_{\pp,\pp'}-q}$ share the same conformal dimension and they are indistinguishable at the level of correlation functions. In other words, we can identify operators with $(r,s)$ and $(\pp' {-}r, \pp{-}s)$. For instance, in the case of the critical Ising model with $(\pp,\pp')=(4,3)$ we have the following Kac table:
\be\arraycolsep=.5em\def\arraystretch{1.5}
\begin{array}{c||c|c|c}
 & \medmath{s=1} & \medmath{s=2} & \medmath{s=3} \\
 \hline
\medmath{r=1} & {\cal O}_{0} = \mathds{1}  & {\cal O}_{\frac{\sqrt{3}}{4}} = \sigma & {\cal O}_{\frac{\sqrt{3}}{2}} = \varepsilon \\ 
\medmath{r=2} & {\cal O}_{-\frac{1}{\sqrt{3}}} = \varepsilon & {\cal O}_{-\frac{1}{4\sqrt{3}}} = \sigma & {\cal O}_{\frac{1}{2\sqrt{3}}} = \mathds{1} \\  
\end{array}
\ee
Here $\mathds{1}$, $\sigma$, and $\varepsilon$ are the usual identity, spin, and energy operators of conformal weight $0$, $\frac{1}{16}$ and $\frac{1}{2}$, respectively.

We will be interested in computing the correlation function of $\mathrm{N}$ such operators. For readability we will simply label the $j$-th vertex operator ${\cal O}_{q_j}(x_j) = e^{i \sqrt{2} q_j \phi(x_j)}$ by its charge $q_j$:
\be\label{MM-correlator}
\langle {\cal O}_{q_1}(x_1) {\cal O}_{q_2}(x_2) \cdots {\cal O}_{q_\mathrm{N}}(x_\mathrm{N})  \rangle_{\pp,\pp'} \, .
\ee
Such a computation might not seem approachable, because we deal with a strongly-interacting system. However, one can simplify this problem conceptually using the \emph{Coulomb gas formalism} \cite{Dotsenko:1984nm,Dotsenko:1984ad}, which is the idea that correlation functions in interacting theories with background charge can be equivalently represented as those in a free theory with insertions of $p$ charged operators integrated over the whole surface. As a result, the correlation functions \eqref{MM-correlator} can be represented as
\be\label{int}
\int_{(\mathbb{CP}^1)^p} \prod_{i=2}^{p+1} \dd^2 z_i\, \langle {\cal O}_{q_1}(x_1) {\cal O}_{q_2}(x_2) \cdots {\cal O}_{q_\mathrm{N}}(x_{\mathrm{N}}) \prod_{i=2}^{p+1} {\cal O}_{q_{\pm}}(z_i)\rangle_{\mathrm{free}}
\ee
up to a constant.
The additional operators are called \emph{screening charges}, and their charges can only take two values, $q_+$ and $q_-$, given by
\be
q_+ = \sqrt{\pp / \pp'} \, , \qquad q_- = -\sqrt{\pp'/\pp} \, ,
\ee
such that $q_+ + q_- = Q_{\pp,\pp'}$.
We will denote the number of screening charges ${\cal O}_{q_\pm}$ by $p_\pm$, such that $p_+ + p_- = p$. These numbers can be determined by imposing the neutrality condition (Ward identity), i.e.\ requiring that the sum of charges equals to the background charge,
\be\label{neutrality}
\sum_{i=1}^{\mathrm{N}} q_i \,+\, p_+ q_+ \,+\, p_- q_- \,=\, Q_{\pp,\pp'} \, .
\ee
As a heuristic, for sufficiently generic $(\pp,\pp')$, reading off the coefficients of the irrational numbers $\sqrt{\pp/\pp'}$ and $\sqrt{\pp'/\pp}$ translates to the following condition for the integers $(r_i,s_i)$ labeling every operator:
\be\label{p-plus-minus}
p_+ = \frac{1}{2}\left(\sum_{i=1}^{\mathrm{N}} r_i - \mathrm{N}+ 2 \right) \, , \qquad p_- = \frac{1}{2}\left(\sum_{i=1}^{\mathrm{N}} s_i - \mathrm{N}+ 2 \right) \, .
\ee
For instance, the four-point correlation function of ${\cal O} = {\cal O}_{q_{(2,1)}}$ operators requires $p_+=3$ and $p_-=1$ and hence can be written as
\begin{align}
\langle {\cal O}(x_1) {\cal O}(x_2) {\cal O}(x_3){\cal O}(x_4) \rangle_{\pp,\pp'} = \int_{(\mathbb{CP}^1)^4} \prod_{i=2}^{5}\dd^2 z_i\, \langle &{\cal O}(x_1) {\cal O}(x_2) {\cal O}(x_3){\cal O}(x_4)\\[-1em]
&{\cal O}_{q_+}(z_2) {\cal O}_{q_+}(z_3) {\cal O}_{q_+}(z_4) {\cal O}_{q_-}(z_5) \rangle_{\mathrm{free}} \nn
\end{align}
since in this case the neutrality condition reads
\be
4q_{(2,1)} + 3q_+ + q_- = Q_{\pp,\pp'} \, .
\ee
However, this representation is not unique. For example, since we can dually represent one of ${\cal O}$ as $\widetilde{\cal O} = {\cal O}_{Q_{\pp,\pp'}-q_{(2,1)}}$, we find a simpler representation
\begin{align}\label{N4-example}
\langle {\cal O}(x_1) {\cal O}(x_2) {\cal O}(x_3) \widetilde{\cal O}(x_4) \rangle_{\pp,\pp'} = \int_{\mathbb{CP}^1} \dd^2 z_2\, \langle &{\cal O}(x_1) {\cal O}(x_2) {\cal O}(x_3) \widetilde{\cal O}(x_4) {\cal O}_{q_+}(z_2)\rangle_{\mathrm{free}} \, ,
\end{align}
given that
\be
3 q_{(2,1)} + (Q_{\pp,\pp'} - q_{(2,1)}) + q_+ = Q_{\pp,\pp'} \, .
\ee
We will return to this example in section~\ref{sec:7.5} once we establish the connection to the results of this paper. (Note that when we use the dual description $\widetilde{\cal O}$, $(r_4,s_4)=(\pp'-2,\pp-1)$ are $\pp,\pp'$-dependent and we can no longer use \eqref{p-plus-minus}, which would otherwise predict $p_{\pm} \geq 1$. The neutrality condition \eqref{neutrality} always holds.)

Of course, the free-theory correlator inside of the integrand of \eqref{int} can be written down explicitly, giving us the explicit formula
\be
\langle {\cal O}_{q_1}(x_1) {\cal O}_{q_2}(x_2) \cdots {\cal O}_{q_\mathrm{N}}(x_\mathrm{N})  \rangle_{\pp,\pp'} = \int_{(\mathbb{CP}^1)^p} \prod_{i=2}^{p+1} \dd^2 z_i\, e^{W+\overline{W}},
\ee
where
\be
W = 2\!\!\sum_{1 \leq i< j}^{\mathrm{N}} q_i q_j \log (x_i {-} x_j) + 2\sum_{i=1}^{\mathrm{N}} \sum_{j=2}^{p+1} q_i Q_j \log (x_i {-} z_j) + 2\!\!\sum_{2 \leq i<j}^{p+1} Q_{i} Q_j \log (z_i {-} z_j)
\label{wpotential}
\ee
is (the holomorphic part of) the potential for interacting charges on a genus-zero surface. Here the screening charges $Q_j$ with $j=2,3,\ldots,p{+}1$ are given by $q_+$ or $q_-$ as determined by the rules explained above.

At this stage, one sees that the computation of correlation functions in minimal models involves complex integrals that are structurally identical to those considered in section~\ref{sec:6}. This relationship is rather well-known \cite{DiFrancesco:1997nk,Mimachi2003} and was previously exploited in the case $p=n{-}4$ in the context of the single-valued map in string perturbation theory \cite{Vanhove:2018elu,Vanhove:2020qtt}. We follow with a summary of implications of our results for such correlation functions.

\subsection{Translation of notation}

Correlation functions in the Coulomb gas formalism involve a total of $\mathrm{N}{+}p$ punctures, out of which $p$ are integrated. In the notation of this paper it means
\be
n = \mathrm{N} {+} p \, ,
\ee
together with
\be
(z_1, z_{p+2}, z_{p+3}, \ldots, z_{n}) = (x_1, x_2, x_3, \ldots, x_{\mathrm{N}})
\ee
and gauge fixing $(x_1,x_{\mathrm{N}-1},x_{\mathrm{N}}) = (0,1,\infty)$. Moreover, the Mandelstam invariants are identified according to
\be
s_{ij} = 2 \tilde{q}_i \tilde{q}_j, \qquad s_{i_1 i_2 \ldots i_m} = 2\!\!\!\sum_{1 \leq j<k \leq m} \!\!\! \tilde{q}_{i_j} \tilde{q}_{i_k} \, ,
\label{identsq}
\ee
where
\be
(\tilde{q}_1, \tilde{q}_2, \tilde{q}_3, \ldots, \tilde{q}_{p+1}, \tilde{q}_{p+2}, \tilde{q}_{p+3}, \ldots, \tilde{q}_{n} ) = (q_1, Q_2, Q_3, \ldots, Q_{p+1}, q_2, q_3, \ldots, q_{\mathrm{N}}) \, .
\ee
In terms of the Koba--Nielsen factor defined in \eqref{conv2} we have
\begin{align}\label{transl}
\lim_{x_{\mathrm{N}}\to\infty}\frac{|x_{\mathrm{N}}|^{4q_{\mathrm{N}}(q_{\mathrm{N}} - Q_{\pp,\pp'})}}{\pi^p}\prod_{1\leq i<j}^{\mathrm{N}-1} |x_i {-} x_j|^{-4q_i q_j} &\langle {\cal O}_{q_1}(x_1) {\cal O}_{q_2}(x_2) \cdots {\cal O}_{q_\mathrm{N}}(x_{\mathrm{N}})  \rangle_{\pp,\pp'} \\
&=  \frac{1}{\pi^p}\! \int_{{\cal C}^{(\mathrm{N}+p,p)}} \prod_{i=2}^{p+1} \dd^2 z_i \, | {\rm KN}^{(\mathrm{N}+p,p)} |^2 \, ,\nn
\end{align}
The right-hand side is in the class of integrals given in \eqref{sph.5} as $\langle \prod_{i=2}^{p+1} \dd \overline{z}_i | \prod_{i=2}^{p+1} \dd z_i \rangle$. Here ${\cal C}^{(\mathrm{N}+p,p)}$ is the configuration space of $p$ points on an $\mathrm{N}$-punctured sphere, as defined in \eqref{C-np}. 
We have inserted the factors of $|x_i {-} x_j|^{-4q_i q_j}$ on the left-hand side of (\ref{transl}) to compensate for the analogous terms with opposite exponents in the correlator of ${\cal O}_j$, and the
inverse factors of $\pi^p$ ensure that the right-hand side can be lined up with the sphere integrals (\ref{sph.5}).
Moreover, one needs to compensate with the correct power of $|x_\mathrm{N}|^2$ before fixing the last operator to $x_{\mathrm{N}} \to \infty$. Notice that correlation functions of different operators, even in distinct $(\pp,\pp')$ models, might have the same functional form once written in terms of formal variables $\tilde{q}_{i}$, as will be illustrated below.

\subsection{Minimal bases for minimal models}

Since correlation functions in CFTs have to be single-valued in the positions of operators $x_i$, we can always analytically continue them to the configurations in which all operators are aligned along a circle in $\CP^1$, i.e., $x_i \in \mathbb{R}$ in our chart (with $x_{\mathrm{N}}=\infty$). From now on we consider only such configurations. This restriction is consequential only in the intermediate steps of the computation, but of course does not affect the correlation functions, which can be freely continued away from such configurations once computed.

Following section~\ref{sec:6.1} we can decompose any $\mathrm{N}$-point correlator as a quadratic sum over $(\mathrm{N}{+}p{-}1)!/(\mathrm{N}{-}1)!$ contour integrals:
\begin{align}
\langle {\cal O}_{q_1}(x_1) {\cal O}_{q_2}(x_2) \cdots {\cal O}_{q_\mathrm{N}}(x_{\mathrm{N}})  \rangle_{\pp,\pp'} = \medmath{\left(\frac{i}{2}\right)^{\!p}} \ 
\sum_{a,b=1}^{\frac{(\mathrm{N}+p-1)!}{(\mathrm{N}-1)!}} e^{i \pi \phi_{ab}}\, &\left(\int_{\gamma_{a}^{(\mathrm{N}+p,p)}} \prod_{i=2}^{p+1} \dd z_i\, e^{W} \right) \nn\\ \times& \left(\int_{\gamma_{b}^{(\mathrm{N}+p,p)}} \prod_{i=2}^{p+1} \dd \overline{z}_i\, e^{\overline{W}}\right),\label{overcomplete-KLT-MM}
\end{align}
where the sums runs over all ways of distributing screening charges among the vertex operators ${\cal O}_{q_i}$. It splits into holomorphic and anti-holomorphic integrals, which we will loosely call \emph{conformal blocks} (in particular, they are not Virasoro conformal blocks). They are multi-valued in $x_i$'s. The phase of the potentials $W$ is defined such that each block is real and the overall phase is stripped away as $e^{i\pi \phi_{cd}}$. It is defined in \eqref{phi-cd}, which is essentially a product of factors $e^{2\pi i \tilde{q}_i \tilde{q}_j}$ for every time the charge $\tilde{q}_i$ crosses $\tilde{q}_j$ when transforming the $c$-th ordering to the $d$-th one, as in figure~\ref{fig:phase}.

As emphasized in the previous section, individual conformal blocks are redundant and one can reduce them to a minimal basis. The size of the basis is a topological invariant of the configuration space ${\cal C}^{(\mathrm{N}+p,p)}$ (the absolute value of its Euler characteristic) and equal to
\be\label{chi-MM}
|\chi({\cal C}^{(\mathrm{N}+p,p)})| = \frac{(\mathrm{N}{+}p{-}3)!}{(\mathrm{N}{-}3)!} = \frac{(\tfrac{1}{2}\sum_{i=1}^{\mathrm{N}}(r_i {+} s_i)-1)!}{(\mathrm{N}{-}3)!} \ ,
\ee
where in the second equality we used \eqref{p-plus-minus}, which is valid for generic $(\pp,\pp')$. Here we also assumed that one of the operators is fixed at infinity.
The physical interpretation is that the stronger the background charge $Q_{\pp,\pp'}$, the more screening charges are necessary to neutralize it, which leads to more ways of sprinkling them among the operators. It might sometimes occur that the combinations $s_{ij} = 2\tilde{q}_i \tilde{q}_j$ or their sums $s_{i_1 i_2 \ldots i_m}$ are integers. In those situations we say that the singularities of ${\cal C}^{(\mathrm{N}+p,p)}$ are no longer ramified (the corresponding hyperplane arrangement is resonant), and the size of the basis might drop.

In this paper we introduced natural bases for both integration cycles and differential forms. The cycles are given by disk integration domains
\be
\gamma_{\vec{A},\vec{i}}^{(\mathrm{N}+p,p)} \quad \text{introduced in }\;\eqref{gen.1} \, ,
\ee
as well as their duals
\be
\beta_{\vec{A},\vec{i}}^{(\mathrm{N}+p,p)} \quad \text{introduced in }\;\eqref{allklt.3} \, .
\ee
For the cocycles we have the Parke--Taylor-like basis
\be
\nu_{\vec{A},\vec{i}}^{(\mathrm{N}+p,p)} \quad \text{introduced in }\;\eqref{nu-forms} \, ,
\ee
and their duals
\be
\omega_{\vec{A},\vec{i}}^{(\mathrm{N}+p,p)} \quad \text{introduced in }\;\eqref{gen.2} \, .
\ee
For example, in the bases of cycles, the KLT-like formula (\ref{overcomplete-KLT-MM}) simplifies to
\be\label{correlator-KLT}
\langle {\cal O}_{q_1}(x_1) {\cal O}_{q_2}(x_2) \cdots {\cal O}_{q_\mathrm{N}}(x_{\mathrm{N}})  \rangle_{\pp,\pp'} = \! \!
\sum_{a=1}^{\frac{(\mathrm{N}+p-3)!}{(\mathrm{N}-3)!}} \! \! \left(\int_{\gamma_{a}^{(\mathrm{N}+p,p)}} \prod_{i=2}^{p+1} \dd z_i\, e^{W}\right) \left(\int_{\beta_{a}^{(\mathrm{N}+p,p)}} \prod_{i=2}^{p+1} \dd \overline{z}_i\, e^{\overline{W}}\right) \, ,
\ee
which can be written in terms of the KLT matrix $S_{\alpha'}(\rho_a|\rho_b)$ given in section~\ref{sec:6.5} or in terms of intersection numbers of cycles according to the prescription in section~\ref{sec:6.2}.

In order to be able to refer to \eqref{overcomplete-KLT-MM} and \eqref{correlator-KLT} as true ``double-copy'' formulae, one would like to give a physical interpretation to individual conformal blocks, so that one set of observables is double-copied to another. One possible interpretation could be as correlations functions of a boundary CFT in the Coulomb gas formalism, perhaps along the lines of \cite{Schulze:1996qm,Kawai:2002vd,Kawai:2002pz}. We leave this idea for future explorations.

The advantage of using these bases is that they produce uniformly transcendental functions. The original correlation functions can be always transformed into the minimal bases via integration-by-parts identities or with intersection numbers of twisted cocycles:
\begin{align}
\langle {\cal O}_{q_1}(x_1) {\cal O}_{q_2}(x_2) \cdots {\cal O}_{q_\mathrm{N}}(x_{\mathrm{N}})  \rangle_{\pp,\pp'} = \sum_{a,b=1}^{\frac{(\mathrm{N}+p-3)!}{(\mathrm{N}-3)!}} &\langle \nu_{a}^{(\mathrm{N}+p,p)} | {\textstyle\prod_{i=2}^{p+1}} \dd z_i  \rangle\, \overline{\langle \omega_{b}^{(\mathrm{N}+p,p)} | {\textstyle\prod_{i=2}^{p+1}} \dd z_i  \rangle} \nn\\
& \! \! \! \! \! \! \! \! \! \! \! \!\times\int_{{\cal C}^{(\mathrm{N}+p,p)}} \prod_{i=2}^{p+1} \dd^2 z_i\, e^{W+\overline{W}}\, \hat{\omega}_a^{(\mathrm{N}+p,p)}\, \overline{\hat{\nu}_b^{(\mathrm{N}+p,p)}} \,  . \label{MM-basis-projection}
\end{align}
However, note that one cannot use the formula \eqref{inter-delta-omega} since not all forms involved are logarithmic. Still, one can use recursion relations for intersection numbers as defined in \cite[Section 3]{Mizera:2019gea}.\footnote{The intersection numbers accessible from 
the infinite families of integration-by-parts identities in \cite{Schlotterer:2016cxa, He:2018pol, He:2019drm} only apply to the $\mathrm{N}=3$ instances of (\ref{MM-basis-projection}). Still, the combinatorial techniques of these references should have an echo at $\mathrm{N}\geq 4$.} 
In this way correlation functions can be expressed in terms of $\mathrm{sv} \, F_{ab}^{(\mathrm{N}+p,p)}$ up to proportionality constants given in \eqref{transl}. Applying this reduction together with the KLT formula \eqref{correlator-KLT} expresses individual conformal blocks in terms of the contour integrals $F_{cb}^{(\mathrm{N}+p,p)}$ and $I_{ca}^{(\mathrm{N}+p,p)}$, see (\ref{sph.6}) and (\ref{J-to-I}) for the latter.

\subsection{Transcendentality properties and the $\pp\to\infty$ limit}

The bases of sphere and disk integrals which can be used
to express correlation functions as discussed above have particularly simple transcendentality properties in their $\alpha'$-expansion. However, under the identifications (\ref{identsq}) of Mandelstam variables and charges, these transcendentality properties only apply to a formal low-charge expansion around $\tilde{q}_i= 0$.
We stress that in the applications to minimal models, the $\tilde{q}_i$ are always fixed real numbers, as given in \eqref{q-r-s}, and hence the expansion in $\tilde{q}_i$ generically can be understood only in a formal sense.

However, there are situations where one might assign a physical meaning to the low-charge limit. Let us consider the $(\pp,2)$ minimal models (with $\pp$ odd). Vertex operators are labeled by $(1,s_i)$ with $1\leq s_i \leq \pp{-}1$ and hence their allowed charges are
\be
q_{(1,s_i)} = \frac{s_i-1}{\sqrt{2\pp}} \, .
\ee
In this situation the background and screening charges are given by
\be
Q_{\pp,2} = \frac{\pp-2}{\sqrt{2\pp}} \, , \qquad q_+ = \sqrt{\pp/2} \, , \qquad q_- = - \sqrt{2/\pp} \, .
\ee
Hence, if we can avoid using the screening charge $q_+$, all the charge pairings $\tilde{q}_i \tilde{q}_j$ would scale as $1/\pp$, and the correlation function in the limit $\pp \to \infty$ would be on the same footing as string-theory amplitudes in the low-energy approximation, $\alpha' \to 0$. This can certainly be done. Let us consider an $\mathrm{N}$-pt function of operators ${\cal O}_{q_{(1,s_i)}}$ and represent the $\mathrm{N}$-th one via its dual ${\cal O}_{q_{(1,\pp-s_i)}}$, i.e.,
\begin{align}
&\langle {\cal O}_{q_{(1,s_1)}}(x_1)\, {\cal O}_{q_{(1,s_2)}}(x_2) \cdots {\cal O}_{q_{(1,s_{\mathrm{N}-1})}}(x_{\mathrm{N}-1})\, {\cal O}_{q_{(1,s_{\mathrm{N}})}}(x_{\mathrm{N}}) \rangle_{\pp,2} \label{p-2-correlator}
\\
&= \langle {\cal O}_{q_{(1,s_1)}}(x_1)\, {\cal O}_{q_{(1,s_2)}}(x_2) \cdots {\cal O}_{q_{(1,s_{\mathrm{N}-1})}}(x_{\mathrm{N}-1})\, {\cal O}_{q_{(1,\pp-s_{\mathrm{N}})}}(x_{\mathrm{N}}) \rangle_{\pp,2} \, .
\nn
\end{align}
We use the second representation in the Coulomb gas formalism. Here the neutrality condition is satisfied if
\be
p_+=0, \qquad p_- = \frac{1}{2}\left( \sum_{i=1}^{\mathrm{N}-1} s_i - s_{\mathrm{N}} - \mathrm{N} + 2 \right) \, ,
\ee
and if $p_-$ is an integer. This leads to a potential $W$ proportional to $1/\pp$: 
\begin{align}
W = \frac{1}{\pp} \bigg(&\sum_{1 \leq i< j}^{\mathrm{N}-1} (s_i{-}1) (s_j{-}1) \log (x_i {-} x_j)  \\
&-4\sum_{i=1}^{\mathrm{N}-1} \sum_{j=2}^{p_- +1} (s_i{-}1) \log (x_i {-} z_j) + 16\!\!\sum_{2 \leq i<j}^{p_- +1} \log (z_i {-} z_j) \bigg) \, , \nn
\end{align}
provided that we fix $x_{\mathrm{N}} = \infty$.
The large-$\pp$ limit of $(\pp,2)$ models coupled to Liouville theory has been recently conjectured to be describing Jackiw--Teitelboim gravity \cite{Saad:2019lba}, which adds further physical motivation for studying such correlation functions.

The individual conformal blocks, once expressed in terms of $F_{cb}^{(\mathrm{N}+p,p)}$, satisfy all the monodromy properties described in previous section as well as the coaction formula from \eqref{conv7}. In addition, once expressed in this basis, the correlation function can be expressed as a single-valued map of a single conformal block:
\begin{align}
&\lim_{x_{\mathrm{N}}\to\infty} \frac{|x_{\mathrm{N}}|^{4q_{\mathrm{N}}(q_{\mathrm{N}} - Q_{\pp,\pp'})}}{\pi^p} \prod_{1 \leq i<j}^{\mathrm{N}-1} |x_i {-} x_j|^{-4q_i q_j} \int_{{\cal C}^{(\mathrm{N}+p,p)}} \prod_{i=2}^{p+1} \dd^2 z_i\, e^{W+\overline{W}}\, \hat{\omega}_a^{(\mathrm{N}+p,p)}\, \overline{\hat{\nu}_b^{(\mathrm{N}+p,p)}} \nn\\
&= \mathrm{sv} \left( \lim_{x_{\mathrm{N}}\to\infty} |x_{\mathrm{N}}|^{2q_{\mathrm{N}}(q_{\mathrm{N}} - Q_{\pp,\pp'})} \prod_{1 \leq i<j}^{\mathrm{N}-1} |x_i {-} x_j|^{-2q_i q_j} \int_{\gamma_b^{(\mathrm{N}+p,p)}}
\prod_{i=2}^{p+1} \dd z_i\, e^{W}\, \hat \omega_a^{(\mathrm{N}+p,p)} \right) \, .
\end{align}
We will return to this relationship in an example computation for $(\pp,2)$ minimal models in the $\pp \to \infty$ limit below.

\subsection{\label{sec:7.5}Example four-point correlators}

In order to illustrate the above formulae on concrete examples we will consider the four-point functions
\be\label{G-x-xbar}
{\cal G}(x,\overline{x}) = \lim_{x_{4}\to\infty} \frac{|x_4|^{4q_4(q_4-Q_{\pp,\pp'})}}{\pi^p} |x|^{-4 q_1 q_2} |1{-}x|^{-4 q_2 q_3} \langle {\cal O}_{q_1}(0)\, {\cal O}_{q_2}(x)\, {\cal O}_{q_3}(1)\, {\cal O}_{q_4}(x_4) \rangle_{\pp,\pp'} \, ,
\ee
where to avoid clutter we expressed it in terms of the cross-ratio $x$. In the intermediate steps we will restrict to $x \in (0,1)$. Let us consider two cases encountered before for four-point functions of $(2,1)$ and $(1,2)$ operators, which both involve a single screening charge, $p=1$. In the case \eqref{N4-example}, we have a single screening charge $q_+$ and
\be\label{case1}
({\mathrm I}):\quad  q_1 = q_2 = q_3 = q_{(2,1)}= -\frac{\sqrt{\pp / \pp'}}{2}, \quad q_4 = Q_{\pp,\pp'} {-} q_{(2,1)} = \frac{3\pp {-} 2\pp'}{2 \sqrt{\pp \pp'}}, \quad q_+ = \sqrt{ \frac{\pp }{ \pp'}} \ .
\ee
On the other hand, we can consider a special case of \eqref{p-2-correlator} with $\pp'=2$ and
\be\label{case2}
(\mathrm{II}):\quad q_1 = q_2 = q_3 = q_{(1,2)}= \frac{\sqrt{\pp'/\pp}}{2}, \quad q_4 = Q_{\pp,\pp'} {-} q_{(1,2)} = -\frac{3\pp' {-} 2\pp}{2\sqrt{\pp \pp'}}, \quad q_- = -\sqrt{ \frac{\pp'}{\pp}} \, ,
\ee
as well as a single screening charge $q_-$. This example of course can be considered also for $\pp' \neq 2$. We can compute these different correlation functions using the same formulae provided that we treat $q_1, q_2, q_3$, and $q_\pm$ as abstract variables and plug in their values only at the end.

Explicitly, ${\cal G}(x,\overline{x})$ is given by the integral
\be
{\cal G}(x,\overline{x}) = \frac{1}{\pi} \int_{\mathbb{C}\setminus\{0,x,1\}} \!\!\! \dd^2z\, |z|^{4q_1 q_\pm} |z-x|^{4q_2 q_\pm} |z-1|^{4q_3 q_\pm} \, ,
\ee
which we can easily express in terms of the following contour integrals obtained by placing the screening charge between the external operators in all possible ways:
\be
\medmath{\begin{pmatrix}
{\cal F}_1(x) \\
{\cal F}_2(x) \\
{\cal F}_3(x) \\
{\cal F}_4(x) \\
\end{pmatrix} =
\begin{pmatrix}
\displaystyle\int_{-\infty}^{0} \!\!\!\dd z\, (-z)^{2q_1 q_\pm} (x-z)^{2q_2 q_\pm} (1-z)^{2q_3 q_\pm} \\[1em]
\displaystyle\int_{0}^{x} \!\dd z\, z^{2q_1 q_\pm} (x-z)^{2q_2 q_\pm} (1-z)^{2q_3 q_\pm} \\[1em]
\displaystyle\int_{x}^{1} \!\dd z\, z^{2q_1 q_\pm} (z-x)^{2q_2 q_\pm} (1-z)^{2q_3 q_\pm} \\[1em]
\displaystyle\int_{1}^{\infty} \!\!\!\dd z\, z^{2q_1 q_\pm} (z-x)^{2q_2 q_\pm} (z-1)^{2q_3 q_\pm}
\end{pmatrix} \, ,}
\ee
where the phase of $W$ is chosen such that  each ${\cal F}_a$ is real and agrees with the convention with absolute values for the Koba--Nielsen factor in \eqref{conv2}.
The overcomplete KLT relation \eqref{overcomplete-KLT-MM} then reads:
\be
{\cal G}(x,\overline{x}) = \medmath{\frac{i}{2}\begin{pmatrix}
{\cal F}_1(x) \\
{\cal F}_2(x) \\
{\cal F}_3(x) \\
{\cal F}_4(x) \\
\end{pmatrix}^{\!\!t}
\ccccb
1 & e^{2\pi i q_1 q_\pm} & e^{2\pi i (q_1+q_2) q_\pm}  & e^{2\pi i (q_1+q_2+q_3) q_\pm}\\
e^{2\pi i q_1 q_\pm} & 1 & e^{2\pi i q_2 q_\pm} & e^{2\pi i (q_2+q_3)q_\pm} \\
e^{2\pi i (q_1+q_2) q_\pm} & e^{2\pi i q_2 q_\pm} & 1 & e^{2\pi i q_3 q_\pm} \\
e^{2\pi i (q_1+q_2+q_3)q_\pm} & e^{2\pi i (q_2+q_3)q_\pm} & e^{2\pi i q_3 q_\pm} & 1
\cccce \overline{\begin{pmatrix}
{\cal F}_1(x) \\
{\cal F}_2(x) \\
{\cal F}_3(x) \\
{\cal F}_4(x) \\
\end{pmatrix}} \, .}
\ee
It is however beneficial to express it in terms of a minimal basis, which according to \eqref{chi-MM} is $|\chi({\cal C}^{(5,1)})|=2$.\footnote{The case \eqref{case1} with $\pp'=3$ is special because it leads to integrands which are not branched at infinity, given that $e^{W|_{\eqref{case1}},\pp'=3} = [z(z-x)(z-1)]^{-\pp/3} \to z^{-\pp}$ as $z \to \infty$. Because of this fact the size of the basis decreases to $1$. It is the same problem as sitting on a factorization channel $s_{25}=0$ in string theory amplitudes, or considering Feynman integrals in integer dimensions, see \cite[Section~4]{Mastrolia:2018uzb}. In those situations one needs to correct KLT relations using the framework of relative twisted cohomologies \cite{matsumoto2019relative}. While in this subsection we ignore this problem to retain generality, we will return to it in sections \ref{sec:7.5.1} and \ref{sec:7.5.2}.} To minimize the number of computations let us pick ${\cal F}_2$ and ${\cal F}_4$ for both holomorphic and antiholomorphic blocks. It leads to a simplification because the two contours do not intersect and hence the intersection matrix is diagonal (another natural choice would be ${\cal F}_1$ and ${\cal F}_3$), cf. \eqref{n-1-int}. We therefore immediately get
\be\label{G-ans}
{\cal G}(x,\overline{x}) = \medmath{\frac{\sin(2\pi q_1 q_\pm) \sin(2\pi q_2 q_\pm)}{\sin(2\pi (q_1{+}q_2)q_\pm)}} |{\cal F}_2(x)|^2 + \medmath{\frac{\sin(2\pi q_3 q_\pm)\sin(2\pi (q_1{+}q_2{+}q_3)q_\pm)}{\sin(2\pi (q_1{+}q_2)q_\pm)}} |{\cal F}_4(x)|^2 .
\ee
Computation of the relevant conformal blocks explicitly gives
\begin{align}
{\cal F}_2(x) &= x^{1+2 \left(q_1+q_2\right) q_\pm}\, \mathrm{B}(1{+}2q_1 q_\pm, 1{+}2q_2q_\pm)\, {}_2F_1 (-2 q_3 q_\pm, 1{+}2 q_1
   q_\pm; 2{+}2(q_1{+}q_2) q_\pm; x ) \, ,  \nn\\
{\cal F}_4(x) &= -\frac{2 q_3 q_\pm\, \mathrm{B} (-2 (q_1{+}q_2{+}q_3) q_\pm, 2 q_3 q_\pm )}{1+2(q_1{+}q_2{+}q_3) q_\pm} \\
&\qquad\qquad\qquad\qquad\qquad\times{}_2F_1(-2 q_2 q_\pm, -1 {-}2 (q_1{+}q_2{+}q_3) q_\pm;-2 (q_1{+}q_2) q_\pm;x ) \, ,\nn
\end{align}
where $\mathrm{B}(a,b)=\frac{\Gamma(a)\Gamma(b)}{\Gamma(a{+}b)}$ is the Euler beta function. This result is in agreement with \cite{Dotsenko:1984ad,Dotsenko:1984nm,Mimachi2003,Vanhove:2018elu}.

Next we analyze the $\pp \to \infty$ behavior of these correlation functions for $\pp'$ fixed and finite. This limit can be qualitatively different, depending on whether charges become small, such as in the case \eqref{case2} where $q_i q_- \to 0$, or large, as is the case in the example \eqref{case1} where $q_i q_+ \to \infty$ (recall that $q_4$ does not enter the expressions directly). While the first case is fairly easy to analyze and leads to interesting connections with transcendentality, the second is more subtle due to the presence of Stokes phenomena similar to those appearing in the $\alpha'\to\infty$ limit of the Veneziano amplitude \cite{Mizera:2019vvs}. We consider examples of these limits below. Before doing so, we give an explicit example where correlation functions can be expressed in terms of elementary functions. In order not to confuse the two cases \eqref{case1} and \eqref{case2}, we will label the correlation function and conformal blocks evaluated on the two sets of charges with superscripts $\mathrm{I}$ and $\mathrm{II}$, respectively.

\subsubsection{\label{sec:7.5.1}Critical Ising model}

Let us consider the example of the critical Ising model with $(\pp,\pp') = (4,3)$. Of course, the Coulomb gas formalism is a hugely wasteful way of computing correlators in this case, since they can be obtained straightforwardly in the free-fermion formulation in a closed form, see for instance \cite[Chapter~12]{DiFrancesco:1997nk}. Instead, we use it as a chance to briefly demonstrate how these simple answers arise from the KLT formula \eqref{G-ans}.

We start with the four-point function of energy operators with $(r,s)=(2,1)$. Using the values of charges from case $\mathrm{I}$ in \eqref{case1} one can see that ${\cal F}_4^{\mathrm{I}}$ does not contribute since its prefactor in \eqref{G-ans} is proportional to
\be
\sin(2\pi(q_1{+}q_2{+}q_3)q_\pm) \big|_{\eqref{case1}} = -\sin(3\pi \pp/\pp')\big|_{\pp'=3} = 0 \, .\label{pp3}
\ee
It is consistent with the size of the basis dropping to $1$ in the case \eqref{case1} with $\pp'=3$, although the fact that the limit $\pp'\to3$ of \eqref{G-ans} was smooth is an accident coming from our choice of bases. As a result we only need
\begin{align}
	{\cal F}_2^{\mathrm{I}}(x) &= x^{1-2\pp/\pp'}\, \mathrm{B}(1{-}\pp/\pp',\, 1{-}\pp/\pp') \,
	{}_2F_1 (1{-}\pp/\pp',\, \pp/\pp';\, 2{-}2 \pp/\pp';\, x ) \big|_{(\pp,\pp')=(4,3)} \nn\\
	&= \frac{\Gamma(-\frac{1}{3})^2}{\Gamma (-\frac{2}{3})} \frac{\left(x^2-x+1\right)}{(1-x)^{5/3} x^{5/3} } \, .
\end{align}
Plugging back into \eqref{G-x-xbar} we find
\be
\lim_{x_4 \to \infty} \langle \varepsilon(0) \varepsilon(x) \varepsilon(1) \varepsilon(x_4) \rangle_{\mathrm{Ising}} = \frac{c_1}{|x_4|^2} \left| \frac{1}{x} - \frac{1}{x{-}1} -1 \right|^2 \, ,
\ee
where $c_1 = -\tfrac{\sqrt{3} \pi}{2}  \Gamma (-\frac{1}{3})^4 / \Gamma (-\frac{2}{3})^2$. Restoring the original coordinates one finds
\be
\langle \varepsilon(x_1) \varepsilon(x_2) \varepsilon(x_3) \varepsilon(x_4) \rangle_{\mathrm{Ising}} = c_1 \left| \mathrm{Pf} \left( \frac{1}{x_i - x_j} \right) \right|^2 \, ,
\label{encorrel}
\ee
where $\mathrm{Pf}(\cdots)$ denotes the Pfaffian of the antisymmetric matrix with entries labelled by $i,j=1,2,3,4$. This is the correct result.

Let us move on to the four-point function of spin operators with $(r,s)=(1,2)$. Using the values of charges from case $\mathrm{II}$ in \eqref{case2} we obtain
\begin{align}
{\cal F}_2^{\mathrm{II}}(x) &= x^{1-2 \pp'/\pp}\, \mathrm{B}(1{-}\pp'/\pp,\, 1{-}\pp'/\pp) \,
{}_2F_1(\pp'/\pp,\, 1{-}\pp'/\pp;\, 2{-}2 \pp'/\pp;\, x )\big|_{(\pp,\pp')=(4,3)} \nn \\
&= \frac{\Gamma(\frac{1}{4})^2}{\sqrt{2 \pi }} \frac{\sqrt{1+\sqrt{1-x}} }{ \sqrt{x(1-x)}} \, ,
\end{align}
as well as
\begin{align}
	{\cal F}_4^{\mathrm{II}}(x) &= \tfrac{\pp'}{\pp-3 \pp'}\, \mathrm{B}(3 \pp'/\pp,\, -\pp'/\pp) \, {}_2F_1(\pp'/\pp,\, 3\pp'/\pp{-}1;\, 2\pp'/\pp;\, x ) \big|_{(\pp,\pp')=(4,3)} \nn \\
	&= \frac{\Gamma(\frac{1}{4})^2}{\sqrt{2\pi }} \frac{\sqrt{1+\sqrt{x}}-\sqrt{1-\sqrt{x}} }{ \sqrt{2x(1-x)}} \, .
\end{align}
Putting everything together according to \eqref{G-ans} (with coefficients $\tfrac{1}{2}$ in front of the two factors) and restoring all the coordinates $x_i$, one finds agreement with the free-fermion computation
\be
\big(\langle \sigma(x_1) \sigma(x_2) \sigma(x_3) \sigma(x_4) \rangle_{\mathrm{Ising}} \big)^2 \;=\; c_2 \!\! \sum_{\substack{e_i = \pm 1\\ \sum_i e_i =0}} \prod_{1\leq i<j\leq 4} |x_i - x_j|^{e_i e_j/2}
\ee
with $c_2 = - \Gamma(\tfrac{1}{4})^8/16$ in the domain $x \in (0,1)$.

\subsubsection{\label{sec:7.5.2}Large-$\pp$ limit for $(2,1)$ four-point correlators}

We now consider the $\pp \to \infty$ limit of the four-point functions of $(2,1)$ operators with charges given in case $\mathrm{I}$ in \eqref{case1} and $\pp' \geq 3$ finite and fixed. Here the situation is qualitatively different to that from the previous subsection because charges blow up. As a consequence, conformal blocks localize on the critical points of the potential $W$. (One cannot easily apply saddle-point analysis directly to the correlator because it is not written in terms of a holomorphic integrand.) There is a large number of critical points located on different sheets of the Riemann surface of $z$.\footnote{Since in this case we have
\be
e^{W|_{\eqref{case1}}} = [z(z-x)(z-1)]^{-\pp/\pp'}
\ee
with a finite $\pp'$ co-prime to $\pp$, the number of sheets is $\pp'^3$. This is because the corresponding Riemann surface of $z$ is $\pp'$-branched around the three points $0$, $x$, $1$ (monodromies around infinity are not independent). Each sheet can be labelled by a point in a $\mathbb{Z}_{\pp'}^3$ lattice counting how many times $z$ winded around each of the branch points. This situation is different to string theory, where $s_{ij}$ are generic non-rational variables and hence the numbers of sheets and critical points are infinite.
} On the first sheet we have
\be
\partial_{z} W(z_\ast) = 2q_+ \left( \frac{q_1}{z_\ast} + \frac{q_2}{z_\ast -x} + \frac{q_3}{z_\ast -1} \right) \bigg|_{\eqref{case1}} = - \frac{\pp}{\pp'} \frac{3 z_\ast^2 -2 (1{+}x) z_\ast + x}{z_\ast (z_\ast-x)(z_\ast-1)} = 0 \, .
\ee
Explicitly, it gives two solutions which we denote by $z_\ast^\pm$,
\be
z_\ast^\pm = \frac{1}{3} \left( 1 + x \pm \sqrt{x^2 - x + 1} \right) \, .
\ee
It is clear that the positions of these critical points depend on the cross-ratio $x$. This is the source of the Stokes phenomenon: the large-$\pp$ asymptotics depends on the value of $x$.

Here we focus on the case $x \in (0,1)$, for which there is exactly one critical point $z_\ast^- \in (0,x)$ and exactly one $z_\ast^+ \in (x,1)$. This is not an accident. By the arguments of \cite{Cachazo:2016ror}, as long as all $\tilde{q}_i \tilde{q}_j$ have the same sign and $x_i$'s are ordered, the problem of computing critical points is equivalent to that of finding stable configurations of mutually-repelling charges on a line. There are exactly $(\mathrm{N}{+}p{-}3)!/(\mathrm{N}{-}3)!$ such configurations corresponding to a single critical point in each bounded chamber of the configuration space ${\cal C}^{(\mathrm{N}+p,p)}$.

While there might be a large number of critical points (with two per sheet), they all give same-magnitude contributions to the large-$\pp$ asymptotics and only differ in the complex phase. These phases typically resum to trigonometric functions. We can exploit the KLT formula with a judicious choice of bases to drastically simplify the computation. In the example at hand, the contours $(0,x)$ and $(x,1)$ are already paths of steepest descent (also known as Lefschetz thimbles) for the potential $W$ at $x \in (0,1)$ and critical points $z_\ast^{-}$ and $z_\ast^+$, respectively. Hence, using ${\cal F}_2^{\mathrm{I}}$ and ${\cal F}_3^{\mathrm{I}}$ as bases, we can compute the asymptotic behavior with only two saddle points from the first sheet, one for each conformal block. For $\pp'>3$, using the intersection numbers computed from \eqref{H-inv-ab} and plugging in \eqref{case1}, we have
\begin{align}
{\cal G}^{\mathrm{I}}(x, \overline{x}) \big|_{\pp'>3} = - \frac{\sin (\pi \pp/\pp')^2}{\sin(3\pi \pp/\pp')} \bigg( & {\cal F}_2^{\mathrm{I}}(x) \overline{{\cal F}_3^{\mathrm{I}}(x)} + {\cal F}_3^{\mathrm{I}}(x) \overline{{\cal F}_2^{\mathrm{I}}(x)}  \label{G-asymp}\\
&+ 2 \cos(\pi \pp/\pp') \left(|{\cal F}_2^{\mathrm{I}}(x)|^2 + |{\cal F}_3^{\mathrm{I}}(x)|^2 \right) \bigg)\, .
\nn
\end{align}
Note that there are no poles or zeros due to the sine factors because $\pp$ and $\pp'$ are co-prime and $\pp'>3$. The case $\pp'=3$ is simpler for the same reason as in the case of the energy correlator (\ref{encorrel}) in the critical Ising model. Namely, even in the basis ${\cal F}_2^{\mathrm{I}}$, ${\cal F}_3^{\mathrm{I}}$ the coefficient of ${\cal F}_3^{\mathrm{I}}$ is zero, as in \eqref{pp3}, 
and hence we have a simplified result,
\be
{\cal G}^{\mathrm{I}}(x, \overline{x}) \big|_{\pp'=3} = - \frac{\sin (\pi \pp/3)^2}{\sin(2\pi \pp/3)}\, |{\cal F}_2^{\mathrm{I}}(x)|^2,\label{G-asymp-pp3}
\ee
which means it only receives contributions from the single critical point $z_\ast^-$. The physical reason for this simplification is that four $(2,1)$ operators can only exchange an identity operator when $\pp'=3$.

At any rate, the asymptotics of the blocks ${\cal F}_2^{\mathrm{I}}$ and ${\cal F}_3^{\mathrm{I}}$ can now be easily computed. The Hessian evaluated at the two critical points is
\be
J_\pm = \partial_z^2 W(z_\ast^\pm) = \frac{\pp}{\pp'} \left(\frac{1}{(z_\ast^\pm)^2}+\frac{1}{(z_\ast^\pm-x)^2}+\frac{1}{(z_\ast^\pm-1)^2}\right)
\ee
and is positive for $x \in (0,1)$. Therefore
\begin{align}
\lim_{\pp \to \infty} {\cal F}_2^{\mathrm{I}}(x) = \frac{1}{\sqrt{2\pi J_-}} [ z_\ast^- (x-z_\ast^-) (1-z_\ast^-) ]^{-\pp/\pp'} \, ,\\
\lim_{\pp \to \infty} {\cal F}_3^{\mathrm{I}}(x) = \frac{1}{\sqrt{2\pi J_+}} [ z_\ast^+ (z_\ast^+ - x) (1-z_\ast^+) ]^{-\pp/\pp'} \, .
\end{align}
They together give the asymptotics of \eqref{G-asymp} and \eqref{G-asymp-pp3} in the case $x \in (0,1)$. The correlator is exponentially suppressed as $\pp \rightarrow \infty$.

\subsubsection{Large-$\pp$ limit for $(1,2)$ four-point correlators}

Let us consider the $\pp \to \infty$ limit of the four-point functions of $(1,2)$ operators with charges given in case $\mathrm{II}$ in \eqref{case2}. While the case $\pp'=2$ is of most interest, we can study arbitrary fixed $\pp' \geq 2$ as long as it remains finite (it is understood that $\pp$ is always co-prime with $\pp'$). Direct expansion of the result in \eqref{G-ans}
gives
\begin{align}
{\cal G}^{\mathrm{II}}(x,\overline{x}) = &-\frac{\pi \pp'}{3\pp}  \left( 1 + |x|^2 + |1{-}x|^2\right) \\
&+ \frac{\pi \pp'^2}{\pp^2} \Big(|x|^2 \log|x|^2 + |1{-}x|^2 \log |1{-}x|^2 -2(1+|x|^2+|1{-}x|^2) \Big) + {\cal O}(1/\pp^{3}) \, . \notag
\label{naiveexp}
\end{align}
One can immediately see that assigning transcendentality weights ${\cal T}(\pp) = 1$ and ${\cal T}(\pp') = {\cal T}(x) = 0$, the result is not uniformly transcendental. This fact can be fixed with a corrected basis of conformal blocks.

To this end, we first recall the differential forms from section~\ref{sec:2}, which serve as building blocks  for the minimal basis. Specializing the Mandelstam variables $s_{12},s_{23},s_{24}$ at $(n,p)=(5,1)$ according to \eqref{case2} we have:
\begin{align}
\omega_1^{(5,1)} = -\frac{\pp'}{\pp} \dd \log z, \qquad &\omega_2^{(5,1)} = \frac{\pp'}{\pp} \dd \log (1{-}z) \, , \\
\nu_1^{(5,1)} = \dd \log \frac{z}{z-x}, \qquad &\nu_2^{(5,1)} = \dd \log \frac{z-x}{z-1} \, .
\end{align}
In order to project ${\cal G}^{\mathrm{II}}(x,\overline{x})$ onto the basis of $\mathrm{sv} F_{ab}^{(5,1)}$ we only need to compute four intersection numbers of the above forms with $\dd z$, giving
\begin{align}
\langle \nu_1^{(5,1)} | \dd z \rangle &= \frac{\pp x}{\pp - 3\pp'} \, , \qquad &\langle \nu_2^{(5,1)} | \dd z \rangle &= \frac{\pp (1-x)}{\pp - 3\pp'} \, ,\\
\overline{\langle \omega_1^{(5,1)} | \dd z \rangle} &= -\frac{\pp'(1+\overline{x})}{3(\pp - 3\pp')} \, , \qquad &\overline{\langle \omega_2^{(5,1)} | \dd z \rangle} &= \frac{\pp'(\overline{x}-2)}{3(\pp - 3\pp')} \, .
\end{align}
Steps needed to reproduce these results were spelled out in \cite[Section~4B]{Mastrolia:2018uzb} in a very similar case. Using the basis expansion formula \eqref{MM-basis-projection} we therefore find
\begin{align}
{\cal G}^{\mathrm{II}}(x,\overline{x}) = \frac{\pi \pp \pp'}{3(\pp - 3\pp')^2}\Big( -&(1{+}\overline{x}) \big(x\, \mathrm{sv} F^{(5,1)}_{11} + (1{-}x)\, \mathrm{sv} F^{(5,1)}_{12} \big) \nn\\
+& (\overline{x}{-}2)\big(x\, \mathrm{sv} F^{(5,1)}_{21} + (1{-}x)\, \mathrm{sv} F^{(5,1)}_{22} \big)
\Big) \, .\label{G-normal}
\end{align}
For completeness let us also give an expression for the basis of conformal blocks in terms of $F_{ab}^{(5,1)}$:
\be
{\cal F}_{a+1}^{\mathrm{II}}(x) = \sum_{b=1}^{2} \langle \nu_b^{(5,1)} | \dd z \rangle\, F_{ab}^{(5,1)} = \frac{\pp}{\pp-3\pp'} \left( x\, F_{a1}^{(5,1)} + (1{-}x)\, F_{a2}^{(5,1)} \right)
\ee
for $a=1,2$. (${\cal G}^{\mathrm{II}}$ can be expressed in terms of ${\cal F}_{2}^{\mathrm{II}}$ and ${\cal F}_{3}^{\mathrm{II}}$ using the same formula as in \eqref{G-asymp} with $\pp \leftrightarrow \pp'$.) One can compute their $1/\pp$-expansion using the formulae explained in section~\ref{sec:3.2} with
\be
s = s_{12} = s_{23} = s_{24} = - \pp'/\pp \, , \qquad z_3 = x \, .
\label{thekinlim}
\ee
More precisely, with the $\ap$-expansion of $F^{(5,1)}_{ab}$ in (\ref{expa.23}) (also see appendix \ref{app:A.2} for the orders of $\ap^{\leq 2}$), the kinematic point (\ref{thekinlim}) gives rise to the following leading orders of their single-valued images
\begin{align}
{\rm sv}\, F^{(5,1)}_{11} &=
1 + 2 s G^{\rm sv}(0;x) + 4 s^2 G^{\rm sv}(0, 0;x) + 
 s^2 G^{\rm sv}(0, 1;x) + 8 s^3 G^{\rm sv}(0, 0, 0;x) \notag \\
 & \ \ \ + 
 2 s^3 G^{\rm sv}(0, 0, 1;x) + 2 s^3 G^{\rm sv}(0, 1, 0;x) + 
 2 s^3 G^{\rm sv}(0, 1, 1;x) + 4 s^3 \zeta_3 + {\cal O}(s^4)\, , \notag \\
{\rm sv}\, F^{(5,1)}_{12} &=
-s G^{\rm sv}(1;x) - 2 s^2 G^{\rm sv}(1, 0;x) - 2 s^2 G^{\rm sv}(1, 1;x) - 
 4 s^3 G^{\rm sv}(1, 0, 0;x) \notag \\
 &\ \ \ - s^3 G^{\rm sv}(1, 0, 1;x) - 
 4 s^3 G^{\rm sv}(1, 1, 0;x) - 4 s^3 G^{\rm sv}(1, 1, 1;x) + {\cal O}(s^4)\, ,
 \\
{\rm sv}\, F^{(5,1)}_{21} &=
-s G^{\rm sv}(0;x) - 2 s^2 G^{\rm sv}(0, 0;x) - 2 s^2 G^{\rm sv}(0, 1;x) - 
 4 s^3 G^{\rm sv}(0, 0, 0;x) \notag \\
 &\ \ \  - 4 s^3 G^{\rm sv}(0, 0, 1;x) - 
 s^3 G^{\rm sv}(0, 1, 0;x) - 4 s^3 G^{\rm sv}(0, 1, 1;x) + 4 s^3 \zeta_3 + {\cal O}(s^4) \,,
\notag \\
{\rm sv}\, F^{(5,1)}_{22} &=
1 + 2 s G^{\rm sv}(1;x) + s^2 G^{\rm sv}(1, 0;x) + 
 4 s^2 G^{\rm sv}(1, 1;x) + 2 s^3 G^{\rm sv}(1, 0, 0;x)  \notag \\
 &\ \ \  + 
 2 s^3 G^{\rm sv}(1, 0, 1;x) + 2 s^3 G^{\rm sv}(1, 1, 0;x) + 
 8 s^3 G^{\rm sv}(1, 1, 1;x) + 12 s^3 \zeta_3 + {\cal O}(s^4)\, .
\notag 
\end{align}
To the weights shown, the single-valued polylogarithms $G^{\rm sv}(\vec{a};z)={\rm sv}\, G(\vec{a};z)$ from Brown's construction \cite{svpolylog} with $\vec{a} \in \{0,1\}^{\times}$ are given by
\begin{align}
G^{\rm sv}(a_1;x)&=G(a_1;x)+\overline{G(a_1;x)} \ , \notag \\
G^{\rm sv}(a_1,a_2;x)&=G(a_1,a_2;x) + G(a_1;x) \overline{G(a_2;x)} + \overline{G(a_2,a_1;x)} \ ,
 \notag \\
G^{\rm sv}(a_1,a_2,a_3;x)&= G(a_1,a_2,a_3;x)
+ G(a_1,a_2;x)\overline{G(a_3;x)} \\
&\ \ \ \
+G(a_1;x)\overline{G(a_3,a_2;x)}
+\overline{G(a_3,a_2,a_1;x)} \, , \notag 
\end{align}
where the explicit expressions for single-valued polylogarithms are given in equations \eqref{svG1} and \eqref{svG2}, also see (\ref{exsvpoly}) for a weight-four example involving a zeta value.
Upon insertion into the correlation function (\ref{G-normal}) with $s= - \frac{ \pp'}{\pp}$, we 
arrive at the following large-charge expansion
\begin{align}
{\cal G}^{\mathrm{II}}(x,\overline{x}) = - \frac{\pi \pp \pp'}{3(\pp - 3\pp')^2}\bigg\{&
1 + |x|^2 + |1{-}x|^2
\notag \\
&\! \! \! \! -  \frac{ 3\pp'}{\pp} \bigg[ 
|x|^2 G^{\rm sv}(0;x) +
|1{-}x|^2 G^{\rm sv}(1;x)
  \bigg]  \label{larcharexp} \\
&\! \! \! \! + \frac{ 3\pp'^2}{\pp^2} \bigg[ 
2 |x|^2 G^{\rm sv}(0, 0;x)
+ x(\overline{x}{-}1)
G^{\rm sv}(0, 1;x)  \notag \\
& \ \ \ \ \ \! \! \! \!  +\overline{x} (x{-}1 )
G^{\rm sv}(1, 0;x) 
 +
 2 |1{-}x|^2
 G^{\rm sv}(1, 1;x)
\bigg] + {\cal O}\bigg( \frac{1}{\pp^3} \bigg)
\bigg\} \, .  \notag
\end{align}
The decomposition \eqref{G-normal} into
uniformly transcendental sphere 
integrals $F^{(5,1)}_{ab}$ exemplifies
a key observation of this section:
In a suitable normalization,
certain four-point correlation functions 
in minimal models furnish another family of physical quantities besides amplitudes \cite{Kotikov:1990kg, ArkaniHamed:2010gh, Henn:2013pwa, Broedel:2013aza, Adams:2018yfj, Broedel:2018qkq, DHoker:2019blr, Mafra:2019xms} and form factors \cite{Brandhuber:2018xzk, Brandhuber:2018kqb} that feature uniform transcendentality. The natural normalization for \eqref{G-normal} is to peel off the prefactor $-\frac{\pi\pp \pp'}{3(\pp - 3\pp')^2}$, and uniform transcendentality then interlocks the transcendental weight of the polylogarithms and MZVs in (\ref{larcharexp}) with the order in $1/ \pp$ in the large-charge expansion. It would be interesting to investigate if more general four- and $n$-point correlation functions exhibit similar transcendentality properties.

\section{Summary and outlook}

In this work we have investigated configuration-space integrals over punctured Riemann spheres with an arbitrary number of integrated and unintegrated punctures $z_j$. Similar to the Koba--Nielsen factor in string tree-level amplitudes, the integrands feature products of $|z_i {-} z_j|^{s_{ij}}$, whose non-integer exponents lead to twisted homologies and cohomologies. The exponents $s_{ij}$ may be either identified with dimensionless Mandelstam invariants $2\alpha' k_i\cdot k_j$ containing the inverse string tension $\alpha'$, or with multiples of the dimensional-regularization parameter of Feynman integrals in spacetime dimensions $\in \mathbb N {-}2\epsilon$.

In this setting, we have given explicit bases of twisted cycles $ \gamma_a^{(n,p)} $
and cocycles $\omega_b^{(n,p)}$, such that the coaction of the period matrix 
$\langle \gamma_a^{(n,p)} | \omega_b^{(n,p)} \rangle$ lines up with the 
master formula (\ref{intr.1}) with coefficients taken from the identity matrix. The coaction applies to the MZVs and multiple polylogarithms in the Taylor expansion of
the period-matrix entries with respect to $s_{ij}$, and we have advanced their structural understanding by
\begin{itemize}[leftmargin=*]
\item introducing a systematic method for obtaining an explicit form of the $s_{ij}$-expansions,
\item decomposing $\langle \gamma_a^{(n,p)} | \omega_b^{(n,p)} \rangle$ into a matrix product
which organizes MZVs and polylogarithms at different arguments into separate factors,
\item pinpointing refined coaction formulae for the individual factors, i.e.\ for generating series of
polylogarithms in different numbers of variables,
\item spelling out the analytic continuations between different orderings of the unintegrated punctures on the real axis.
\end{itemize}

The integrals $\langle \gamma_a^{(n,p)} | \omega_b^{(n,p)} \rangle$ over paths in the configuration space ${\cal C}^{(n,p)}$ are related to complex integrals of $\overline{\omega_a^{(n,p)}} \omega_b^{(n,p)}$ over all of ${\cal C}^{(n,p)}$. Specifically, these complex ${\cal C}^{(n,p)}$-integrals are expressed both as single-valued versions or as complex bilinears in $\langle \gamma_a^{(n,p)} | \omega_b^{(n,p)} \rangle$. In this way, we generalize the KLT formula
and the single-valued map between open- and closed-string tree amplitudes beyond $p=n{-}3$, i.e.\ to more general integrals with
an arbitrary number of unintegrated punctures. Moreover, our results for the complex ${\cal C}^{(n,p)}$-integrals yield a new perspective on double-copy structures of correlation functions in minimal models, generalizing earlier $p=n{-}4$ results on KLT relations,
the single-valued map and the momentum-kernel formalism
\cite{Dotsenko:1984ad,Dotsenko:1984nm, Vanhove:2018elu}.

The discussion in this work calls for a generalization from the Riemann sphere to higher-genus surfaces and elliptic flavors of MZVs and multiple polylogarithms. Following the string-theory nomenclature, the associated twisted homologies are governed by the loop-level monodromy relations \cite{Tourkine:2016bak, Hohenegger:2017kqy, Casali:2019ihm, Casali:2020knc} between integration cycles some but not all of which are realized in open-string scattering. On the cohomology side, candidate bases for integration-by-parts inequivalent forms of open-string integrals were proposed in \cite{Mafra:2019ddf, Mafra:2019xms} and \cite{Broedel:2019gba, Broedel:2020tmd} for one and two unintegrated punctures, respectively. Their complex integrals over punctured tori have been applied to yield differential equations and iterated-integral representations of non-holomorphic modular forms \cite{Gerken:2019cxz, Gerken:2020yii} and single-valued elliptic polylogarithms \cite{Ramakrish, DHoker:2015wxz, DHoker:2020aex}.

It would be interesting to validate the conjectural bases of genus-one cohomologies by a rigorous treatment in the framework of twisted de Rham theory. Based on the elliptic symbol calculus \cite{Broedel:2018iwv}, it remains to translate the KZB-type differential equations in the references into coaction formulae, as for instance initiated in Section~7.2 of \cite{Mafra:2019xms} and Section~4.5 of \cite{Broedel:2020tmd}. 
It would furthermore be rewarding to identify echoes of the braid matrices at genus zero \cite{Mizera:2019gea} in the differential operators of these KZB equations, and to find their general form for an arbitrary number of unintegrated punctures.

Relatedly, the interplay of the KLT double copy and the single-valued map at genus zero raises a variety of questions at genus one and beyond including the following ones: Is there a KLT-type reformulation for the different approaches to single-valued elliptic MZVs in \cite{Brown:2017qwo, Brown:2017qwo2, Gerken:2020xfv} involving complex bilinears of open-string quantities? What is the single-valued map of elliptic multiple polylogarithms in one or several variables, and how do these structures arise in string amplitudes or Feynman integrals? What is the loop-level echo of the connection between minimal-model correlators, closed-string Koba--Nielsen integrals and their conformal-block decomposition at genus zero?

Finally, it would be rewarding to draw inspiration from string amplitudes beyond genus one and Feynman integrals beyond elliptic polylogarithms to classify iterated integrals on more general surfaces and to explore their differential and algebraic structures. In particular, there is a variety of further interesting testing grounds for and applications of the coaction principle in field-theory and string amplitudes -- both within and outside the current reach of genus-one integrals.

\acknowledgments
We are grateful to Samuel Abreu, Claude Duhr, Lorenz Eberhardt, Einan Gardi, Martijn Hidding, Daniel Kapec, Nils Matthes and Bram Verbeek for combinations of inspiring discussions and collaboration on related topics. S.M.\ thanks Uppsala University for hospitality during parts of this project. O.S.\ is grateful to Trinity College Dublin for hospitality during early stages of this project. This research was supported by the Munich Institute for Astro- and Particle Physics (MIAPP) of the DFG cluster of excellence ``Origin and Structure of the Universe.'' S.M.\ gratefully acknowledges the funding provided by Frank and Peggy Taplin as well as the grant DE-SC0009988 from the U.S.\ Department of Energy. R.B. is supported by the European Research Council under grant ERC-CoG-647356 (CutLoops). C.R.\ and O.S.\ are supported by the European Research Council under ERC-STG-804286 UNISCAMP.

\appendix

\section{Further details on the $\ap$-expansion}
\label{app:A}

This appendix complements the discussion of section \ref{sec:3} on the $\alpha'$-expansion
of the integrals $F^{(n,p)}_{ab}$.

\subsection{Monodromy relations for $F^{(5,1)}$}
\label{app:A.1}

In this appendix, we infer the formal initial value at $z_3= 0$ for the five-point integrals $F_{2a}^{(5,1)}$ with cycles and cocycles in (\ref{51.2}) and (\ref{51.3}) from monodromy relations \cite{BjerrumBohr:2009rd, Stieberger:2009hq}. On the integration contour $\gamma_{2}^{(5,1)}=\gamma_{13 \underline245}^{(5,1)} = \{z_3<z_2<1 \}$ for $z_2$ under consideration, the limit $z_3 \rightarrow 0$ does not commute with the $\alpha'$-expansion. Hence, the goal of this appendix is to infer the latter from integrals over $\gamma_{1}^{(5,1)}=\gamma_{1 \underline2345}^{(5,1)}= \{0<z_2<z_3 \}$ and $\gamma_{134 \underline25}^{(5,1)}=\{1<z_2<\infty \}$ where these processes do commute.

\begin{figure}
\begin{center}
\begin{tikzpicture}[scale = 1.6,line width=0.30mm]
\draw[->](-3.5,0) -- (3.6,0)node[above]{$\Re(z_2)$};
\draw[->](0,-0.5) -- (0,3.1)node[left]{$\Im(z_2)$};
\draw(0,0.05)node{$\bullet$}node[below]{$z_1{=}0\ \ \ $};
\draw(0.8,0.05)node{$\bullet$}node[below]{$\phantom{0}z_3\phantom{0}$};
\draw(1.7,0.05)node{$\bullet$}node[below]{$z_4{=}1$};
\draw(2.5,0.05)node{$\bullet$}node[below]{$+\infty$};
\draw(-2.5,0.05)node{$\bullet$}node[below]{$-\infty$};
\draw[blue,dashed] (2.5,0.1) arc (0:180:2.5cm);
\draw[red] (-2.5,0.1) -- (2.5,0.1);
\draw[red] (-1.25,0.3)node{$\gamma^{(5,1)}_{\underline21345}$};
\draw[red] (0.4,0.3)node{$\gamma^{(5,1)}_{1\underline2345}$};
\draw[red] (1.25,0.3)node{$\gamma^{(5,1)}_{13\underline245}$};
\draw[red] (2.1,0.3)node{$\gamma^{(5,1)}_{134\underline25}$};
\end{tikzpicture}
\end{center}
\caption{The closed contour ${\cal C}$ relevant to the monodromy relations (\ref{monodr.1}) at $(n,p)=(5,1)$ consists of subsets
$\gamma^{(5,1)}_{\underline21345},\, \gamma^{(5,1)}_{1\underline2345},\, \gamma^{(5,1)}_{13\underline245},\, \gamma^{(5,1)}_{13\underline245}$ of the real line drawn in red while the dashed semicircle simply indicates that $z_5 \rightarrow  \pm \infty$ are identified on
the Riemann sphere and does not contribute to the integral (\ref{monodr.1}).}
\label{figcirc}
\end{figure}
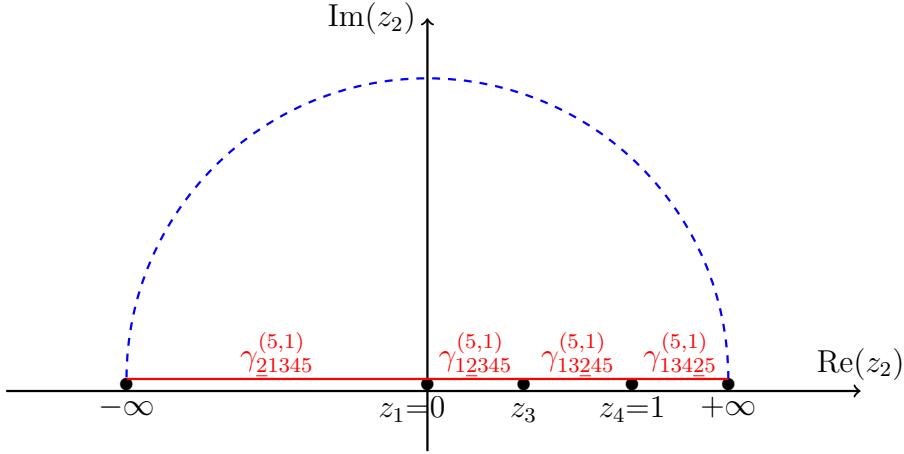

Following the techniques in \cite{Plahte:1970wy, BjerrumBohr:2009rd, Stieberger:2009hq}, we apply Cauchy's theorem to exploit the vanishing of the integrals
\begin{align}
0 &= \oint_{{\cal C}} (-z_2)^{s_{12}} (z_3{-}z_2)^{s_{23}} (1{-}z_2)^{s_{24}} \omega_a^{(5,1)}
\label{monodr.1} \\
&= 
 \int_{\gamma_{\underline{2} 1345}^{(5,1)}}  \! \! {\rm KN}^{(5,1)}  \omega_a^{(5,1)} +
e^{\pm i \pi s_{12}} \int_{\gamma_{1\underline{2} 345}^{(5,1)}}  \! \! {\rm KN}^{(5,1)}  \omega_a^{(5,1)}  \notag \\
&\ \ \ \ \ \ +
e^{\pm i \pi ( s_{12}{+}s_{23})} \int_{\gamma_{13\underline{2} 45}^{(5,1)}} {\rm KN}^{(5,1)}  \omega_a^{(5,1)} +
e^{\pm i \pi ( s_{12}{+}s_{23}{+}s_{24})} \int_{\gamma_{134\underline{2} 5}^{(5,1)}} {\rm KN}^{(5,1)}  \omega_a^{(5,1)} \, .
\notag
\end{align}
The meromorphic but multivalued integrand in the first line is tailored to match the Koba--Nielsen factor ${\rm KN}^{(5,1)}
= |z_2|^{s_{12}} |z_3{-}z_2|^{s_{23}} |1{-}z_2|^{s_{24}}$ up to phase factors composed of $e^{\pm i \pi s_{jk}}$.
These phases arise from relating $(-x)^{s} = e^{\pm i \pi  s} |x|^s$ for negative $x$ and are therefore piecewise constant on the components of the contour ${\cal C}= \gamma_{\underline{2} 1345}^{(5,1)} + \gamma_{1\underline{2} 345}^{(5,1)} + \gamma_{13\underline{2} 45}^{(5,1)} +  \gamma_{134\underline{2} 5}^{(5,1)}$ depicted in figure \ref{figcirc}. The sign of the phases depends on the choice of branches, so we can view (\ref{monodr.1}) as comprising two monodromy relations. Their difference 
\begin{align}
0&= \sin(\pi s_{12} )F_{1a}^{(5,1)}  +
\sin\big(\pi  ( s_{12}{+}s_{23} ) \big) F_{2a}^{(5,1)} +
\sin\big(  \pi ( s_{12}{+}s_{23}{+}s_{24}) \big) \int_{\gamma_{134\underline{2} 5}^{(5,1)}} {\rm KN}^{(5,1)}  \omega_a^{(5,1)}
\label{monodr.2}
\end{align}
involves the entries $F_{1a}^{(5,1)}$ and $F_{2a}^{(5,1)}$ by the identifications of the contours in (\ref{51.2}). For the third integral with $z_3$-independent integration limits in $\gamma_{134\underline{2} 5}^{(5,1)}= \{1<z_2<\infty\}$, we can commute the $z_3\rightarrow 0$ limit with its $\alpha'$-expansion and import (\ref{apexp.36}) for the combination of gamma functions in
\begin{align}
\lim_{z_3\rightarrow 0}  \int_{\gamma_{134\underline{2} 5}^{(5,1)}} {\rm KN}^{(5,1)}  \omega_a^{(5,1)} &= \int_1^{\infty} \dd z_2\, |z_2|^{s_{12}{+}s_{23}} |1{-}z_2|^{s_{24}} \, \bigg( \frac{s_{12}}{z_{21}} , \, \frac{(s_{12}{+}s_{23}) }{z_{21}} \bigg)_a
\label{monodr.3} \\
&=- \frac{(s_{12},\, s_{12}{+}s_{23})_a }{s_{12}{+}s_{23}{+}s_{24}} \, \frac{ \Gamma(1{-}s_{12}{-}s_{23}{-}s_{24} ) \Gamma(1{+}s_{24}) }{ \Gamma(1{-}s_{12}{-}s_{23}) } \, .
\notag
\end{align}
By inserting (\ref{monodr.3}) into the $z_3 \rightarrow0$ limit of (\ref{monodr.2}) and exploiting that 
\beq
\lim_{z_3\rightarrow 0}  F_{1a}^{(5,1)} = (1,0)_a \frac{ \Gamma(1{+}s_{12}) \Gamma(1{+}s_{23}) }{\Gamma(1{+}s_{12}{+}s_{23}) }
\, , \ \ \ \ \ \ \sin(\pi x) = \frac{ \pi x }{\Gamma(1{+}x) \Gamma(1{-}x)} \, ,
\eeq
we can solve for the $z_3\rightarrow0$ limit of $F_{2a}^{(5,1)}$:
\begin{align}
\lim_{z_3 \rightarrow 0} F_{2a}^{(5,1)} &= - \frac{ \sin\big(  \pi ( s_{12}{+}s_{23}{+}s_{24}) \big) }{\sin\big(  \pi ( s_{12}{+}s_{23}) \big)} \lim_{z_3\rightarrow 0}  \int_{\gamma_{134\underline{2} 5}^{(5,1)}} \!\!\!\!{\rm KN}^{(5,1)}  \omega_a^{(5,1)}
 - \frac{ \sin(\pi s_{12}) }{\sin\big(  \pi ( s_{12}{+}s_{23}) \big)} \lim_{z_3 \rightarrow 0} F_{1a}^{(5,1)} 
 \notag
 \\
 &= \bigg( \frac{s_{12} }{s_{12}{+}s_{23} } , \, 1 \bigg)_a
 \frac{\Gamma(1{+}s_{12}{+}s_{23}  ) \Gamma(1{+}s_{24}) }{ \Gamma(1{+}s_{12}{+}s_{23} {+}s_{24}) }
 -  \frac{s_{12}  (1,0)_a }{s_{12}{+}s_{23} } \frac{ \Gamma(1{-}s_{12}{-}s_{23}) \Gamma(1{+}s_{23}) }{\Gamma(1{-}s_{12}) }
\end{align}
This completes the derivation of the second line of (\ref{expa.21}).

The same type of arguments applies to the initial values of $F^{(n,1)}_{ab}$ with
$n\geq 6$. At $n=6$, for instance, the monodromy relation
\begin{align}
0&= \sin(\pi s_{12} )F_{1a}^{(6,1)}  +
\sin\big(\pi  ( s_{12}{+}s_{23} ) \big) F_{2a}^{(6,1)} +
\sin\big(  \pi ( s_{12}{+}s_{23}{+}s_{24}) \big) F_{3a}^{(6,1)}  \notag \\
&\ \ \ \ \ \ +
\sin\big(  \pi ( s_{12}{+}s_{23}{+}s_{24}{+}s_{25}) \big) \int_{\gamma_{1345\underline{2} 6}^{(6,1)}} {\rm KN}^{(6,1)}  \omega_a^{(6,1)}
\label{monodr.4}
\end{align}
involving one contour $\gamma^{(6,1)}_{1345 \underline 2 6} = \{ 1<z_2<\infty\}$
outside the basis (\ref{61.2}) can be used to infer the $z_3\rightarrow 0$ limit
of $F_{2a}^{(6,1)}$ which does not commute with $\alpha'$-expansion. The 
$z_3\rightarrow 0$ limit of the integrals $\gamma^{(6,1)}_{1345 \underline 2 6} $
can be identified with the kinematic limits of $F^{(5,1)}_{ab}$ seen in (\ref{charg}) and (\ref{in61.21}).
In this way, we arrive at the initial values (\ref{general.29}) and (\ref{expa.29})
as well as their $n$-point generalizations to be given in appendix \ref{app:A.3}.

\subsection{$\alpha'$-expansion of $F^{(5,1)}$}
\label{app:A.2}

The method of section \ref{sec:3.2} to obtain the $\alpha'$-expansion of 
$F^{(5,1)}$ via (\ref{expa.23}) gives rise to the leading orders
\begin{align}
F^{(5,1)}_{11} &=  1 +
s_{12} G({0}; z_3) + s_{23} G({0}; z_3)
+(s_{12}{+}s_{23})^2 G({0, 0}; z_3)  \notag \\
&\ \ \ \
+ s_{12} s_{24} G({0, 1}; z_3) - s_{12} s_{23} \zeta_2
+{\cal O}(s_{ij}^3)
\notag \\
F^{(5,1)}_{12} &=  
-s_{24} G({1}; z_3)
-  s_{23} s_{24} G({1, 1}; z_3) - s_{24}^2 G({1, 1}; z_3) \notag \\
&\ \ \ \ -(s_{12}{+}s_{23}) s_{24} G({1, 0}; z_3)  +{\cal O}(s_{ij}^3)
\label{apexp.8} \\
F^{(5,1)}_{21} &=  -s_{12} G({0}; z_3)
-s_{12}^2 G({0, 0}; z_3) - s_{12} s_{23} G({0, 0}; z_3) - 
 s_{12} s_{23} G({0, 1}; z_3)  \notag \\
 &\ \ \ \ - s_{12} s_{24} G({0, 1}; z_3)  - s_{12} s_{23} \zeta_2 - 
 s_{12} s_{24} \zeta_2
 +{\cal O}(s_{ij}^3)
\notag \\
F^{(5,1)}_{22} &=  1 
+(s_{23}{+}s_{24}) G({1}; z_3) 
 + (s_{23}{+}s_{24})^2 G({1, 1}; z_3)  \notag \\
 &\ \ \ \ +s_{12} s_{24} G({1, 0}; z_3)
 - (s_{12}{+}s_{23}) s_{24} \zeta_2 
 +{\cal O}(s_{ij}^3)\, .
\notag 
\end{align}
%

\subsection{The explicit form of $ \mathbb P^{(n,1)}$ and $ \mathbb M^{(n,1)} $}
\label{app:A.3}

The derivation of the $z_3\rightarrow 0$ asymptotics of $F^{(5,1)}_{ab}$ and $F^{(6,1)}_{ab}$ in (\ref{expa.21}) and (\ref{in61.21}) from 
monodromy relations generalizes to
\begin{align}
F^{(7,1)}_{ab}(z_3\rightarrow 0,z_4,z_5)  &= \ccccb
|z_4|^{s_{24}} |z_5|^{s_{25}} F_{11}^{(4,1)} &0 &0 &0 \\
\frac{ s_{12} \hat F^{(6,1)}_{11}}{s_{12}{+}s_{23}} - K^{(7,1)} &\hat F^{(6,1)}_{11} &\hat F^{(6,1)}_{12} &\hat F^{(6,1)}_{13} \\
\frac{ s_{12} \hat F^{(6,1)}_{21}}{s_{12}{+}s_{23}}  &\hat F^{(6,1)}_{21} &\hat F^{(6,1)}_{22} &\hat F^{(6,1)}_{23} \\
\frac{ s_{12} \hat F^{(6,1)}_{31}}{s_{12}{+}s_{23}}  &\hat F^{(6,1)}_{31} &\hat F^{(6,1)}_{32} &\hat F^{(6,1)}_{33}
\cccce  \label{inn1.20} 
\end{align}
and more generally
\begin{align}
F^{(n,1)}_{ab}(z_3 \rightarrow 0,z_4,\ldots,z_{n-2})  &= \cccccb
\Big( \prod_{k=4}^{n-2} |z_k|^{s_{2k}} \Big) F_{11}^{(4,1)} &0 &0 &\ldots &0 \\
\frac{ s_{12} \hat F^{(n-1,1)}_{11}}{s_{12}{+}s_{23}} - K^{(n,1)} &\hat F^{(n-1,1)}_{11} &\hat F^{(n-1,1)}_{12} &\ldots &\hat F^{(n-1,1)}_{1,n-4} \\
\frac{ s_{12} \hat F^{(n-1,1)}_{21}}{s_{12}{+}s_{23}}  &\hat F^{(n-1,1)}_{21} &\hat F^{(n-1,1)}_{22} &\ldots &\hat F^{(n-1,1)}_{2,n-4} \\
\vdots &\vdots &\vdots & &\vdots \\
\frac{ s_{12} \hat F^{(n-1,1)}_{n-4,1}}{s_{12}{+}s_{23}}  &\hat F^{(n-1,1)}_{n-4,1} &\hat F^{(n-1,1)}_{n-4,2} &\ldots &\hat F^{(n-1,1)}_{n-4,n-4}
\ccccce
 \, ,
\end{align}
where the $(a,b)=(2,1)$ entries involve
\begin{align}
K^{(7,1)} &= \frac{ \sin (\pi s_{12}) }{\sin (\pi(s_{12}{+}s_{23})) }|z_4|^{s_{24}}  |z_5|^{s_{25}} F^{(4,1)} 
 \label{inn1.21}\\
 &= |z_4|^{s_{24}}  |z_5|^{s_{25}} \frac{ s_{12} }{s_{12}{+}s_{23}} \frac{ \Gamma(1{+}s_{23}) \Gamma(1{-}s_{12}{-}s_{23}) }{\Gamma(1{-}s_{12}) } \notag \\
K^{(n,1)} &= \frac{ \sin (\pi s_{12}) }{\sin (\pi(s_{12}{+}s_{23})) }  \Big( \prod_{k=4}^{n-2} |z_k|^{s_{2k}} \Big) F^{(4,1)}  \label{inn0.2}\\
&=\Big( \prod_{k=4}^{n-2} |z_k|^{s_{2k}} \Big) \frac{ s_{12} }{s_{12}{+}s_{23}} \frac{ \Gamma(1{+}s_{23}) \Gamma(1{-}s_{12}{-}s_{23}) }{\Gamma(1{-}s_{12}) }\, .
\notag
\end{align}
The hat notation instructs to change the arguments of $F^{(n-1,1)}_{ab}$ to (cf.\ (\ref{charg}))
\beq
 s_{12} \rightarrow s_{12}{+}s_{23} \, , \ \ \ \ s_{2,j} \rightarrow s_{2,j+1} 
\, , \ \ \ \ z_{k} \rightarrow z_{k+1}  
 \label{inn1.3} 
\eeq
for $j=3,4,\ldots,n{-}2$ and $k=3,4,\ldots,n{-}3 $. The four-point integral ($d^{(4,1)}=1$) yields
the standard Euler beta function (\ref{apexp.36}).
The formal $z_j \rightarrow 0$ limits (\ref{expa.21}), (\ref{expa.29}) and their generalizations
 involve the following differences of beta functions,
\begin{align}
W_3(s_{12},s_{23},s_{24}) &= \frac{ F^{(4,1)}(s_{12}{+}s_{23},s_{24}) - F^{(4,1)}(s_{23},{-}s_{12}{-}s_{23}) }{s_{12}{+}s_{23}} \notag \\
&= -(s_{23}{+}s_{24}) \zeta_2 + \big[  (s_{12}{+}  s_{23} {+} s_{24} )s_{24} - s_{12} s_{23} \big] \zeta_3 + {\cal O}(s_{ij}^3) \notag \\
W_4(s_{12},s_{23},s_{24},s_{25}) &= \frac{ F^{(4,1)}(s_{12}{+}s_{23}{+}s_{24},s_{25}) - F^{(4,1)}(s_{24},{-}s_{12}{-}s_{23}{-}s_{24}) }{s_{12}{+}s_{23}{+}s_{24}} \notag \\
&= -(s_{24}{+}s_{25}) \zeta_2 +   ( s_{12} {+} s_{23} {+} s_{24} {+} s_{25} ) s_{25}  \zeta_3 \label{apexp.37} \\
& \ \ \ \  - (s_{12} {+} s_{23}) s_{24}\zeta_3 + {\cal O}(s_{ij}^3)  \notag\\
W_5(s_{12},s_{23},\ldots,s_{26}) &= \frac{ F^{(4,1)}(s_{12}{+}s_{23}{+}s_{24}{+}s_{25},s_{26}) - F^{(4,1)}(s_{25},{-}s_{12}{-}s_{23}{-}s_{24}{-}s_{25}) }{s_{12}{+}s_{23}{+}s_{24}{+}s_{25}} \notag \\
&=  -(s_{25}{+}s_{26}) \zeta_2 +  (s_{12} {+} s_{23} {+} s_{24} {+} s_{25}{+} s_{26}) s_{26} \zeta_3 \notag \\
&\ \ \ \  - (s_{12} {+} s_{23} {+} s_{24}) s_{25}   \zeta_3 + {\cal O}(s_{ij}^3)\, ,\notag
\end{align}
and more generally
\beq
W_j(s_{12},s_{23},\ldots,s_{2,j+1})=  \frac{ F^{(4,1)}(s_{12}{+}\sum_{i=3}^j s_{2i}
,s_{2,j+1}) - F^{(4,1)}(s_{2j},{-}s_{12}{-}
\sum_{i=3}^j s_{2i}) }{s_{12}{+}s_{23}{+}\ldots {+}s_{2j}} \, .
\label{apexp.38}
\eeq
This notation yields the compact representations:
\begin{align}
 &~\lim_{z_3\rightarrow 0} F^{(5,1)}(z_3) = \ccb F^{(4,1)}(s_{12},s_{23}) & 0 \\
s_{12} \, W_3&F^{(4,1)}(s_{12}{+}s_{23},s_{24}) \cce \,,
 \label{apexp.39a}  
\\
& \lim_{z_3,z_4\rightarrow 0} F^{(6,1)}(z_3,z_4) = \cccb F^{(4,1)}(s_{12},s_{23}) & 0 &0 \\
s_{12} \, W_3&\! \! \! F^{(4,1)}(s_{12}{+}s_{23},s_{24}) \! \! \! &0 \\
s_{12}\, W_4 &(s_{12}{+}s_{23}) \, W_4 &\! \! \! F^{(4,1)}(s_{12}{+}s_{23}{+}s_{24},s_{25}) \ccce\,,
 \label{apexp.39b}  
\end{align}
\begin{align}
&\lim_{z_3,z_4,z_5\rightarrow 0} F^{(7,1)}(z_3,z_4,z_5) = 
 \label{apexp.39c}  
\\
&
\left(\begin{array}{ccccc} F^{(4,1)}(s_{12},s_{23}) \! \! & 0 &0 &0  \\
s_{12} \, W_3&\! \! F^{(4,1)}(s_{12}{+}s_{23},s_{24}) \! \! &0 &0 \\
s_{12}\, W_4 &(s_{12}{+}s_{23}) \, W_4 &\! \! F^{(4,1)}(s_{12}{+}s_{23}{+}s_{24},s_{25}) \! \! &0 \\
s_{12}\, W_5 &(s_{12}{+}s_{23}) \, W_5 &(s_{12}{+}s_{23}{+}s_{24}) \, W_5 &\! \! F^{(4,1)}(s_{12}{+}s_{23}{+}s_{24}{+}s_{25},s_{26}) 
\end{array} \right)\,. \notag
\end{align}
These expressions follow from the initial values (\ref{inn1.20})
and generalize as follows to higher multiplicity:
\beq
\lim_{z_k \rightarrow 0} F_{ab}^{(n,1)} = \left\{ \begin{array}{cl}
0 &: \ b>a \\
F^{(4,1)}\Big(s_{12}+\sum_{m=3}^{b+1} s_{2,m},s_{2,b+2} \Big) &: \ b=a \\
\Big(s_{12}+\sum_{m=3}^{b+1} s_{2,m} \Big) W_{a+1} &: \ b<a
\end{array} \right.
 \label{apexp.40} 
\eeq
%

\subsection{The explicit form of $\mathbb P^{(6,2)}$ and $\mathbb M^{(6,2)}$}
\label{app:A.4}

\begin{figure}
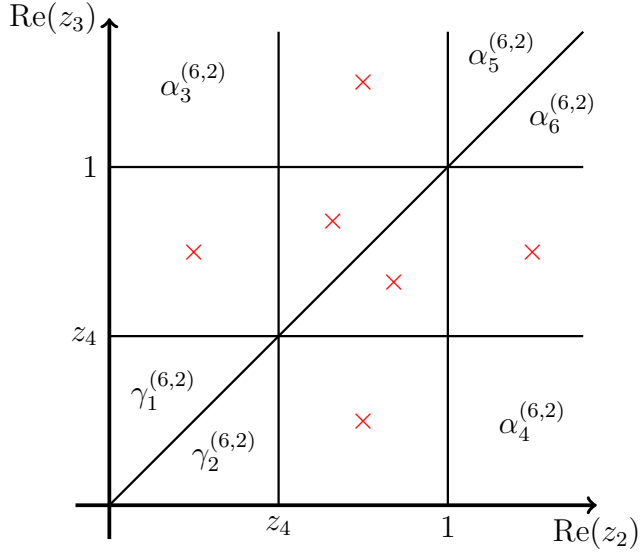

  \begin{center}
\tikzpicture [scale=0.9,line width=0.50mm]
\draw[->](-0.5,0) -- (7.2,0) node[below]{$\Re(z_2)$};
\draw[->](0,-0.5) -- (0,7.2) node[left]{$\Re(z_3)$};
\draw[line width=0.30mm](2.5,0)node[below]{$z_4$} -- (2.5,7) ;
\draw[line width=0.30mm](5,0)node[below]{$1$} -- (5,7) ;
\draw[line width=0.30mm](0,2.5)node[left]{$z_4$} -- (7,2.5) ;
\draw[line width=0.30mm](0,5)node[left]{$1$} -- (7,5) ;
\draw[line width=0.30mm](0,0)--(7,7);
\draw(0.8,1.7)node{$\gamma^{(6,2)}_1$};
\draw(1.7,0.8)node{$\gamma^{(6,2)}_2$};
\draw(3.3,4.2)node[red]{$\times$};
\draw(4.2,3.3)node[red]{$\times$};
\draw(5.8,6.7)node{$\alpha^{(6,2)}_5$};
\draw(6.7,5.8)node{$\alpha^{(6,2)}_6$};
\draw(3.75,1.25)node[red]{$\times$};
\draw(6.25,1.25)node{$\alpha^{(6,2)}_4$};
\draw(1.25, 3.75)node[red]{$\times$};
\draw(1.25, 6.25)node{$\alpha^{(6,2)}_3$};
\draw(3.75,6.25)node[red]{$\times$};
\draw(6.25, 3.75)node[red]{$\times$};
\endtikzpicture
    \caption{We will determine the initial conditions for $F^{(6,2)}_{ab}$ from
    the depicted six-dimensional basis of contours $\gamma_1^{(6,2)},\gamma_2^{(6,2)},
    \alpha_3^{(6,2)},\alpha_4^{(6,2)},\alpha_5^{(6,2)},\alpha_6^{(6,2)}$. For these contours, the
    $z_4\rightarrow 0$ limit commutes with the $\alpha'$-expansion which is not the case for
    the contours marked with $\textcolor{red}{\times}$ such as $\gamma_{j}^{(6,2)}$
    with $j=3,4,5,6$ in (\ref{bascont}).}
    \label{basiccyl}
  \end{center}
\end{figure}

We shall finally give the key steps towards the $\ap$-expansion
of the integrals $F^{(6,2)}_{ab}$ over the basis forms
\begin{align}
\hat \omega_1^{(6,2)} &= \frac{ s_{12}}{z_{12}}  \bigg(  \frac{ s_{13}}{z_{13}} 
+ \frac{ s_{23}}{z_{23}} \bigg) \, ,  &\hat \omega_3^{(6,2)} &= \frac{ s_{12}}{z_{12}}  \bigg(  \frac{ s_{13}}{z_{13}} + \frac{ s_{23}}{z_{23}} + \frac{ s_{43}}{z_{43}} \bigg) 
\notag \\
\hat\omega_5^{(6,2)} &=
\bigg(\frac{ s_{12}}{z_{12}} + \frac{ s_{42}}{z_{42}} \bigg) \bigg(  \frac{ s_{13}}{z_{13}} + \frac{ s_{23}}{z_{23}} + \frac{ s_{43}}{z_{43}} \bigg) \, , &\hat \omega_{2k}^{(6,2)} &= \hat \omega_{2k-1}^{(6,2)} \, \Big|_{2\leftrightarrow 3} \, , \ \ k=1,2,3
\end{align}
according to (\ref{p2expl}), with $2\leftrightarrow 3$ referring
to the subscripts of both $z_{ij}$ and $s_{ij}$. Again, the basis of integration 
contours
\begin{align}
\gamma_1^{(6,2)} &= \{0< z_2 < z_3<z_4\} \, , 
&\gamma_3^{(6,2)} &= \{0< z_2 <z_4 < z_3<1\}  \notag \\
\gamma_5^{(6,2)} &= \{z_4< z_2 < z_3<1 \} \, ,
& \gamma_{2k}^{(6,2)} &= \gamma_{2k-1}^{(6,2)} \, \Big|_{2\leftrightarrow 3}\, , \ \ k=1,2,3
\label{bascont}
\end{align}
contains four cases $\gamma_j^{(6,2)}$ at $j=3,4,5,6$
where the $\ap$-expansion does not commute with the $z_4\rightarrow 0$ limit.
Similar to the strategy for the $(n,p)=(5,1)$ case in appendix \ref{app:A.1}, we use
monodromy relations to relate
these problematic contours to auxiliary ones
\begin{align}
\alpha^{(6,2)}_3 &= \{(z_2,z_3) \in \mathbb R^2 \ | \ 0<z_2<z_4 \ \te{and} \ 1<z_3<\infty \} \notag \\
\alpha^{(6,2)}_5 &= \{(z_2,z_3) \in \mathbb R^2 \ | \ 1<z_2<z_3<\infty \}
\\
\alpha^{(6,2)}_4 &= \alpha^{(6,2)}_3 \, \Big|_{2\leftrightarrow 3} \, , \ \ \ \ 
\alpha^{(6,2)}_6 = \alpha^{(6,2)}_5 \, \Big|_{2\leftrightarrow 3}
\notag
\end{align}
depicted in figure \ref{basiccyl}. These $\alpha^{(6,2)}_j$ are engineered
to have commutative limits $z_4\rightarrow 0$ and $\ap \rightarrow0$
and will therefore serve as a crucial tool to assemble the initial values
$\mathbb P^{(6,2)}$ and $\mathbb M^{(6,2)}$ in (\ref{apexp.35}).

We will make use of the monodromy relations
\begin{align}
\gamma_3^{(6,2)} &= -\frac{\alpha_3^{(6,2)}
   \sin \left(\pi  \left(s_{13}{+}s_{23}{+}s_{34}{+}s_{35}\right)\right)+\gamma_1^{(6,2)} \sin \left(\pi  \left(s_{13}{+}s_{23}\right)\right)+\gamma_2^{(6,2)} \sin
   \left(\pi  s_{13}\right)}{\sin \left(\pi  \left(s_{13}{+}s_{23}{+}s_{34}\right)\right)}\, , \notag \\
\gamma_5^{(6,2)} &= \frac{-\sin \left(\pi  s_{12}\right) \sin \left(\pi  s_{34}\right) \gamma_1^{(6,2)} + \sin \left(\pi  s_{13}\right) \sin \left(\pi 
   \left(s_{12}{+}s_{13}{+}s_{23}{+}s_{34}\right)\right) \gamma_2^{(6,2)}}{\sin \left(\pi  \left(s_{13}{+}s_{23}{+}s_{34}\right)\right) \sin \left(\pi 
   \left(s_{12}{+}s_{13}{+}s_{23}{+}s_{24}{+}s_{34}\right)\right)}  \nn\\
   &\ \ \ \ + \frac{\sin \left(\pi  s_{12}\right) \sin
   \left(\pi 
   \left(s_{13}{+}s_{23}{+}s_{34}{+}s_{35}\right)\right) \alpha_3^{(6,2)}}{\sin
   \left(\pi  \left(s_{13}{+}s_{23}{+}s_{34}\right)\right) \sin
   \left(\pi  \left(s_{12}{+}s_{24}\right)\right)}   \label{bigmonodr}\\
   &\ \ \ \ + \frac{\sin \left(\pi 
   \left(s_{13}{+}s_{23}{+}s_{34}{+}s_{35}\right)\right) \sin
   \left(\pi  s_{25}\right) \alpha_5^{(6,2)}}{\sin \left(\pi 
   \left(s_{12}{+}s_{13}{+}s_{23}{+}s_{24}{+}s_{34}\right)\right) \sin
   \left(\pi  \left(s_{12}{+}s_{24}\right)\right)}  \nn\\
   &\ \ \ \ + \frac{\sin \left(\pi 
   \left(s_{12}{+}s_{23}{+}s_{24}{+}s_{25}\right)\right) \sin
   \left(\pi 
   \left(s_{12}{+}s_{13}{+}s_{23}{+}s_{24}{+}s_{34}{+}s_{35}\right)\right
   ) \alpha_6^{(6,2)}}{\sin \left(\pi 
   \left(s_{12}{+}s_{13}{+}s_{23}{+}s_{24}{+}s_{34}\right)\right) \sin
   \left(\pi  \left(s_{12}{+}s_{24}\right)\right) }  \nn
\end{align}
and two similar relations for $\gamma_4^{(6,2)} $ and $\gamma_6^{(6,2)} $ that
are obtained from relabelling the Mandelstam invariants via $2\leftrightarrow3$
and exchanging $\gamma_1^{(6,2)}  \leftrightarrow \gamma_2^{(6,2)} $
as well as $\alpha_{3}^{(6,2)}  \leftrightarrow \alpha_{4}^{(6,2)} $ and 
$\alpha_{5}^{(6,2)}  \leftrightarrow \alpha_{6}^{(6,2)} $ on the right-hand sides 
of (\ref{bigmonodr}).

\subsubsection{$z_4\rightarrow 0$ limits on the $\alpha^{(6,2)}_i$ contours}

The integrals $F^{(6,2)}_{\alpha_i, b} = \langle \alpha^{(6,2)}_i | \omega_b^{(6,2)} \rangle$
with $i=3,4,5,6$ can be shown to have the $z_4\rightarrow 0$ limits
\begin{align}
\lim_{z_4\rightarrow 0} F^{(6,2)}_{\alpha_3, b}&=
- \frac{ s_{13}{+}s_{23}{+}s_{34} }{ s_{13}{+}s_{23}{+}s_{34} {+}s_{35}}
 \frac{ \Gamma(1{+}s_{12}) \Gamma(1{+}s_{24}) }{ \Gamma(1{+}s_{12}{+}s_{24})}   \frac{ \Gamma(1{-}s_{13}{-}s_{23}{-}s_{34}{-}s_{35}) \Gamma(1{+}s_{35}) }{ \Gamma(1{-}s_{13}{-}s_{23}{-}s_{34})} \notag \\
 &\ \ \ \ \times
 \bigg( \frac{ s_{13}{+}s_{23} }{s_{13}{+}s_{23}{+}s_{34}},\
 \frac{ s_{13}}{s_{13}{+}s_{23}{+}s_{34}},\ 1,\  0, \ 0, \ 0\bigg)_b  \\
 \lim_{z_4\rightarrow 0} F^{(6,2)}_{\alpha_4, b}&= - \frac{ s_{12}{+}s_{23}{+}s_{24} }{ s_{12}{+}s_{23}{+}s_{24} {+}s_{25}}
 \frac{ \Gamma(1{+}s_{13}) \Gamma(1{+}s_{34}) }{ \Gamma(1{+}s_{13}{+}s_{34})}   \frac{ \Gamma(1{-}s_{12}{-}s_{23}{-}s_{24}{-}s_{25}) \Gamma(1{+}s_{25}) }{ \Gamma(1{-}s_{12}{-}s_{23}{-}s_{24})} \notag \\
 &\ \ \ \ \times
 \bigg( \frac{ s_{12} }{s_{12}{+}s_{23}{+}s_{24}},\
 \frac{ s_{12}{+}s_{23}}{s_{12}{+}s_{23}{+}s_{24}},\ 0 ,\ 1,\ 0, \ 0\bigg)_b 
  \label{in62.4}
\end{align}
as well as
\begin{align}
\lim_{z_4\rightarrow 0} F^{(6,2)}_{\alpha_{5,6}, b}&=  \bigg( u_{11} \hat F^{(5,2)}_{\alpha_{5,6},1}+ u_{12} \hat F^{(5,2)}_{\alpha_{5,6},2} , \  
 u_{21} \hat F^{(5,2)}_{\alpha_{5,6},1}+ u_{22} \hat F^{(5,2)}_{\alpha_{5,6},2}, \notag \\
 & \ \ \ \  \frac{ s_{12}}{s_{12}{+}s_{24}} \hat F^{(5,2)}_{\alpha_{5,6},1},\ 
\frac{ s_{13}}{s_{13}{+}s_{34}} \hat F^{(5,2)}_{\alpha_{5,6},2},\ 
\hat F^{(5,2)}_{\alpha_{5,6},1},  \ \hat F^{(5,2)}_{\alpha_{5,6},2}\bigg)_b \, .
 \label{in62.6}
\end{align}
The gamma functions in (\ref{in62.4}) stem from the unique 
component of $F^{(4,1)}$ in (\ref{apexp.36}), and
the coefficients $u_{ij}$ in (\ref{in62.6}) are given by
\beq
u_{ij} = \ccb \frac{ s_{12}(s_{123}{+}s_{24}) }{(s_{12}{+}s_{24})(s_{123}{+}s_{24}{+}s_{34})} 
& \frac{ - s_{12}s_{34} }{(s_{13}{+}s_{34})(s_{123}{+}s_{24}{+}s_{34})} 
\\ \frac{ - s_{13}s_{24} }{(s_{12}{+}s_{24})(s_{123}{+}s_{24}{+}s_{34})}  
& \frac{ s_{13}(s_{123}{+}s_{34}) }{(s_{13}{+}s_{34})(s_{123}{+}s_{24}{+}s_{34})} \cce_{ij} \, .
\label{5ptdisk4}
\eeq
Furthermore, the integration contours of the $\hat F^{(5,2)}_{\alpha_{5,6},i}$ on 
the right-hand side of (\ref{in62.6}) can be reduced to a basis of $\gamma_{1,2}^{(5,2)}$
via monodromy relations
\begin{align}
\medmath{\hat F^{(5,2)}_{\alpha_5,i}} &\medmath{= \frac{
  \sin(\pi  (s_{13}{+}s_{34})) \sin(\pi  (s_{123}{+}s_{24}{+}s_{34}{+}s_{35})) \hat F^{(5,2)}_{2,i}  - \sin(\pi  s_{35}) \sin(\pi  (s_{12}{+}s_{24})) \hat F^{(5,2)}_{1,i} }{\sin (\pi  (s_{13}{+}s_{23}{+}s_{34}{+}s_{35})) \sin (\pi 
   (s_{123}{+}s_{24}{+}s_{25}{+}s_{34}{+}s_{35}))} }\nn\\
\medmath{\hat F^{(5,2)}_{\alpha_6,i}} &\medmath{= \frac{\sin (\pi  (s_{12}{+}s_{24})) \sin (\pi 
   (s_{123}{+}s_{24}{+}s_{25}{+}s_{34})) \hat{F}_{1,i}^{(5,2)} -\sin (\pi  s_{25}) \sin (\pi 
   (s_{13}{+}s_{34})) \hat{F}_{2,i}^{(5,2)} }{\sin (\pi  (s_{12}{+}s_{23}{+}s_{24}{+}s_{25})) \sin (\pi 
   (s_{123}{+}s_{24}{+}s_{25}{+}s_{34}{+}s_{35}))}} \, .
\label{in62.7}
\end{align}
Finally, the hat denotes the following replacement of the arguments of
$F^{(5,2)}$,
\beq
\hat{F}^{(5,2)}_{ab}(s_{12},s_{13},s_{23},s_{24},s_{34}) =F^{(5,2)}_{ab}(s_{12}{+}s_{24},s_{13}{+}s_{34},s_{23},s_{25},s_{35})\, ,
\eeq
which can be traced back to the $z_4 \rightarrow 0$ behavior of the Koba--Nielsen factor
\beq
\lim_{z_4 \rightarrow 0} {\rm KN}^{(6,2)}
= |z_2|^{s_{12}+s_{24}} |z_3|^{s_{13}+s_{34}} |z_{23}|^{s_{23}}|1{-}z_2|^{s_{25}}|1{-}z_3|^{s_{35}}\, .
\eeq

\subsubsection{Assembling the initial value}

By combining the monodromy relations (\ref{bigmonodr}) with the $z_{4} \rightarrow 0$
limits of $F^{(6,2)}_{\alpha_i,b}(z_4)$, we arrive at the following initial values
of $F^{(6,2)}_{ab}(z_4)$ in the basis of (\ref{bascont}):  
\begin{align}
&\mathbb P^{(6,2)} \mathbb M^{(6,2)} = \lim_{z_4\rightarrow0} F^{(6,2)}=
\label{Fn6p2initVal}
\\
&
\medmath{\left( \begin{array}{cccccc}  F^{(5,2)}_{11} & F^{(5,2)}_{12} & 0 & 0 & 0 & 0 \\ 
F^{(5,2)}_{21} & F^{(5,2)}_{22} & 0  & 0 & 0 & 0 \\
H^{(6,2)}_{11} & H^{(6,2)}_{12} & F^{(4,1)}(s_{12},s_{24}) F^{(4,1)}(s_{35},s_{13}{+}s_{23}{+}s_{34}) \! \! \! \! \! \! \! \! \! \! & 0 & 0 & 0 \\
H^{(6,2)}_{21} & H^{(6,2)}_{22}  & 0 &\! \! \! \! \! \! \! \! \! \! F^{(4,1)}(s_{13},s_{34}) F^{(4,1)}(s_{25},s_{12}{+}s_{23}{+}s_{24}) & 0 & 0 \\
J^{(6,2)}_{11} & J^{(6,2)}_{12} & K^{(6,2)}_{11} & \frac{s_{13} }{s_{13}+s_{34}}  \hat{F}^{(5,2)}_{12}& \hat{F}^{(5,2)}_{11} & \hat{F}^{(5,2)}_{12} \\
J^{(6,2)}_{21} & J^{(6,2)}_{22} & \frac{s_{12}}{s_{12}+s_{24}}  \hat{F}^{(5,2)}_{21} &  K^{(6,2)}_{22} & \hat{F}^{(5,2)}_{21} & \hat{F}^{(5,2)}_{22}
\end{array} \right)\,.
}
\notag
\end{align}
The entries $H^{(6,2)}_{1j}$ are given by
\begin{align}
H^{(6,2)}_{11} &=   \frac{s_{13}{+}s_{23}}{s_{13}{+}s_{23}{+}s_{34}} F^{(4,1)}(s_{12},s_{24}) F^{(4,1)}(s_{35},s_{13}{+}s_{23}{+}s_{34}) \\
&\ \ \ \ -\frac{\sin(\pi(s_{13}{+}s_{23}))}{\sin(\pi(s_{13}{+}s_{23}{+}s_{34}))}F^{(5,2)}_{11}-\frac{\sin(\pi s_{13})}{\sin(\pi(s_{13}{+}s_{23}{+}s_{34}))}F^{(5,2)}_{21}  \notag \\
H^{(6,2)}_{12} &= 
\frac{s_{13}}{s_{13}{+}s_{23}{+}s_{34}} F^{(4,1)}(s_{12},s_{24}) F^{(4,1)}(s_{35},s_{13}{+}s_{23}{+}s_{34}) \\
&\ \ \ \ -\frac{\sin(\pi(s_{13}{+}s_{23}))}{\sin(\pi (s_{13}{+}s_{23}{+}s_{34}))} F^{(5,2)}_{12} - \frac{\sin(\pi s_{13})}{\sin(\pi (s_{13}{+}s_{23}{+}s_{34}))} F^{(5,2)}_{22} \, , \notag
\end{align}
while the entries $H^{(6,2)}_{2j}$ can be obtained from $H^{(6,2)}_{1j}$ by relabeling $2\leftrightarrow 3$ at the level of the Mandelstam variables throughout and are thus given by
\begin{align}
H^{(6,2)}_{21} &=\frac{s_{12}}{s_{12}{+}s_{23}{+}s_{24}} F^{(4,1)}(s_{13},s_{34}) F^{(4,1)}(s_{25},s_{12}{+}s_{23}{+}s_{24}) \\ 
&\ \ \ \ -\frac{\sin(\pi(s_{12}{+}s_{23}))}{\sin(\pi (s_{12}{+}s_{23}{+}s_{24}))} F^{(5,2)}_{21} - \frac{\sin(\pi s_{12})}{\sin(\pi (s_{12}{+}s_{23}{+}s_{24}))} F^{(5,2)}_{11}  \notag \\
H^{(6,2)}_{22} &=
\frac{s_{12}{+}s_{23}}{s_{12}{+}s_{23}{+}s_{24}} F^{(4,1)}(s_{13},s_{34}) F^{(4,1)}(s_{25},s_{12}{+}s_{23}{+}s_{24}) \\
&\ \ \ \ -\frac{\sin(\pi(s_{12}{+}s_{23}))}{\sin(\pi(s_{12}{+}s_{23}{+}s_{24}))}F^{(5,2)}_{22}-\frac{\sin(\pi s_{12})}{\sin(\pi(s_{12}{+}s_{23}{+}s_{24}))}F^{(5,2)}_{12} 
 \, . \notag 
\end{align}
The entries $J^{(6,2)}_{1j}$ are given by
\begin{align}
    J^{(6,2)}_{11} =\, & u_{11} \hat{F}^{(5,2)}_{11}+u_{12} \hat{F}^{(5,2)}_{12}  \label{exentry}\\
     &- \frac{\sin(\pi s_{12})}{\sin(\pi(s_{12}{+}s_{24}))}\frac{s_{13}{+}s_{23}}{s_{13}{+}s_{23}{+}s_{34}} F^{(4,1)}(s_{12},s_{24}) F^{(4,1)}(s_{35},s_{13}{+}s_{23}{+}s_{34}) \notag \\
     &+ \frac{\left[ - \sin(\pi s_{12})\sin(\pi s_{34}) F^{(5,2)}_{11} + \sin(\pi s_{13})\sin(\pi (s_{123}{+}s_{34})) F^{(5,2)}_{21}
     \right]}{\sin(\pi(s_{13}{+}s_{23}{+}s_{34})) \sin(\pi(s_{123}{+}s_{24}{+}s_{34}))} 
\notag \\
    J^{(6,2)}_{12} =\, & u_{21} \hat{F}^{(5,2)}_{11}+u_{22} \hat{F}^{(5,2)}_{12} \\
     &- \frac{\sin(\pi s_{12})}{\sin(\pi(s_{12}{+}s_{24}))}\frac{s_{13}}{s_{13}{+}s_{23}{+}s_{34}} F^{(4,1)}(s_{12},s_{24}) F^{(4,1)}(s_{35},s_{13}{+}s_{23}{+}s_{34}) \notag \\
     &+ \frac{\left[ - \sin(\pi s_{12})\sin(\pi s_{34}) F^{(5,2)}_{12} + \sin(\pi s_{13})\sin(\pi (s_{123}{+}s_{34})) F^{(5,2)}_{22}
     \right]}{\sin(\pi(s_{13}{+}s_{23}{+}s_{34})) \sin(\pi(s_{123}{+}s_{24}{+}s_{34}))} \, , \notag
\end{align}
where the $u_{ij}$ are defined in (\ref{5ptdisk4}).
The entries $J^{(6,2)}_{2j}$ can again be obtained from $J^{(6,2)}_{1j}$ by relabeling $2\leftrightarrow 3$:
\begin{align}
    J^{(6,2)}_{21} =\, &J^{(6,2)}_{12}\big|_{(2 \leftrightarrow 3)} = u_{12} \hat{F}^{(5,2)}_{22}+u_{11} \hat{F}^{(5,2)}_{21} \\
     &- \frac{\sin(\pi s_{13})}{\sin(\pi(s_{13}{+}s_{34}))}\frac{s_{12}}{s_{12}{+}s_{23}{+}s_{24}} F^{(4,1)}(s_{13},s_{34}) F^{(4,1)}(s_{25},s_{12}{+}s_{23}{+}s_{24}) \notag \\
     &+ \frac{\left[ - \sin(\pi s_{13})\sin(\pi s_{24}) F^{(5,2)}_{21} + \sin(\pi s_{12})\sin(\pi (s_{123}{+}s_{24})) F^{(5,2)}_{11}
     \right]}{\sin(\pi(s_{12}{+}s_{23}{+}s_{24})) \sin(\pi(s_{123}{+}s_{24}{+}s_{34}))} 
\notag \\
    J^{(6,2)}_{22}=\, & J^{(6,2)}_{11}\big|_{(2 \leftrightarrow 3)} = u_{22} \hat{F}^{(5,2)}_{22}+u_{21} \hat{F}^{(5,2)}_{21} \\
     &- \frac{\sin(\pi s_{13})}{\sin(\pi(s_{13}{+}s_{34}))}\frac{s_{12}{+}s_{23}}{s_{12}{+}s_{23}{+}s_{24}} F^{(4,1)}(s_{13},s_{34}) F^{(4,1)}(s_{25},s_{12}{+}s_{23}{+}s_{24}) \notag\\
     &+ \frac{\left[ - \sin(\pi s_{13})\sin(\pi s_{24}) F^{(5,2)}_{22} + \sin(\pi s_{12})\sin(\pi (s_{123}{+}s_{24})) F^{(5,2)}_{12}
     \right]}{\sin(\pi(s_{12}{+}s_{23}{+}s_{24})) \sin(\pi(s_{123}{+}s_{24}{+}s_{34}))} \, .
     \notag
\end{align}
Lastly, the entries  $K^{(6,2)}_{ii}$ related by $2 \leftrightarrow 3$ are given by
\begin{align}
K^{(6,2)}_{11}&=\frac{s_{12} \hat{F}^{(5,2)}_{11}}{s_{12}{+}s_{24}}-\frac{\sin(\pi s_{12})}{\sin(\pi(s_{12}{+}s_{24}))}F^{(4,1)}(s_{12},s_{24})F^{(4,1)}(s_{35},s_{13}{+}s_{23}{+}s_{34})
\notag
\\
K^{(6,2)}_{22}&= \frac{s_{13} \hat{F}^{(5,2)}_{22}}{s_{13}{+}s_{34}}-\frac{\sin(\pi s_{13})}{\sin(\pi(s_{13}{+}s_{34}))}F^{(4,1)}(s_{13},s_{34})F^{(4,1)}(s_{25},s_{12}{+}s_{23}{+}s_{24}) \, .
\end{align}
With the known $\alpha'$-expansions of the four- and five-point integrals $F^{(4,1)}$
and $F^{(5,2)}_{ab}$ in open-string tree amplitudes, one can expand (\ref{Fn6p2initVal}) to any 
desired order. While the expansion of $F^{(4,1)}$ is given by (\ref{apexp.36}),
all-order results for $F^{(5,2)}$ can for instance be obtained from the methods in \cite{Broedel:2013aza, Boels:2013jua, Puhlfuerst:2015gta},
and certain orders are available for download from the website \cite{wwwap}.

\subsubsection{Further comments}
\label{appA4det}

Several entries of the initial value (\ref{Fn6p2initVal}) feature spurious poles such
as $(s_{12}{+}s_{24})^{-1}$ and $(s_{123}{+}s_{24}{+}s_{34})^{-1}$ within the individual 
terms of (\ref{exentry}). It is a strong consistency check of both the assembly of the initial value
and the $\ap$-expansion of the $F^{(5,2)}_{ab}$ that each order of $\lim_{z_4\rightarrow0} F^{(6,2)}$ in $\alpha'$ conspires to polynomials in $s_{ij}$. The coefficient of $\zeta_{2}$, for instance, has the following entries in the first, third and fifth line,
\begin{align}
(P_2^{(6,2)})_{1a} &=
\Big({-}s_{12} s_{23} {-} s_{12} s_{24} {-} s_{12} s_{34} {-} s_{13} s_{34} {-} s_{23} s_{34}, s_{13} s_{24}, 0, 0, 0, 
  0\Big)_a\,,
\notag \\
(P_2^{(6,2)})_{3a} &=
\Big(s_{12} s_{23} {-} s_{13} s_{34} {-} s_{23} s_{34} {-} s_{13} s_{35} {-} 
   s_{23} s_{35}, -s_{13} (s_{23} {+} s_{34} {+} s_{35}), \notag \\
   &\ \ \ \  -s_{12} s_{24} {-} s_{13} s_{35} {-} s_{23} s_{35} {-} 
   s_{34} s_{35}, 0, 0, 0\Big)_a\,,
 \\
(P_2^{(6,2)})_{5a} &= \Big( -s_{12} (s_{23} {+} s_{24} {+} s_{25} {+} s_{34} {+} s_{35}), 
  s_{13} (s_{23} {+} s_{24} {+} s_{25} {+} s_{34} {+} s_{35}), \notag \\
  &\ \ \ \  -s_{12} (s_{23} {+} s_{24} {+} s_{25} {+} s_{35}), 
  s_{13} s_{25}, \notag \\
  &\ \ \ \  -s_{12} s_{23} {-} s_{23} s_{24} {-} s_{12} s_{25} {-} s_{24} s_{25} {-} s_{12} s_{35} {-} 
   s_{13} s_{35} {-} s_{234} s_{35}, 
  s_{25} (s_{13} {+} s_{34})\Big)_a\,, \notag
\end{align}
while the remaining entries can be reconstructed from relabelling $2 \leftrightarrow 3$. The explicit form of the matrices $P_w^{(6,2)},M_w^{(6,2)}$ up to and including $w=9$ and the braid matrices in (\ref{62.5}) below can be found in an ancillary file within the \texttt{arXiv} submission of this work available at \url{https://arxiv.org/src/2102.06206/anc/AncillaryPsMsAndEsUpTo9.txt}.

In contrast to the initial values (\ref{apexp.40}) of the $F^{(n,1)}_{ab}$ which boil down to Riemann zeta values $\zeta_k$, the $F^{(n,p)}_{ab}$ with $p\geq 2$ involve irreducible MZVs at depth $\geq 2$ starting with $\zeta_{3,5}$. The MZVs in the $\ap$-expansion of $F^{(5,2)}$ known from string amplitudes \cite{Stieberger:2009rr, Schlotterer:2012ny} propagate to the initial value of $F^{(6,2)}$ as spelt out above. We have verified up to and including $\alpha'^8$ that the 
initial values (\ref{Fn6p2initVal}) obey the coaction principle (\ref{expa.16}), e.g.\ that the coefficient of $\zeta_{3,5}$ in $\mathbb P^{(6,2)}   \mathbb M^{(6,2)}$ is given by $\frac{1}{5}[ M^{(6,2)}_5 , M^{(6,2)}_3 ]$.
Moreover, the $\alpha'$-expansion at finite $z_4$ (cf.\ (\ref{coprop.8})),
\beq
F^{(6,2)}(z_4) = \mathbb P^{(6,2)}   \mathbb M^{(6,2)}  \mathbb G^{(6,2)}_{\{0,1\}}(z_4) \, ,
\label{coprop.8ex}
\eeq
involves the series in polylogarithms $\mathbb G^{(6,2)}_{\{0,1\}}(z_4) $
in (\ref{apexp.26}) that depends on the transposes 
$E_{0,z_4}^{(6,1)} = (e^{(6,2)}_{41})^t$
and $E_{1,z_4}^{(6,1)} = (e^{(6,2)}_{45})^t$ of the braid matrices
\begin{align}
e^{(6,2)}_{41} &=  \left( \begin{array}{cccccc}
s_{123}{+}s_{24}{+}s_{34} &0 &-s_{13}{-}s_{23} &-s_{12} &-s_{12} &s_{12} \\
0 &s_{123}{+}s_{24}{+}s_{34}  &-s_{13} &-s_{12}{-}s_{23} &s_{13} &-s_{13} \\
0 &0 &s_{12}{+}s_{24} &0 &-s_{12} &0 \\
0 &0 &0 &s_{13}{+}s_{34} &0 &-s_{13} \\
0&0&0&0&0&0 \\
0&0&0&0&0&0
\end{array} \right)\,, \notag \\
e^{(6,2)}_{45} &=   \left( \begin{array}{cccccc}
0&0&0&0&0&0 \\
0&0&0&0&0&0 \\
-s_{35} &0 &s_{35}{+}s_{34} &0 &0 &0 \\
0 &-s_{25} &0 &s_{25}{+}s_{24} &0 &0 \\
-s_{35} &s_{35} &-s_{23}{-}s_{25} &-s_{35} &s_{235}{+}s_{24}{+}s_{34} &0 \\
s_{25} &-s_{25} &-s_{25} &-s_{23}{-}s_{35} &0 &s_{235}{+}s_{24}{+}s_{34} \end{array} \right) \, .
\label{62.5}
\end{align}
We have checked that the combination  $\mathbb M^{(6,2)}  \mathbb G^{(6,2)}_{\{0,1\}}(z_4)$
obeys the formulation (\ref{C01.32}) of the coaction principle up to and including $\alpha'^{10}$. Moreover, since the $F_{ab}^{(6,2)}$ are the simplest instance where MZVs beyond depth one and polylogarithms coexist in the $\alpha'$-expansion,we highlight the following crosschecks at the $\alpha'^9$-order:
The coefficients of $G(1;z_4) \otimes  \zeta_{3} \zeta_{5}$ and $G(1;z_4) \otimes \zeta_{3,5}$ in $\Delta \mathbb G^{(6,2)}_{\{0,1\}}(z_4)$ are indeed given by $[[(e_{45}^{(6,2)})^t, M_5^{(6,2)}], M_3^{(6,2)}]$ and $ \tfrac{1}{5}[[M_3^{(6,2)}, M_5^{(6,2)}],  (e_{45}^{(6,2)})^t]$, respectively, in agreement with (\ref{C01.25}). 

Finally, the $F^{(5,2)}$ in five-point string amplitudes exhibit a first dropout among the MZVs at weight 18, which is due to the vanishing of $[ [M_3^{(5,2)}, M_5^{(5,2)}],[M_3^{(5,2)}, M_7^{(5,2)}]]$ \cite{Schlotterer:2012ny, Drummond:2013vz}. By their assembly from $(n,p)=(4,1),(5,2)$ integrals in (\ref{Fn6p2initVal}), the $F^{(6,2)}$ must share this dropout, and we have cross-checked its consistency with the coaction principle by verifying $[ [M_3^{(6,2)}, M_5^{(6,2)}],[M_3^{(6,2)}, M_7^{(6,2)}]]=0$.

Note that the soft limit $s_{24},s_{34}\rightarrow 0$ of $e^{(6,2)}_{41}, e^{(6,2)}_{45}$ in (\ref{62.5}) reproduces the five-point instances of the arguments of the $6\times6$ Drinfeld associator to assemble the $\ap$-expansion of $F^{(5,2)}$ \cite{Broedel:2013aza, AKtbp}.


\section{Braid group, monodromies and analytic continuation}
\label{app:B}

\subsection{Obtaining  $\mathbb X^{(n,p)}(g)$ for any $g \in S_{n-p}$}

In section \ref{sec:5} we have determined the analytic continuation of the $F^{(n,p)}$-integrals from $z_i<z_{i+1}$ to $z_{i+1}<z_i$ for unintegrated punctures $i=p{+}2,\ldots,n{-}2$. The group action of such braid operations $\sigma_{i,i+1}$ of neighboring punctures was explicitly given by matrices $\mathbb X(\sigma_{i,i+1})$ in (\ref{braidGeneralNP2}).
In this appendix we will discuss the composition of group operations $\mathbb X(g_1 g_2)$ for $g_1,g_2\in B_\mathrm{N}$ to reduce more general analytic continuations to the $\mathbb X(\sigma_{i,i+1})$, and mainly refer to \cite{kohno2002conformal} for facts about the braid group. This will also be relevant to show that $\mathbb X$ is indeed compatible with the group structure and that we can recover monodromies by doing the same braiding operation twice.

It is convenient to remember that there exists a canonical projection,
\beq
\textrm{proj} \colon \quad B_\mathrm{N} \,\rightarrow\, S_\mathrm{N}
\eeq
given by forgetting the details of how the punctures braid around each other.\footnote{One can also conveniently perform this map by replacing the generators of the braid group, $\sigma_{i,i+1}$ by transpositions $(i,i{+}1)\in S_\mathrm{N}$.} 
Let us call $g^{\rm pr} \in S_\mathrm{N}$ the image of an element $g \in B_\mathrm{N}$ under this projection. Then, we can rewrite the content of (\ref{braidGeneralNP}) as follows:
\be
 \tilde{\mathcal G}^{(n,p)} \left(\sigma^{\textrm{pr}}_{i,i+1}\left(z_{p+2},\ldots, z_{n-2}\right)   \right) = \mathbb X^{(n,p)}(\sigma_{i,i+1}) \mathcal{G}^{(n,p)}\left(z_{p+2},\ldots, z_{n-2}\right) \ ,
\label{eqBraidActionGenerator}
\ee
where the permutation $\sigma^{\textrm{pr}}_{i,i+1}$ acts on the indices of the punctures  $z_i$. In (\ref{eqBraidActionGenerator}) we can describe the braiding due to the element $\sigma^{-1}_{i,i+1}$ by changing the sign in the exponential in $\mathbb X^{(n,p)}(\sigma_{i,i+1})$. For $g\in B_{n-p}$, we can generalize (\ref{eqBraidActionGenerator}) to
\beq
\tilde{\mathcal G}^{(n,p)} \left(g^\textrm{pr}\left(z_{p+2},\ldots, z_{n-2}\right)   \right) = \mathbb X^{(n,p)}(g) \mathcal{G}^{(n,p)}\left(z_{p+2},\ldots, z_{n-2}\right) \, ,
\label{eqBraidActionAnyElement}
\eeq
where we can define  $\mathbb X^{(n,p)}(g)$ recursively for the following formula for composing two group elements, $g_1$,  $g_2 \in B_{n-p}$:\footnote{We are using a convention of composition of braidings and permutations consistent with $ \sigma^\textrm{pr}_{34}\sigma^{\textrm{pr}}_{45} = (34)(45)=(345)$.}
\begin{align}
\tilde{\mathcal G}^{(n,p)} \big( (g_1^\textrm{pr} g_2^\textrm{pr}) \left(z_{p+2},\ldots, z_{n-2}\right)   \big)&=
\tilde{\mathcal G}^{(n,p)} \big( g_1^\textrm{pr} \left( g_2^\textrm{pr} \left(z_{p+2},\ldots, z_{n-2}\right) \right)  \big)
\label{eqBraidActionTwoElements}\\
&= g_1^\textrm{pr}\left( \mathbb{X}^{(n,p)}(g_2) \right) \mathbb{X}^{(n,p)}(g_1)  \mathcal{G}^{(n,p)}\left(z_{p+2},\ldots, z_{n-2}\right) \, . \notag
\end{align}
The permutation $g_1^\textrm{pr}$ acts on the indices of the braid matrices in $\mathbb{X}^{(n,p)}(g_2) $, and not in any way on the signs of the exponentials in this expression. From equation (\ref{eqBraidActionTwoElements}) we can read off a formula for $\mathbb X^{(n,p)}(g_1 g_2)$:
\beq
\mathbb X^{(n,p)}(g_1 g_2)=g_1^\textrm{pr}\left( \mathbb{X}^{(n,p)}(g_2) \right) \mathbb{X}^{(n,p)}(g_1) \, .
\eeq
Because we can decompose any $g\in B_{n-p}$ into generators $\sigma_{i,i+1}$ of the braid group, and we know the form of $\mathbb{X}^{(n,p)}(\sigma_{i,i+1})$, we have obtained a prescription to compute any $\mathbb{X}^{(n,p)}(g)$.

As a sanity check, we should verify that $\mathbb X ^{(n,p)}$ satisfies the equations of the presentation of the braid group, (\ref{eqPresentationBraidGroup}). The first of these equations, exemplified in, $\mathbb X^{(n,p)}(\sigma_{3,4}\sigma_{5,6}) =\mathbb X^{(n,p)}(\sigma_{5,6}\sigma_{3,4}) $ follows easily from the algebra of braid matrices. We found the second of these equations, exemplified by $\mathbb X^{(n,p)}(\sigma_{3,4}\sigma_{4,5}\sigma_{3,4}) =\mathbb X^{(n,p)}(\sigma_{4,5}\sigma_{3,4}\sigma_{4,5}) $, harder to prove in general, but checked explicitly that it holds for $(n,p)=(7,1)$ up to weight $\ap^8$.

\subsection{Example: Monodromies from braiding twice in $(n,p)=(5,1)$}
\label{appbtwo}

The monodromies of $\mathcal{G}^{({n,p})}$ can be identified by the kernel of $\textrm{proj}$, which is a normal subgroup of $B_{n-p}$, called the pure braid group, $PB_{n-p}$. From our description of the generators of $B_{n-p}$, the simplest elements to describe in $PB_{n-p}$ are the squares of the generators, $\sigma^2_{i,i+1}$. Using $\mathbb{X}^{(5,1)}(\sigma_{1,3}) = \textrm{exp}(i \pi E^{(5,1)}_{3,1})$, the effect of braiding twice is described in a way consistent with \eqref{eqn5p1monodromy}:\footnote{Note that in the $(n,p)=(5,1)$ case, there is no puncture $z_2$ after integration, so punctures $z_1=0$ and $z_3$ are neighbors. Furthermore, notice that we are braiding a puncture $\mathrm{SL}(2,\mathbb{C})$-fixed to 0,  which could cause some problems due to regularization of terms $G(\vec{u};0)$ with $\vec{u} \in \{1,z \}^\times$. This is not a big problem if the end result lies in $PB_{n-p}$.}
\beq
\mathbb{X}^{(5,1)} (\sigma^2_{1,3}) = \mathcal{M}_{0,z_3} \, .
\eeq

\subsection{Example: Analytic continuation from two braidings}

For concreteness, we shall discuss an example with $(n,p)=(7,1)$ and suppress this superscript. We will analytically continue from the integration domain $0<z_3<z_4<z_5<1$ into $0<z_4<z_5<z_3<1$ via two generators $\sigma_{i,i+1}$, where $z_{i+1}$ is again taken counterclockwise around $z_i$ in both cases. Expanding (\ref{eqBraidActionTwoElements}) for $(g_1,g_2)=(\sigma_{3,4},\sigma_{4,5})$:
\begin{align}
\tilde{\mathcal G}(z_4,z_5,z_3) &= \mathbb{X}(\sigma_{4,5}) \big|_{3 \leftrightarrow 4} \mathbb{X}(\sigma_{3,4}) \mathcal G (z_3,z_4,z_5)  \label{analyticContinuation3punctures} 
\\
&=\Phi(E_{51}{+}E_{45},E_{35})\exp{(i \pi E_{35})}\Phi(E_{35},E_{31}{+}E_{34}) \notag 
\\
& \ \ \ \ \times \Phi(E_{41},E_{34})\exp{(i \pi E_{34})}\Phi(E_{34},E_{31}) \mathcal G(z_3,z_4,z_5) \notag \, .
\end{align}
We have checked this equation to be consistent with the conventions of \texttt{PolyLogTools} \cite{Duhr2019PolylogTools} when 
the respective arguments obey $\arg(z_3)>\arg(z_4)>\arg(z_5)$.
We have  performed such checks for several terms up to and including $\ap^5$, i.e. for MPLs up to weight 5. One can generate equations valid in other regions of $\{z_3,z_4,z_5\}$ by just changing the signs of the exponentials in (\ref{analyticContinuation3punctures}), or equivalently, by using the inverses of the braid operations, $\sigma^{-1}_{3,4}$ or $\sigma^{-1}_{4,5}$. For instance, the analogue of (\ref{analyticContinuation3punctures}) for the braiding $(\sigma^{-1}_{3,4},\sigma_{4,5})$ is consistent in the region where
$\arg(z_4)>\arg(z_3)>\arg(z_5)$.

\subsection{Initial values in an alternative fibration basis of polylogarithms}
\label{tildepms}

We shall here spell out examples of the modified initial conditions of the $F^{(6,1)}$ in section \ref{initbraid} that arise from a change of fibration basis for the polylogarithms in (\ref{61ancont.1}). More specifically, the simplest instances of the matrices $\tilde P_w^{(6,1)}$ and $\tilde M_{2k+1}^{(6,1)}$ in (\ref{newini.2}) read
\begin{align}
\tilde P_1^{(6,1)} &= \cccb 0 &s_{24} &0
\\0 & -(s_{23} {+} s_{24}) &0 \\
0 &s_{23} &0 \ccce = - (e_{34}^{(6,1)} )^t\,,
\notag \\
\tilde P_2^{(6,1)} &= \cccb
 -s_{12} (s_{23} {+} 2 s_{24}) &-(2 s_{12} {-} s_{23} {-} 3 s_{24}) s_{24} &0 \\
 s_{12} (s_{23} {+} s_{24})
 &-3 s_{23}^2 {+} s_{12} s_{24} {-} 7 s_{23} s_{24} {-} 3 s_{24}^2 & 0 \\
 -s_{12} ( s_{23}{+}s_{2,345})
 &\ \ 3 s_{23}^2 {-} s_{12} s_{24} {+} 2 s_{23} s_{24} {-} s_{13,2} s_{25} \ \ &-s_{134,2}s_{25}
\ccce\,,
 \\
\tilde P_3^{(6,1)} &=
\cccb
s_{12} s_{24} (s_{23} {+} s_{24}) &(s_{12} {-} s_{23} {-} s_{24}) s_{24}^2 &0 \\
-s_{12} (s_{23} {+} s_{24})^2
& \ \ s_{2,34 } (s_{23}^2 {-} s_{12} s_{24} {+} 
3 s_{23} s_{24} {+} s_{24}^2)\ \ &0 \\
s_{12} s_{23} (s_{23} {+} s_{24}) &
-s_{23} (s_{23} s_{2,34} {-} s_{12} s_{24} ) &0
\ccce\,,
\notag \\
\tilde M_3^{(6,1)} &=
\cccb
s_{12} (
s_{134,2}s_{23}{+} s_{24}^2)
&s_{24} ( s_{12} s_{24} {-}s_{134,2}s_{23})
     &0 \\
s_{12} (s_{12} s_{24} {-} s_{134,2} s_{23})
&s_{24} (s_{12}^2 {+} s_{134,2} s_{23}  )
& 0 \\
s_{12} (s_{23}^2  
{+}s_{1345,2}s_{25}
{-} s_{12} s_{24})
& \ \ s_{13,2}   s_{25}  s_{1345,2}
 {-}(s_{12}^2{+} s_{23}^2) s_{24} \ \
&s_{134,2}s_{25}s_{1345,2}
\ccce \, ,
\notag 
\end{align}
where we use the shorthand $s_{ij \ldots k,2} = s_{i2}{+}s_{j2}{+}\ldots{+}s_{k2}$.
Note that $\tilde P_2^{(6,1)}$ and $\tilde M_{3}^{(6,1)}$ evidently differ from the
earlier $P_2^{(6,1)}$ and $ M_{3}^{(6,1)}$
\begin{align}
 P_2^{(6,1)} &=- \cccb
s_{12} s_{23} &0 &0 \\
s_{12} (s_{23} {+} s_{24})&(s_{12} {+} s_{23}) s_{24}&0 \\ 
s_{12} (s_{24} {+} s_{25})&\ \ (s_{12} {+} s_{23}) (s_{24} {+} s_{25}) \ \
&(s_{12} {+} s_{23} {+} s_{24}) s_{25}
 \ccce\,,
 \\
 M_3^{(6,1)} &= \cccb s_{12} s_{23} (s_{12} {+} s_{23})&0& 0 \\
 s_{12} ( 
 s_{134,2} s_{24}{-}s_{12} s_{23})& 
 s_{13,2}s_{24} 
 s_{134,2}& 0
    \\ 
     s_{12} (s_{1345,2} s_{25} - s_{13,2}s_{24})
     &\ \ s_{13,2}  (s_{1345,2} s_{25}-s_{13,2}s_{24}) \ \
     &s_{134,2}
     s_{25} s_{1345,2}
     \ccce\,,
\notag 
\end{align}
which are extracted from the initial conditions (\ref{general.29}) and tailored to an $\ap$-expansion in the fibration bases of (\ref{apexp.27}).

\linespread{1.1}

\bibliographystyle{JHEP}
\bibliography{references}

\end{document}